\documentstyle[sprocl,epsf,rotating]{article}
\input epsf
\input axodraw.sty 
\def\beq{\begin{equation}}
\def\eeq{\end{equation}}

\def\beqn{\begin{eqnarray}}
\def\eeqn{\end{eqnarray}}
 
\def\qsq{q^2}

\newcommand {\mwsq}  {m_W^2}

\newcommand {\mhsq}  {m_H^2}

\newcommand {\call}  {{\cal L}}

\newcommand {\kapgam}  {\kappa_\gamma}
\newcommand {\kapz}    {\kappa_Z}
\newcommand {\lamgam}  {\lambda_\gamma}
\newcommand {\lamz}    {\lambda_Z}

\newcommand {\gonez}   {g_1^Z}

\newcommand {\loglammz} {\log\bigg(\frac{\Lambda^2}{m_Z^2}\bigg)}
\begin{document}
\title{ INTRODUCTION TO  ELECTROWEAK SYMMETRY BREAKING 
}
\author{ S.~Dawson}
\address{  Physics Department,\\
               Brookhaven National Laboratory,\\  Upton, NY 11973}
\maketitle
\abstracts{ 
An introduction to the physics of electroweak
symmetry breaking is
given.  We discuss Higgs boson production in $e^+e^-$  and 
hadronic collisions and survey search techniques at present and future
accelerators.    Indirect limits on the Higgs boson mass
from triviality arguments, vacuum
stability,  and precision electroweak  measurements are presented.
An
 effective Lagrangian, valid when there is no low mass Higgs
boson, is used to discuss the physics of a strongly interacting
electroweak symmetry breaking sector.
Finally, the Higgs bosons of the minimal supersymmetric model are 
considered, along with the resulting differences in phenomenology
from the Standard Model.
}

\section{Introduction}
The search for the Higgs boson has become  a major focus of
all particle accelerators. In the simplest version of the
electroweak theory, the Higgs boson serves both to give 
the $W$ and $Z$ bosons their masses and to give the fermions
mass.   It is thus a vital part of the theory.
   In these lectures, we will introduce
the Higgs boson of the Standard 
Model of electroweak interactions.\cite{hhg,bag}

Section 2 contains a derivation of the Higgs mechanism, with particular
attention to the choice of gauge.  In Section 3 we 
 discuss  indirect limits on the Higgs boson mass coming
from theoretical arguments and from precision measurements at the LEP
and LEP2 
colliders.   The production of the Standard Model Higgs boson is
then summarized in Sections 4 - 8, beginning with a discussion
of the Higgs boson branching ratios in Section 4.
Higgs production in $e^+e^-$ collisions at LEP and LEP2
 and in hadronic
collisions at the Tevatron and the LHC
 are discussed in Sections 5 and 6, with an emphasis on
the potential for discovery in the different channels.

  Section 7 contains a derivation of the effective
$W$ approximation and a discussion of Higgs production through vector
boson fusion at the LHC.
The potential for a Higgs boson discovery at
a very  high energy $e^+e^-$ collider, ($\sqrt{s}>500~GeV$) is discussed in
Section 8.

Suppose the Higgs boson is not discovered in an
$e^+ e^-$ collider or at the LHC?  Does this mean the
Standard Model with a Higgs boson must be abandoned?  In
Section 9, we discuss the implications of a very heavy Higgs boson,
($M_h>> 800~GeV$).  In this regime the $W$ and $Z$ gauge bosons
are strongly interacting and new techniques must be used.  We present
an effective Lagrangian valid for the case where $M_h>>\sqrt{s}$.

Section 10  contains 
a list of some of the objections which many theorists
 have to the minimal Standard Model
with a single Higgs boson.  One of the most popular alternatives to the  
minimal Standard Model is to make the theory supersymmetric.
The Higgs sector of the minimal supersymmetric model (MSSM)
is surveyed
in Section 11. 
We end with some conclusions in
Section 12.

\section{The Higgs Mechanism}
\subsection{Abelian Higgs Model}
 
The central question of electroweak physics is :``Why are the $W$ and $Z$
boson masses non-zero?"  The measured values,
$M_W=80~GeV$ and $M_Z=91~GeV$, are
 far from zero and cannot be considered as 
small effects.
  To see that this is a problem, we consider
a $U(1)$ gauge theory with a single gauge field, the photon.  The Lagrangian
is simply\cite{quigg}
\beq
{\cal L}=-{1\over 4} F_{\mu \nu}F^{\mu\nu},
\eeq
where
\beq F_{\mu\nu}=\partial_\nu A_\mu-\partial _\mu A_\nu.
\eeq
The statement of local $U(1)$ gauge invariance is that the Lagrangian
is invariant under the transformation:$A_\mu(x)
\rightarrow A_\mu(x)-
\partial_\mu \eta(x)$ for any $\eta$ and $x$.   Suppose we now add a
mass term for the photon to the Lagrangian,
\beq
{\cal L}=-{1\over 4} F_{\mu \nu}F^{\mu\nu}+{1\over 2}m^2 A_\mu A^\mu
.
\eeq
It is easy to see that the mass term violates the local gauge invariance.
It is thus  the $U(1)$
gauge invariance which requires the photon to be massless.

We can extend the model by adding a single complex scalar field
with charge $-e$
which couples to the photon.
The Lagrangian is now,
\beq
{\cal L}=-{1\over 4} F_{\mu\nu} F^{\mu\nu}+\mid D_\mu\phi\mid^2
-V(\phi),
\eeq
where
\beqn D_\mu & =&\partial_\mu -i e A_\mu \nonumber \\
V(\phi) &=& \mu^2 \mid \phi\mid^2+\lambda(\mid \phi\mid^2)^2 . \eeqn
$V(\phi)$ is the most general renormalizable potential allowed by
the $U(1)$  gauge invariance.

This Lagrangian is invariant under global $U(1)$ rotations, $\phi
\rightarrow e^{i\theta}\phi$, and also  under local gauge transformations:
\beqn
A_\mu(x) &\rightarrow & A_\mu(x)-\partial _\mu \eta(x) \nonumber \\
\phi(x) &\rightarrow & e^{-i e \eta(x)} \phi(x).\\
\eeqn

\begin{figure}[t]
\vskip -.5in 
\centering   
\epsfxsize=3.in
\leavevmode
\epsffile{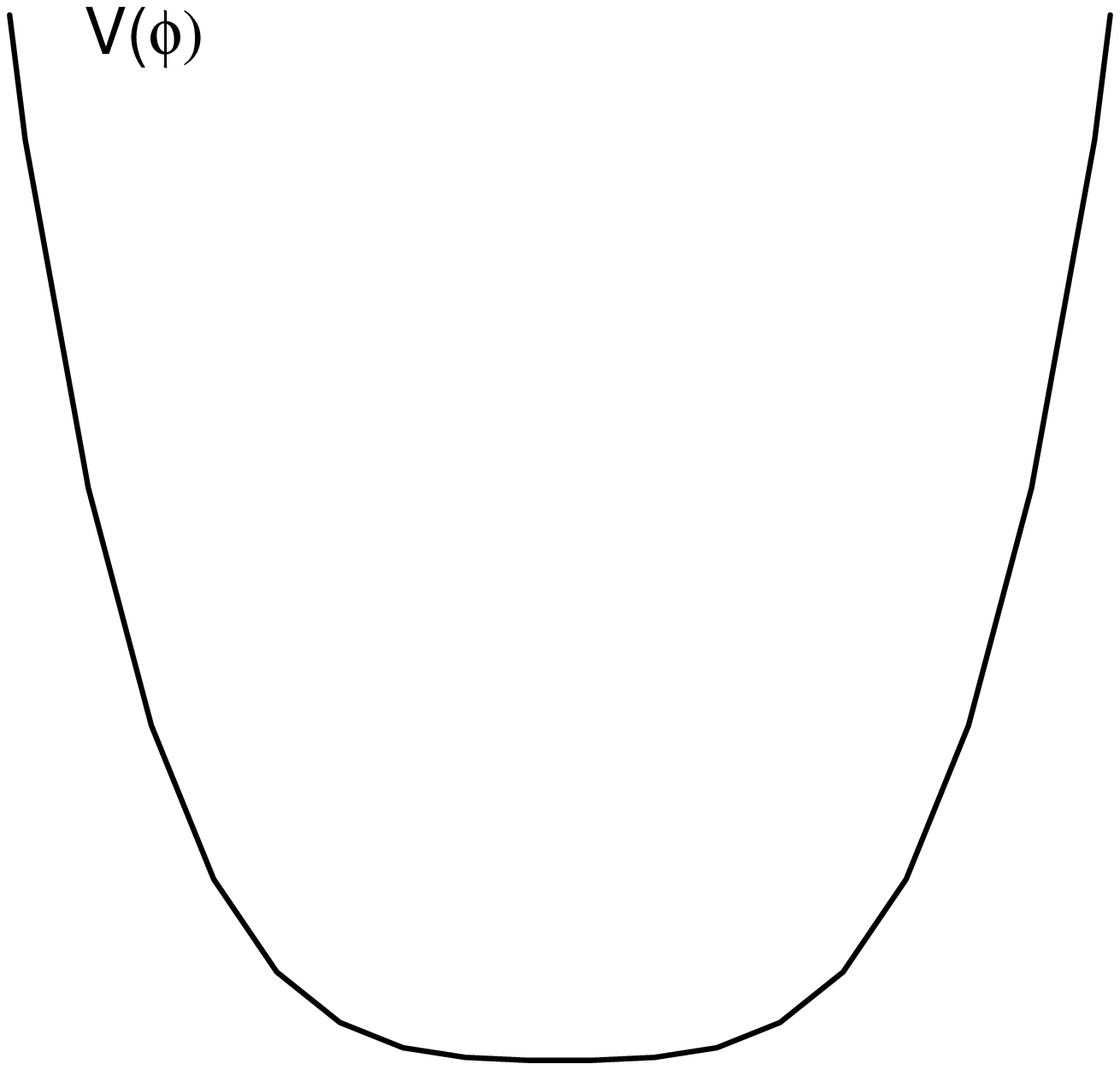}
\vskip -.2in 
\caption{Scalar potential with $\mu^2>0$.}
\end{figure}  

\begin{figure}[t]
\vskip -.5in 
\centering
\epsfxsize=4.in
\leavevmode
\epsffile{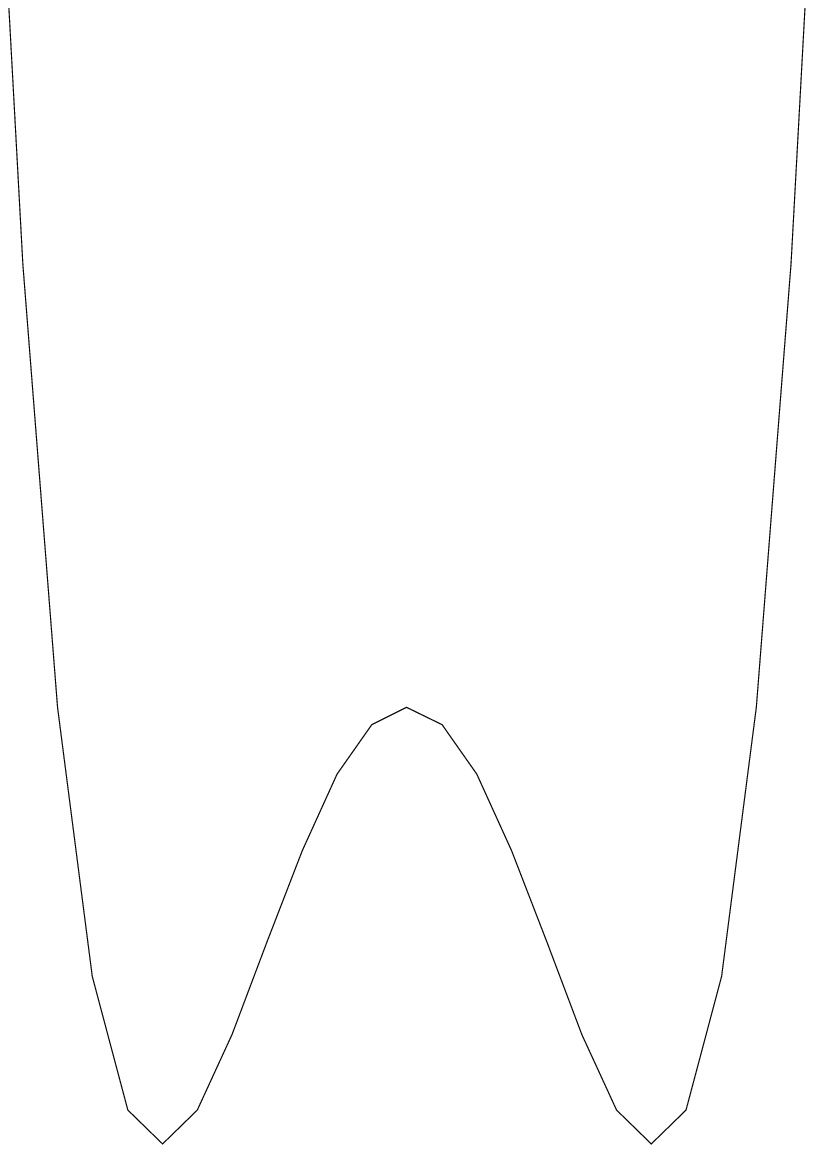}
\vskip -.5in 
\caption{Scalar potential with $\mu^2<0$.}
\end{figure}

There are now two possibilities for the theory.\footnote{We
assume $\lambda>0$. If $\lambda<0$, the potential is unbounded
from below and has no state of minimum energy.}  If $\mu^2>0$ the
potential has the shape shown in Fig. 1 and preserves the symmetries
of the Lagrangian.  
The state of lowest energy is that with $\phi=0$, the vacuum state.
The theory is simply quantum electrodynamics
with a massless photon and a charged
scalar field $\phi$ with mass $\mu$.

The alternative scenario
 is more interesting.  In this case $\mu^2<0$ and the 
potential can be written as, \cite{quigg,ablee}
\beq V(\phi)=-\mid \mu^2\mid \mid \phi\mid^2 +
\lambda (\mid \phi\mid^2)^2,
\eeq
which has the Mexican hat shape shown in Fig. 2.  In  this case
the minimum energy state is not at $\phi=0$ but rather at
\beq
\langle \phi\rangle=\sqrt{-{\mu^2\over 2 \lambda}}
\equiv {v\over\sqrt{2}}.
\eeq
$\langle \phi\rangle$ is called the vacuum expectation value (VEV)
of $\phi$.
Note that the direction in which the vacuum is chosen is 
arbitrary, but it is conventional to choose it to lie along
the direction of the real part of $\phi$.
The VEV then clearly breaks the global $U(1)$ symmetry.  
 
It is convenient to rewrite $\phi$ as
\beq \phi\equiv {1\over \sqrt{2}}e^{i {\chi\over v}} 
\biggl(v+h\biggr),
\label{phidef}
\eeq
where $\chi$ and $h$ are real fields which have no VEVs.
If we substitute Eq.~\ref{phidef}  back
 into the original Lagrangian, 
the interactions in terms of the fields with no VEVs
can be found,
\beqn
{\cal L} & = &
-{1\over 4} F_{\mu\nu}F^{\mu\nu} - e v A_{\mu}\partial^{\mu}\chi
+{e^2 v^2\over 2} A_{\mu} A^\mu \nonumber \cr
&&+{1\over 2} \biggl( \partial_\mu h \partial^\mu h
+ 2 \mu^2 h^2\biggr)
+{1\over 2} \partial_\mu\chi\partial^\mu\chi\nonumber \\
&&+(h,~\chi {\rm ~interactions}).
\label{ints}
\eeqn
Eq.~\ref{ints}
 describes a theory with a photon of mass $M_A=ev$, a scalar field
$h$ with mass-squared $-2 \mu^2>0$, and a massless scalar field $\chi$.
The mixed $\chi-A$ propagator is confusing, however.  This term can be
removed by making a gauge transformation:
\beq
A^\prime_\mu\equiv A_\mu -{1\over e v} \partial_\mu\chi.
\label{gauget}
\eeq
After making the gauge transformation of Eq.~\ref{gauget}
 the $\chi$ field disappears from the theory and we
say that it has been ``eaten'' to give the photon mass.
This is called the Higgs mechanism and   
the $\chi$ field is often called a Goldstone 
boson.\cite{gold}
 In the gauge of Eq. \ref{gauget}
the particle content of the theory
 is  apparent; a massive photon and
a scalar field $h$, which we call
a Higgs boson.
The Higgs mechanism can be summarized by saying that the spontaneous
breaking of a gauge theory by a non-zero VEV results in the
disappearance of a Goldstone boson and its
transformation into the longitudinal component of a massive 
gauge boson.

It is instructive to count the number of
degrees of freedom (dof).  Before the 
spontaneous symmetry breaking there was a massless photon (2 dof) and
a complex scalar field (2 dof) for a total of
 4 degrees of freedom.\footnote{Massless gauge fields have 2 transverse
degrees of freedom, while a massive gauge field has an additional
longitudinal degree of freedom.}
After the spontaneous symmetry breaking there is a massive photon
(3 dof) and a real scalar, $h$, (1 dof) for the same total
number of  degrees
of freedom.

At this point let us consider the gauge dependance of these
results.  The gauge choice above with the transformation 
$A_\mu^\prime=A_\mu -{1\over e v }\partial_\mu\chi$ is called the
unitary gauge.  This gauge has the advantage that the particle
spectrum is obvious and there is no $\chi$ field.
The unitary gauge, however, has the disadvantage that the 
photon 
propagator, $\Delta_{\mu\nu}(k)$,
 has bad high energy behaviour,
\beq
\Delta_{\mu\nu}(k)=-{i\over k^2-M_A^2}\biggl( g_{\mu\nu}-{k^{\mu}k^{\nu}
\over M_A^2}\biggr).
\eeq
In the unitary gauge, scattering cross sections have contributions which
grow with powers of $k^2$ (such as $k^4$, $k^6$, etc.) which cannot
be removed by the conventional mass, coupling constant, and wavefunction
renormalizations.
More convenient gauges are the $R_{\xi}$ gauges
which are obtained by adding the gauge fixing term to
the Lagrangian,\cite{ablee}
\beq {\cal L}_{GF}=-{1\over 2 \xi}\biggl(\partial_\mu A^\mu+\xi
e v \chi\biggl)^2.
\label{rtsi}
\eeq
Different choices for $\xi$ correspond to different gauges.
In the limit $\xi\rightarrow \infty$ the unitary gauge is recovered.
Note that after integration by parts
 the cross term in Eq. ~\ref{rtsi}  exactly cancels the mixed
$\chi \partial_\mu A^\mu$ term of Eq. \ref{ints}.
The gauge boson propagator in $R_\xi$ gauge is given by
\beq \Delta_{\mu\nu}(k)=-{i\over k^2-M_A^2}
\biggl( g_{\mu\nu}-{(1-\xi)k_{\mu}k_{\nu}\over k^2-\xi M_A^2}\biggr).
\eeq
 In the $R_\xi$ gauges the $\chi$ field is part of the spectrum
and has mass $M_{\chi}^2=\xi M_A^2$.  Feynman gauge corresponds to the
choice $\xi=1$  and has a massive Goldstone boson, $\chi$, 
while Landau gauge has $\xi=0$ and  the Goldstone boson
$\chi$ is massless with no
coupling to the physical Higgs boson.
The Landau gauge is often the most convenient for calculations involving
the Higgs boson since there is no coupling to the unphysical $\chi$ field.

\subsection{Weinberg-Salam Model}

It is now straightforward to obtain the usual Weinberg-Salam model
of electroweak interactions.\cite{ws}
The Weinberg- Salam model is an $SU(2)_L \times U(1)_Y$ gauge theory containing
three $SU(2)_L$ gauge bosons, $W_\mu^i$, $i=1,2,3$, and one $U(1)_Y$
gauge boson, $B_\mu$, with  kinetic energy terms,
\beq
{\cal L}_{\rm KE} =-{1\over 4}W_{\mu\nu}^i W^{\mu\nu i}
-{1\over 4} B_{\mu\nu} B^{\mu\nu}\quad ,
\eeq
where
\beqn
W_{\mu\nu}^i&=& \partial_\nu W_\mu^i-\partial _\mu W_\nu^i
+g \epsilon^{ijk}W_\mu^j W_\nu^k \quad ,
\nonumber \\
B_{\mu\nu}&=&\partial_\nu B_\mu-\partial_\mu B_\nu\quad .
\eeqn
Coupled to the gauge fields is a complex scalar $SU(2)$
doublet, $\Phi$,
\beq
\Phi
= \left(\begin{array}{c}
 \phi^+   \\
 \phi^0   \end{array}\right)
\eeq
with a  scalar potential given by
\beq
 V(\Phi)=\mu^2 \mid \Phi^\dagger\Phi\mid +\lambda
\biggl(\mid \Phi^\dagger \Phi\mid\biggr)^2\quad ,
\label{wspot}
\eeq
($\lambda>0$).
This is the most general renormalizable and $SU(2)_L$ invariant
potential allowed.

Just as in the Abelian model of Section 2.1, the state of minimum
energy for $\mu^2<0$ is not at $\Phi=0$ and the scalar field develops
a VEV.
The direction of the minimum in $SU(2)_L$ space is not determined
since the potential depends only on 
the combination 
$\Phi^\dagger \Phi$
and we arbitrarily choose
\beq
\langle \Phi\rangle
= {1\over\sqrt{2}} \left(\begin{array}{c}
 0   \\
 v   \end{array}\right)\quad .
\label{vevdef}
\eeq
With this choice the scalar doublet has $U(1)_Y$ charge
(hypercharge) $Y_\Phi=1$ and the
 electromagnetic charge is\footnote{The $\tau_i$ are
 the Pauli matrices with $Tr(\tau_i\tau_j)
=2\delta_{ij}$.}
\beq
Q={(\tau_3 +Y)\over 2} \,\,.
\label{qdef}
\eeq
  Therefore,
\beq 
Q \langle \Phi\rangle
= 0
\eeq
and electromagnetism is unbroken by the scalar VEV.
The VEV of Eq. \ref{vevdef} hence yields the desired symmetry breaking
scheme, 
\beq
SU(2)_L\times U(1)_Y\rightarrow U(1)_{EM}.
\eeq

It is now straightforward to see how the Higgs mechanism
generates masses for the $W$ and $Z$ gauge bosons in the
same fashion as a mass was generated for the
photon in the Abelian Higgs model of Section 2.1.  The
contribution of the scalar doublet to the Lagrangian is,
\beq
{\cal L}_s=(D^\mu \Phi)^\dagger (D_\mu \Phi)-V(\Phi)\quad ,
\label{scalepot}
\eeq
where\footnote{ Different choices for the gauge kinetic energy
and the covariant derivative depend on whether $g$ and $g^\prime$
are chosen positive or negative.  There is no physical consequence
of this choice.}
\beq
D_\mu=\partial_\mu +i {g\over 2}\tau\cdot W_\mu+i{g^\prime\over 2}
B_\mu Y.
\eeq
In unitary gauge there are no Goldstone
bosons and only the physical Higgs scalar remains in the spectrum
after the spontaneous symmetry breaking has occurred.
Therefore the scalar doublet in unitary gauge can be written as
\beq
\Phi={1\over \sqrt{2}}\left(\begin{array}{c}  0 \\
 v+h\end{array}\right)                  \quad ,
\eeq
which gives the contribution to the gauge boson masses
from the scalar kinetic energy term of Eq. \ref{scalepot},
\beq
{1\over 2} (0 ,v )
\biggl({1\over 2}g \tau\cdot W_\mu
+{1\over 2} g^\prime B_\mu
\biggr)^2 \left(\begin{array}{c}  0 \\  v \end{array}
\right).
\eeq
The physical gauge fields are then two
charged fields, $W^\pm$, and two  neutral gauge
bosons, $Z$ and $\gamma$.  
\beqn
W^{\pm}_\mu&=&
{1\over \sqrt{2}}(W_\mu^1 \mp i W_\mu^2)\nonumber \\
Z^\mu&=& {-g^\prime B_\mu+ g W_\mu^3\over \sqrt{g^2+g^{\prime~2}}}
\nonumber \\
A^\mu&=& {g B_\mu+ g^{\prime} W_\mu^3\over \sqrt{g^2+g^{\prime~2}}}.
\label{masseig}
\eeqn
The gauge bosons   obtain  masses
 from the Higgs mechanism:
\beqn
M_W^2 &=& {1\over 4} g^2 v^2\nonumber \\
M_Z^2 &=& {1\over 4} (g^2 + g^{\prime~2})v^2\nonumber \\
M_A& = & 0.
\eeqn

Since the massless photon must couple with electromagnetic
strength, $e$, the coupling constants 
define the weak mixing angle $\theta_W$,
\beqn
e&=& g \sin\theta_W \nonumber \\
e&=& g^\prime \cos\theta_W
\quad .
\eeqn

It is 
instructive to count the degrees of freedom after the spontaneous
symmetry breaking has occurred.  We began with a complex scalar 
$SU(2)_L$ doublet
$\Phi$ with four degrees of freedom, a massless $SU(2)_L$ gauge field,
$W_i$, with six degrees of freedom and a massless $U(1)_Y$ gauge field,
$B$, with 2 degrees of freedom for a total of $12$.  After the
spontaneous symmetry breaking there remains a physical real scalar field
$h$ ($1$ degree of freedom),  massive $W$ and $Z$ fields ($9$
degrees of freedom), and a massless photon ($2$ degrees of freedom).
We say that the scalar degrees of freedom have been ``eaten'' to
give the $W^\pm$ and $Z$ gauge bosons their longitudinal components.

Just as in the case of the Abelian Higgs model, if we go to
a gauge other than unitary gauge, there will be Goldstone
bosons in the spectrum and the scalar field can be written,
\beq
\Phi={1\over \sqrt{2}}
 e^{i{\omega\cdot\tau\over 2 v}}
\left(\begin{array}{c}  0 \\
 v+h\end{array}\right).
\eeq 
In the Standard Model, there are three Goldstone bosons,
${\vec \omega}=(\omega^\pm,z)$, with masses $M_W$ and $M_Z$ in
the Feynman gauge.  These Goldstone bosons will play an important
role in our understanding of the physics of a very heavy Higgs
boson, as we will discuss in Section 9.

Fermions can easily be included in the theory and we will consider
the electron and its neutrino as an example.  
It is convenient to write the fermions in terms of their left-
and right-handed projections,
\beq
\psi_{L,R}={1\over 2}(1\mp\gamma_5)\psi
\,\, .
\eeq
From the four-Fermi theory of weak interactions, we know that the
$W$-boson couples only to left-handed fermions and so we construct the 
$SU(2)_L$ doublet, 
\beq
L_L=
\left(\begin{array}{c}
\nu_L\\ e_L \end{array}\right)
.
\eeq
From Eq. \ref{qdef}, the hypercharge of the lepton doublet
must be $Y_L=-1$. 
Since the neutrino is (at least approximately) massless, it can have only
one helicity state which is taken to be $\nu_L$.  Experimentally,
we know that
right-handed fields do not interact
with the $W$ boson, and so the right-handed electron, $e_R$, must be an
$SU(2)_L$ singlet and  so has $Y_{e_R}=-2$.  Using these hypercharge
assignments, the leptons can be coupled
in a gauge invariant manner
 to the $SU(2)_L\times
U(1)_Y$ gauge fields,
\beq
{\cal L}_{lepton}
=i {\overline e}_R \gamma^\mu
\biggl(\partial_\mu+i{g^\prime\over 2}Y_e B_\mu\biggr)e_R+
i{\overline L}_L\gamma^\mu
\biggl(\partial_\mu+i {g\over 2}\tau\cdot W_\mu
+i{g^\prime\over 2}Y_L
B_\mu\biggr)L_L\,\,.
\eeq
All of the known fermions can be accommodated in the Standard
Model in an identical manner as was done for the leptons. 
The $SU(2)_L$ and $U(1)_Y$ charge assignments of
the first generation of fermions 
are given in Table 1. 
\begin{table}[t]
\begin{center}
{Table 1: Fermion Fields of the Standard Model}
\vskip6pt
\renewcommand\arraystretch{1.2}
\begin{tabular}{|lccr|}
\hline
\multicolumn{1}{|c}{Field}& SU(3)& $SU(2)_L$& $U(1)_Y$
\\
\hline
$Q_L=\left(\begin{array}{c}
u_L\\ d_L \end{array}\right)$
   &    $3$          & $2$&  $~{1\over 3}$
\\
$u_R$ & $3$ & $1$& ${4\over 3}$
\\
$ d_R$ & $ 3$ & $1$&  $~-{2\over 3}$
\\
$L_L=\left(\begin{array}{c}
\nu_L\\ e_L \end{array}\right)
$  & $1$             & $2$& $~-1$
\\
$e_R$ & $1$             & $1$& $~-2$ 
\\
$\Phi= \left(\begin{array}{c}
\phi^+\\ \phi^0 \end{array}\right)
$  & $1$             & $2$& $1$ 
\\
\hline
\end{tabular}
\end{center}
\end{table}

A fermion mass term takes the form
\beq
{\cal L}_{mass}=-m{\overline{\psi}}\psi=-m
\biggl({\overline{\psi}}_L\psi_R+
{\overline{\psi}}_R\psi_L\biggr) 
\,\, .
\label{massterm}
\eeq
As is obvious from Table 1, the left-and right-handed
fermions transform differently under $SU(2)_L$ and $U(1)_Y$ and so
gauge invariance forbids a term like Eq. \ref{massterm}.  
 The Higgs
boson can be used to give the fermions mass, however. The
gauge invariant  Yukawa coupling of the
Higgs boson to the up and down quarks  is
\beq
{\cal L}_d=-\lambda_d {\overline Q}_L \Phi d_R + h.c.\quad ,
\eeq
This gives the effective coupling
\beq
-\lambda_d {1\over\sqrt{2}}
({\overline u}_L,~ {\overline d}_L)\left(
\begin{array}{c}  0 \\
v+ h \end{array} \right) d_R + h.c.
\eeq
which can be seen to yield a mass term for the down quark if
we make the identification
\beq 
\lambda_d = {m_d \sqrt{2}\over v}.
\eeq
In order to generate a mass term for the up quark we use the 
fact that $\Phi^c \equiv - i \tau_2 \Phi^*$ is an $SU(2)_L$
doublet and  we can write the $SU(2)_L$ invariant coupling
\beq
{\cal L}_u=-\lambda_u {\overline Q}_L \Phi^c u_R + h.c.
\eeq
which generates a mass term for the up quark.  Similar couplings
can be used to generate mass terms for the charged leptons.
Since the neutrino has no right handed partner, it remains 
massless.  

For the multi-family case, the Yukawa couplings, $
\lambda_d$ and $\lambda_u$, become $N_F \times N_F$ matrices
(where $N_F$ is the number of families).  Since the  fermion
mass matrices
and Yukawa matrices  are proportional, the interactions of the
Higgs boson with the fermion mass eigenstates are flavor diagonal
and
 the Higgs boson does not mediate flavor changing interactions.
(In models with extended Higgs sectors, this need not be the case.)

By expressing the fermion kinetic energy in terms of the gauge
boson mass eigenstates of Eq. \ref{masseig},
 the charged and neutral weak current interactions
of the fermions
can be found.    A complete set of Feynman rules for the interactions
of the fermions and gauge bosons of the Standard Model is given in
Ref. 3.

The parameter $v$ can be found  from 
the charged current for $\mu$ decay,
$\mu\rightarrow e {\overline \nu}_e \nu_\mu$, as shown in 
Fig. 3.    The interaction strength for muon decay is measured
very accurately to be $G_F=1.16639\times 10^{-5}~GeV^{-2}$ and
can be used to determine $v$.
\begin{figure}
\centering
\begin{picture}(200,100)  
\ArrowLine(0,100)(50,90)
\ArrowLine(50,90)(100,100)
\Photon(50,90)(65,50){3}{6}
\ArrowLine(60,50)(110,70) 
\ArrowLine(110,30)(60,50)
\put(110,95){$\nu_\mu$} 
\put(0,92){$\mu$} 
\put(63,67){$W$}
\put(115,65){$e$}
\put(115,25){$\nu_e$}
\end{picture}
\caption{Determination of the vacuum
expectation value $v$ from $\mu$ decay.}
\end{figure}
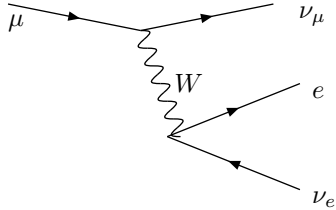
Since the momentum carried by the $W$ boson is of order $m_\mu$ it
can be neglected in comparison with $M_W$ and we make the identification
\beq
{G_F\over \sqrt{2}}={g^2\over 8 M_W^2}={1\over 2 v^2},
\eeq
which gives the result
\beq
v=(\sqrt{2} G_F)^{-1/2} = 246~GeV\,\,.
\eeq

One of the most important points about the Higgs mechanism is
that all of the couplings of the Higgs boson to fermions and
gauge bosons are completely determined in terms of coupling
constants and fermion masses. The potential of Eq. \ref{wspot}
had two free parameters, $\mu$ and $\lambda$.  We can trade
these for
\beqn
v^2&=&-{\mu^2\over 2 \lambda}\nonumber \\
M_h^2&=& 2 v^2 \lambda.
\label{mhdef}
\eeqn
  There are no remaining adjustable
parameters and so Higgs production  and decay processes
can be computed unambiguously in terms of the Higgs mass alone.

\section{Indirect Limits on the Higgs Boson Mass}

Before we discuss the experimental searches for
the Higgs boson, it is worth considering some theoretical
constraints on the Higgs boson mass.      Unfortunately, these
constraints can often be evaded by postulating
the existence of some unknown
new physics which enters into the theory at a mass scale above that
of current experiments, but below the Planck scale. 
Never the less, in the minimal Standard Model where there is no new physics
between the electroweak scale and the Planck scale, there exist
both upper and lower bounds on the Higgs boson mass.

\subsection{Triviality}

Bounds on the Higgs boson  mass have been  deduced on the grounds
of {\it triviality}.\cite{triv,chivtriv}
  The basic argument goes as follows:
Consider a pure scalar theory in which the potential is given 
by\footnote{$\mu^2<0$, $\lambda>0$.}
\beq
V(\Phi)=\mu^2\mid \Phi^\dagger \Phi\mid
 +\lambda (\mid \Phi^\dagger \Phi\mid )^2
\label{potsm}
\eeq
where the quartic coupling is
\beq
\lambda={M_h^2\over 2 v^2}.
\label{lamdef}
\eeq
This is the scalar sector of the Standard Model with no gauge bosons
or fermions.
The quartic coupling, $\lambda$,
changes with the effective energy scale $Q$ due to
the self interactions of the scalar field:
\beq
{d \lambda \over dt}={3 \lambda^2\over 4 \pi^2},
\label{lams}
\eeq
where $t\equiv \log(Q^2/Q_0^2)$ and $Q_0$ is some reference
scale. (The reference scale is often
taken to be $v$ in the Standard Model.)
 Eq. \ref{lams} is easily
solved,
\beqn
{1\over \lambda(Q)}&=&{1\over \lambda(Q_0)}-{3\over 4 \pi^2}
\log\biggl({Q^2\over Q^2_0}\biggr),
\nonumber \\
\lambda(Q)&=& {\lambda(Q_0)\over 
\biggl[1-{3\lambda(Q_0)\over 4 \pi^2}\log({Q^2\over Q^2_0})
\biggr]}.
\label{lampol}
\eeqn
Hence if we measure $\lambda$ at some energy scale, we can predict
what it will be at all other energy scales. 
From Eq. ~\ref{lampol} we see that $\lambda(Q)$ blows up 
as $Q\rightarrow \infty$
 (called the Landau pole).  Regardless of how small $\lambda(Q_0)$
is, $\lambda(Q)$ will eventually 
become infinite at some large $Q$.
Alternatively, $\lambda(Q_0)\rightarrow 0$ as
$Q\rightarrow 0$ with $\lambda(Q)>0$.  Without the $\lambda\Phi^4$
interaction of Eq.~\ref{potsm}
the theory becomes a non-interacting theory at low energy, termed a
{\it trivial} theory. 

To obtain a bound on the Higgs mass we require that the quartic
coupling be finite,
\beq
{1\over \lambda(\Lambda)}>0, 
\eeq
where $\Lambda$ is some large scale where new physics
enters in.
Taking the reference
scale  $Q_0=v$, and substituting Eq. \ref{lamdef} gives an 
approximate upper bound
on the Higgs mass,
\beq
M_h^2 < {8 \pi^2 v^2\over 3 \log (\Lambda^2/v^2)}
\quad .
\eeq
Requiring that there be no new physics before $10^{16}~GeV$ yields
the approximate upper bound on the Higgs boson mass,
\beq
M_h < 160~GeV.
\eeq
As the scale $\Lambda$ becomes smaller, the limit on the Higgs
mass becomes progressively weaker and for $\Lambda\sim 3~TeV$, the
bound is roughly $M_h < 600~GeV$. 
 Of course, this picture is valid only if the one
loop evolution equation of Eq.
\ref{lams} is an accurate description of the theory at
large $\lambda$.  For large $\lambda$, however, higher order or
non-perturbative corrections
 to the evolution equation must be included.\cite{quirrev}

Lattice gauge theory calculations have used
similar techniques to obtain
a bound on the Higgs mass.
As above, they consider
a purely scalar theory and require that the scalar self coupling
$\lambda$ remain finite for all scales less than some
cutoff which is arbitrarily chosen to be $2 \pi M_h$.
This gives a limit\cite{hasen}
\beq
M_h({\rm lattice})< 640~GeV. 
\eeq
 The lattice results are relatively
insensitive to the value of the cutoff chosen.  Note that this bound
is in rough
agreement with that found above for $\Lambda\sim 3~TeV$.

 Of course everything we have done so far is for a theory with only
scalars.  The physics changes dramatically
 when we couple the theory to fermions
and gauge bosons.
Since the Higgs coupling to fermions is proportional to the Higgs
boson mass, the most relevant fermion is the top quark.  Including
the top quark and the gauge bosons, Eq. ~\ref{lams} becomes\cite{rge}
\beq
{d \lambda \over d t}={1\over 16 \pi^2}\biggl[
12 \lambda^2 + 12 \lambda g_t^2 - 12 g_t^4-{3\over 2}\lambda(3 g^2
+g^{\prime~2})+{3\over 16} (2 g^4 +(g^2+g^{\prime~2})^2)\biggr]
\label{renorm}
\eeq
where 
\beq
g_t\equiv -{ M_t\over v} \quad.
\eeq
For a heavy Higgs boson , $\lambda>g_t,g,g^{\prime}$, and
the dominant contributions to the running of $\lambda$ are,
\beq
{d \lambda \over d t}\sim{\lambda\over 16 \pi^2}\biggl[
12 \lambda + 12  g_t^2 -{3\over 2}(3 g^2
+g^{\prime~2})\biggr].
\eeq
There is a critical value of 
the quartic coupling
$\lambda$ which depends on the top
quark mass,
\beq
\lambda_c\equiv{1\over 8} ( 3 g^2 +g^{\prime~2})-g_t^2\quad .
\eeq
The evolution of the quartic coupling stops when
 $\lambda=\lambda_c$.\cite{cab}
 If
$M_h>M_{h}^c\equiv \sqrt{2 \lambda_c} v$ then  the quartic coupling
becomes infinite at some scale  and the theory is non-perturbative. 
If we require that the theory be perturbative (i.e., 
the Higgs quartic coupling be finite) at all energy scales
below some unification scale ($\sim 10^{16}~GeV$)
then an upper bound on the Higgs mass
is obtained as a function of the top quark mass.
To obtain a numerical value for the Higgs mass limit, the evolution 
of the gauge coupling constants and the Higgs Yukawa coupling must
also be included.
  For $M_t=175~GeV$
this bound is $M_h < 170~GeV$. \cite{cab}
  If a Higgs boson were found which was
heavier than this bound, it would require that there be some new
physics below the unification scale.
The bound on the Higgs boson mass
 as a function of the cut-off scale
from the requirement that the quartic coupling $\lambda(\Lambda)$
 be finite
is shown  as the upper curve in Fig. \ref{trivlims}.

\subsection{Vacuum Stability}

A bound on the Higgs mass can also be derived by the requirement that 
spontaneous symmetry breaking actually occurs;\cite{linde} that is,
\beq
V(v)< V(0).
\label{vacstab}
\eeq
This bound is essentially equivalent to the requirement
that $\lambda$ remain positive at all scales $\Lambda$.
(If $\lambda$ becomes negative, the potential is unbounded from
below and has no state of minimum energy.)
For small $\lambda$, Eq. ~\ref{renorm} becomes,
\beq
{d \lambda \over d t}={1\over 16 \pi^2}\biggl[
- 12 g_t^4+{3\over 16} (2 g^4 +(g^2+g^{\prime~2})^2)\biggr]
\quad .
\label{renormsmall}
\eeq
This is easily solved to find,
\beq
\lambda(\Lambda)=\lambda(v)+{1\over 16 \pi^2}\biggl[
- 12 g_t^4+{3\over 16} (2 g^4 +(g^2+g^{\prime~2})^2)\biggr]
\log\biggl({\Lambda^2\over v^2}\biggr)
\quad .
\eeq
Requiring
$\lambda(\Lambda)>0$ gives the bound on the Higgs boson mass,
\beq
M_h^2>{v^2\over 8 \pi^2}\biggl[
- 12 g_t^4+{3\over 16} (2 g^4 +(g^2+g^{\prime~2})^2)\biggr]
\log\biggl({\Lambda^2\over v^2}\biggr)
\quad .
\eeq
 A more careful 
analysis along the
same lines as above \cite{vacbounds} using the $2$ loop renormalization group
improved effective potential\footnote{The renormalization
group improved effective potential sums all potentially large
logarithms, $\log(Q^2/v^2)$.} and the
running of all couplings gives the requirement from vacuum 
stability if we require that the Standard Model be valid
up to scales of order 
$10^{16} GeV$,\cite{quirrev,sher}
\beq
M_h(GeV)> 130.5 + 2.1(M_t-174)
\quad .
\label{sherlim}
\eeq
If the Standard Model is only valid to $1~TeV$, then the limit
of Eq.~\ref{sherlim} becomes,
\beq
M_h(GeV)> 71 + .74(M_t-174)
\quad .
\eeq

\begin{figure}[t]
\vskip -.5in 
\centering
\epsfxsize=4.in
\leavevmode
\epsffile{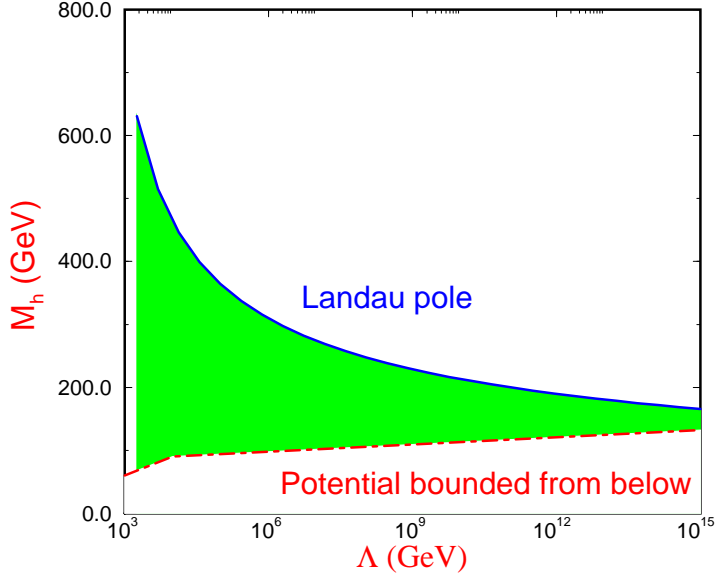}
\caption{Theoretical limits on the Higgs boson mass. The allowed
region is shaded.  The region above the solid line (labelled Landau
pole) is forbidden because the quartic coupling is infinite.
The region below the dot-dash line is forbidden because the
quartic coupling is negative, causing the potential to be
unbounded from below. }
\label{trivlims}
\end{figure}

We see that when $\lambda$ is small (a light Higgs boson) radiative
corrections
from the top quark and gauge couplings
 become important and lead to a lower limit on the Higgs
boson mass from the requirement of vacuum stability,
$\lambda(\Lambda)>0$.  If $\lambda$
is large (a heavy Higgs boson) then triviality
arguments, (${1\over\lambda(\Lambda)}
>0$), lead to an upper bound on the Higgs mass. The allowed 
region for the Higgs mass from these considerations
is shown in Fig. \ref{trivlims} as a
function of the scale of new physics, $\Lambda$.
If the Standard Model is valid up to $10^{16}~GeV$, then the allowed
region for the Higgs boson mass is restricted to be between
about $130~GeV$
and $170~GeV$.
A Higgs boson with a mass outside this region would be a signal for new
physics.

\subsection{Bounds from Electroweak Radiative Corrections}

The Higgs boson enters into one loop
radiative  corrections in the Standard 
Model and so precision electroweak measurements
can bound the Higgs boson mass.
  For example,  the
$\rho $ parameter gets a contribution from the 
Higgs boson\cite{rhoh}\footnote{This result is scheme dependent.  Here
$\rho\equiv M_W^2/M_Z^2\cos^2\theta_W(M_W)$, where $\cos\theta_W$ is
a running parameter calculated at an energy scale of $M_W$.}
\beq
\rho=1-{11 g^2\over 96 \pi^2}\tan^2\theta_W \log
\biggl({M_h\over M_W}\biggr).
\eeq
Since the dependence on the Higgs boson mass is only logarithmic,
the
limits derived on the Higgs boson from this method are relatively
weak.  In contrast, the top quark contributes quadratically to
many electroweak observables such as the $\rho$ parameter. 

 It is straightforward to demonstrate 
that at one loop all  electroweak
parameters have at most a
logarithmic dependance on $M_h$.\cite{abl}  This fact has been
glorified by the name of the ``screening theorem''.\cite{veltman}
  In general, 
electroweak radiative corrections involving the Higgs boson take the form,
\beq
g^2\biggl( \log{M_h\over M_W}+g^2 {M_h^2\over M_W^2} 
\cdots \biggr).
\eeq
That is, effects quadratic in the Higgs boson  mass are always screened
by an additional power of $g$ relative to the lower order logarithmic
effects and so radiative corrections involving the Higgs
boson can never be large.\cite{einhorn}
  
From precision measurements at LEP and SLC of  electroweak observables,
the direct measurements of $M_W$ and $M_t$
at the Tevatron, and the measurements
of $\nu$ scattering experiments,  there is
 the bound on the Higgs boson mass coming from the 
effect of radiative corrections,\cite{kar}
\beq
M_h< 280~ GeV \quad (95\%~ {\rm~ confidence~ level}).
\label{ewmhlim}
\eeq
This bound does not include the Higgs boson direct search experiments
and applies only in the minimal Standard Model.
Since the bound of Eq. \ref{ewmhlim} arises from loop corrections, it
can be circumvented by unknown new physics which
contributes to the loop corrections.

\begin{figure}[t]
\vskip -.5in 
\centering
\epsfxsize=4.in
\leavevmode
\epsffile{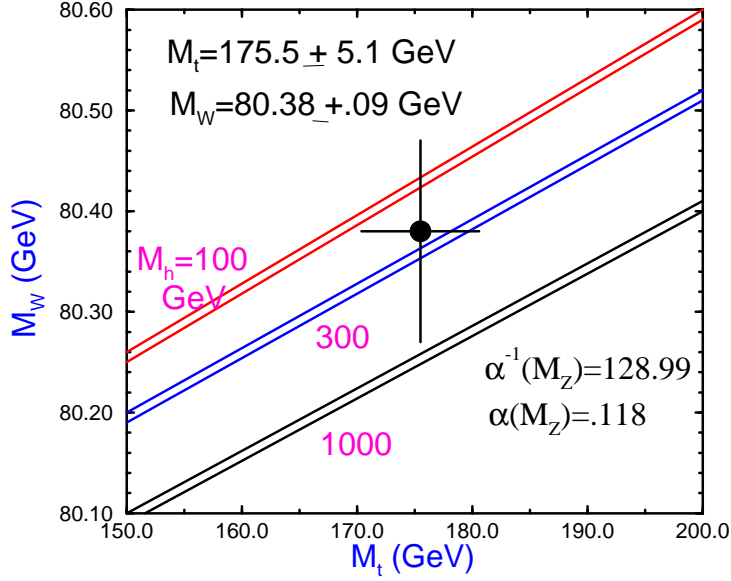}
\caption{Limits on the Higgs boson mass from the
Tevatron measurements of $M_W$ and $M_t$ for
various values of the Higgs boson mass.}
\label{tevres}
\end{figure}

The relationship between $M_W$ and $M_t$ arising from
radiative corrections depends sensitively on
the Higgs boson mass. The radiative corrections to $M_W$ can
be written as,
\beq
\biggl({M_W^2\over M_Z^2}\biggr)=1-{\pi\alpha\over\sqrt{2}
G_F M_W^2(1-\delta r)}       \quad ,
\eeq
where $\delta r$ is a function of $M_t^2$ and $\log(M_h)$.  The 
results using the Tevatron measurements of $M_W$ and $M_t$ are shown in
Fig. \ref{tevres} and clearly prefer a relatively light Higgs boson,
in agreement with the global fit of Eq. \ref{ewmhlim}.

While the Higgs boson mass remains  a free parameter, the combination
of limits from electroweak radiative corrections and the triviality
bounds of the previous section suggest that the Higgs boson may
be relatively light, in the few hundred GeV range.

\section{Higgs Branching Ratios}

In the Higgs sector, the Standard Model is extremely predictive,
with all couplings, decay widths, and production cross sections
given in terms of the unknown Higgs boson mass.
The measurements of the various Higgs decay channels will
serve to discriminate between the Standard Model
and other models with more complicated Higgs sectors
which may have different decay chains and Yukawa
couplings.  It
is hence vital that we have reliable predictions for the
branching ratios in order to 
verify the correctness of the Yukawa couplings
of the Standard Model.\cite{spirrev}

In this section, we review the Higgs boson branching ratios in the Standard
Model.
\subsection{Decays to Fermion Pairs}

  The dominant
decays of a Higgs boson
with a mass  below the $W^+W^-$ threshold  are into fermion-
antifermion pairs. In the Born approximation,
 the width into charged lepton pairs is 
\beq
\Gamma(h\rightarrow l^+l^-)={G_FM_l^2\over 
4\sqrt{2}\pi} M_h \beta_l^3
\quad ,
\eeq
where
$\beta_l\equiv \sqrt{1-4M_l^2/M_h^2}$ is the velocity of the
final state leptons.  
The Higgs  boson
decay into quarks is enhanced by the color
factor $N_c=3$ and also receives significant QCD corrections,
\beq
\Gamma(h\rightarrow q {\overline q})={3G_FM_q^2\over 4 \sqrt{2} \pi}
M_h \beta_q^3\biggl(1+{4\over 3}{\alpha_s\over \pi}\Delta_h^{QCD}
\biggr)                              
\quad ,
\eeq
where the QCD correction factor, $\Delta_h^{QCD}$,  can be found in
Ref. 22.
The Higgs boson clearly decays predominantly into the heaviest fermion
kinematically allowed. 
  
  A
large portion of the QCD corrections can be absorbed by
expressing the decay width in terms of a
running  quark mass, $M_q(\mu)$, evaluated at the
scale $\mu=M_h$.  The QCD
corrected decay width can then be written as,\cite{hbqcd}
\beq
\Gamma(h\rightarrow q {\overline q})=
{3G_F\over 4 \sqrt{2} \pi} M_q^2(M_h^2)
M_h \beta_q^3 \biggl(1+5.67{\alpha_s(M_h^2)\over \pi}+\cdots
\biggr)                             , 
\eeq
where $\alpha_s(M_h^2)$ is defined in the ${\overline{MS}}$ scheme with
$5$ flavors and $\Lambda_{\overline{MS}}=150~GeV$.  
The ${\cal O}(\alpha_s^2)$ corrections are also known
in the limit $M_h>>M_q$.\cite{h2l}

For $10~GeV < M_h < 160~GeV$,
the most important fermion decay mode is $h\rightarrow
b {\overline b}$. 
In 
leading log QCD, the running of the $b$ quark mass is,
\beq
M_{b}(\mu^2)=M \biggl[{\alpha_s(M^2)
\over \alpha_s(\mu^2)}
\biggr]^{(-12/23)}
\biggl\{1+{\cal O}(\alpha_s^2)\biggr\} ,  
\label{bscale}  
\eeq
where $M_b(M^2)\equiv M$ implies that the running mass at the
position of the propagator pole is equal to the location of the pole.
For $M_b(M_b^2)=4.2~GeV$, this yields an
effective value $M_b((M_h=100~GeV)^2) = 3~GeV$.
 Inserting the QCD corrected mass into  the expression
for the width thus leads to a
significantly smaller rate than that found using $M_b=4.2~GeV$.  
For a Higgs boson in the $100~GeV$ range, 
the ${\cal O}(\alpha_s)$ corrections decrease the decay width 
for $h\rightarrow b {\overline b}$ by about a factor of two.

The electroweak radiative corrections to $h\rightarrow q {\overline q}$
are not significant and amount to only a few 
percent correction.\cite{ewh}
These can be neglected in comparison with the much larger 
QCD corrections.  

\begin{figure}[t]
\centering   
\epsfxsize=3.5in
\leavevmode
\epsffile{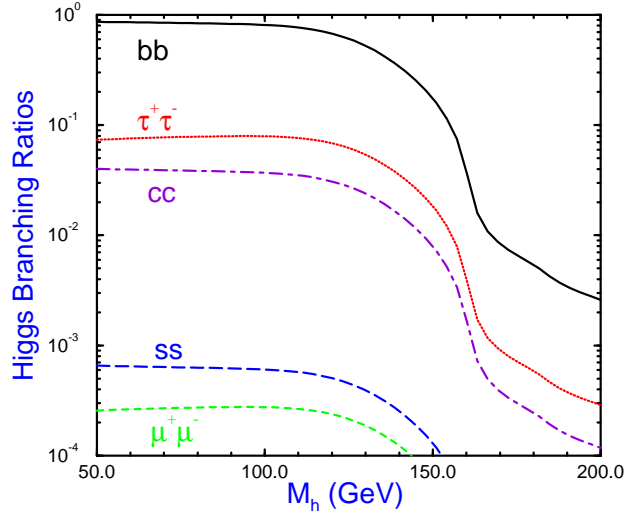}
\caption{Branching ratios of the Standard Model Higgs boson
 to fermion-antifermion pairs, including 
QCD radiative corrections. 
The radiative corrections were computed using the program 
HDECAY.$^{25}$}
\label{brferm}
\end{figure}

The branching ratios for the dominant decays to fermion-
antifermion pairs are shown in Fig. \ref{brferm}.\footnote{
A convenient FORTRAN code for computing the QCD radiative
corrections to the Higgs boson decays is HDECAY,
which is documented in Ref. 25.}  The decrease 
in the $h\rightarrow f {\overline f}$  branching ratios at $M_h\sim
150~GeV$ is due to the turn-on of the $WW^*$ decay channel,
where $W^*$ denotes a virtual $W$.
For most of the region   below the $W^+W^-$ threshold, the Higgs
decays almost entirely to $b {\overline b}$ pairs, although it
is possible that the decays to $\tau^+\tau^-$ will
  be useful in the experimental searches.
The other fermionic Higgs boson 
decay channels are almost certainly too small to be
separated from the backgrounds.

Even including the QCD corrections, the rates
roughly
scale with the fermion masses and the color factor, $N_c$,
\beq
{\Gamma(h\rightarrow b {\overline b})\over
\Gamma(h\rightarrow \tau^+\tau^-)}\sim
{3 M_b^2(M_h^2)\over M_\tau^2},
\eeq  
and so a measurement of the branching ratios could
serve to verify the correctness of the Standard Model
couplings. 
The largest uncertainty is in the value of $\alpha_s$, which
affects the running $b$ quark mass, as in Eq. \ref{bscale} .

\subsection{Decays to Gauge Boson Pairs}

The Higgs boson can also decay to gauge boson pairs.
At tree level, the decays $h\rightarrow W^+W^-$ and $h\rightarrow
ZZ$ are possible, while at one-loop 
the decays $h\rightarrow gg,\gamma\gamma,
~ \gamma Z$
occur.
 
The decay widths of the Higgs boson to 
physical $W^+W^-$ or $ZZ$
pairs    are
 given by,
\beqn
\Gamma(h\rightarrow W^+W^-) =&
{G_F M_h^3\over 8 \pi\sqrt{2}}\sqrt{
1-r_W}(1-r_W+{3\over 4}r_W^2)
\nonumber \\
\Gamma(h\rightarrow ZZ)=&
{G_F M_h^3\over 16 \pi\sqrt{2}}\sqrt{
1-r_Z}(1-r_Z+{3\over 4}r_Z^2)
,
\eeqn
where $r_V\equiv 4 M_V^2/M_h^2$. 

\begin{figure}[t]
\centering   
\epsfxsize=4.in
\leavevmode
\epsffile{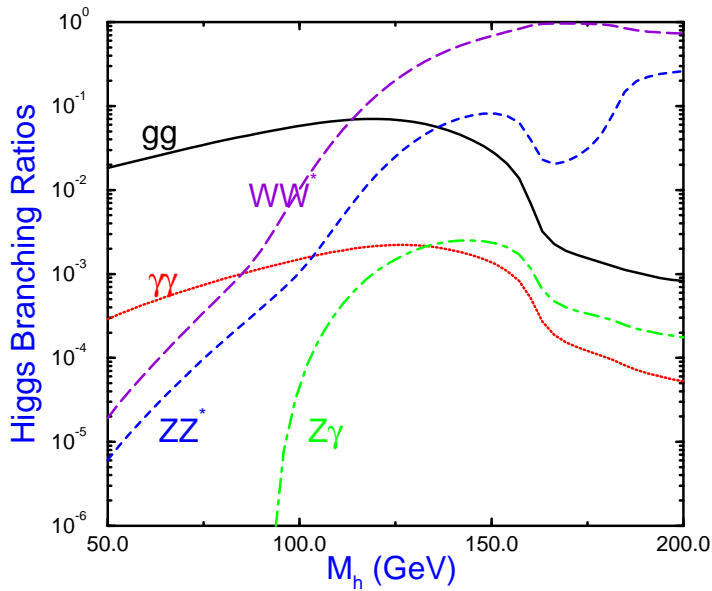}
\caption{Branching ratios  of the Standard Model Higgs
boson to gauge boson pairs, including QCD
radiative corrections.  The rates to $WW^*$ and $ZZ^*$ must be
multiplied by the appropriate branching ratios for $W^*$ and $Z^*$
decays to $f {\overline f} $ pairs.
The radiative corrections were computed using the program 
HDECAY.$^{25}$}
\label{gagwid}
\end{figure}

Below the $W^+W^-$ and $ZZ$
thresholds, the Higgs boson can also decay to vector
boson pairs $V V^*$, ($V=W^\pm,Z$),
 with one of the gauge bosons off-shell. 
The widths, summed over all available channels for $V^*\rightarrow
f {\overline f}$ are: ~\cite{wkwm}
\beqn
\Gamma(h\rightarrow Z Z^*)&=&
{g^4 M_h\over 2048 (1-s_W)^2\pi^3}
\biggl(7-{40\over 3}s_W+{160\over 9}s_W^2
\biggr)
  F\biggl({M_Z\over M_h}\biggr)
\nonumber \\
\Gamma(h\rightarrow W W^*)&=&{3 g^4 M_h\over 512 \pi^3} F\biggl({M_W\over M_h}
\biggr)
\quad ,
\eeqn
where $s_W\equiv \sin^2\theta_W$ and 
\beqn
F(x)&\equiv & -\mid 1-x^2\mid
\biggl({47\over 2} x^2-{13\over 2} +{1\over x^2}\biggr)
-3\biggl(1-6 x^2+4 x^4\biggr)
\mid \ln (x) \mid 
\nonumber \\  && 
+3{1-8 x^2+20 x^4\over \sqrt{4 x^2-1}}
\cos ^{-1}\biggl( {3 x^2-1\over 2 x^3}\biggr)
\quad .
\eeqn
These widths can be significant when the Higgs boson mass
approaches the real $W^+W^-$ and $ZZ$ thresholds, as can be seen
in Fig. \ref{gagwid}. The $WW^*$ and $ZZ^*$ branching ratios grow
rapidly with increasing Higgs mass and above $2M_W$
the rate for $h\rightarrow W^+W^-$ is close to 1.   
The decay width to $ZZ^*$ is roughly an order of magnitude
smaller than the decay width to $WW^*$ over
most of the Higgs mass range due to the smallness
of the neutral current couplings as compared to the
charged current couplings.

The decay of the Higgs boson to gluons arises through
fermion loops,\cite{hhg}
\beq
\Gamma_0(h\rightarrow gg)={ G_F \alpha_s^2 M_h^3
\over 64 \sqrt{2}\pi^3}
\mid \sum_q F_{1/2}(\tau_q)\mid^2
\eeq
where $\tau_q\equiv
4 M_q^2/M_h^2$ and  $F_{1/2}(\tau_q)$ is defined to be,
\beq
F_{1/2}(\tau_q) \equiv -2\tau_q\biggl[1+(1-\tau_q)f(\tau_q)\biggr]
\quad .
\label{etadef}
\eeq
The function $f(\tau_q)$ is given by,
\beq
f(\tau_q)=\left\{\begin{array}{ll}
\biggl[\sin^{-1}\biggl(\sqrt{1/\tau_q}\biggr)\biggr]^2,&\hbox{if~} 
\tau_q\ge 1
\\
-{1\over 4}\biggl[\log\biggl({x_+\over x_-}\biggr)
-i\pi\biggr]^2,
&\hbox{ if~}\tau_q<1,
\end{array} 
\right .
\label{fundef}
\eeq
with
\beq
x_{\pm}=1\pm\sqrt{1-\tau_q}
.
\eeq
In the limit in which the quark mass is much less than the Higgs boson mass,
(the relevant limit for the $b$ quark),
\beq
F_{1/2}\rightarrow {2 M_q^2\over M_h^2}\log^2\biggl(
{M_q\over M_h}\biggr)
.
\eeq 
A Higgs boson decaying to $b {\overline b}$ will therefore be extremely
narrow.
On
the other hand, for a heavy quark, $\tau_q\rightarrow\infty$,
 and $F_{1/2}(\tau_q)$ approaches
a constant,
\beq
F_{1/2}\rightarrow -{4\over 3}
.
\label{f12}
\eeq
It is clear that the dominant contribution to the gluonic decay of the 
Higgs boson is from the top quark loop and from possible new generations of
heavy fermions.   A measurement of this rate would serve to count
the number of heavy fermions since the 
effects of the heavy fermions do not
decouple from the theory.

The QCD radiative corrections
from  $h\rightarrow ggg$ and $h\rightarrow
g q {\overline q}$
to the hadronic decay of the Higgs boson
are large and they typically increase the
width by more than $50\%$.  The radiatively corrected width can be
approximated by
\beq
\Gamma(h\rightarrow ggX)=
\Gamma_0(h\rightarrow gg)\biggl[
1+C{\alpha_s(\mu)\over \pi}\biggr]
\quad ,
\eeq
where $C={215\over 12}-
{23\over 6}\log(\mu^2/M_h^2)$, for $M_h<2 M_t$.\cite{hggg}
The radiatively corrected branching ratio for $h\rightarrow
ggX$ is the solid curve in Fig. \ref{gagwid}.

The decay $h\rightarrow Z \gamma$ is not useful phenomenologically,
so we will not discuss it here although
  the complete expression for the
branching ratio can be found in Ref. 28.  
On the other hand, the decay $h\rightarrow \gamma \gamma$ is an
important mode for the Higgs search at the LHC. 
At lowest order, the branching
ratio is, \cite{hgg} 
\beq
\Gamma(h\rightarrow \gamma\gamma)={\alpha^2 G_F\over 128\sqrt{2} \pi^3}
M_h^3\mid \sum_i N_{ci} Q_i^2 F_i(\tau_i)\mid^2 
\eeq
where the sum is over fermions and  $W^\pm$ bosons with
$F_{1/2}(\tau_q)$ given in Eq. 23, and 
\beq
F_W(\tau_W)= 2+3\tau_W[1+(2-\tau_W)f(\tau_W)]
\quad .
\label{fdef}
\eeq 
$\tau_W=4 M_W^2/M_h^2$, $N_{Ci}=3$ for quarks
and $1$ otherwise, and $Q_i$ is the electric charge in units of $e$.
The function $f(\tau_q)$ is given in Eq. \ref{fundef}. 
The $h\rightarrow \gamma\gamma$ decay channel clearly
probes the possible existence of heavy charged particles.  
(Charged scalars, such as those existing in supersymmetric models,
would also contribute to the rate.)\cite{hhg}

\begin{figure}[t]
\centering   
\epsfxsize=4.in
\leavevmode
\epsffile{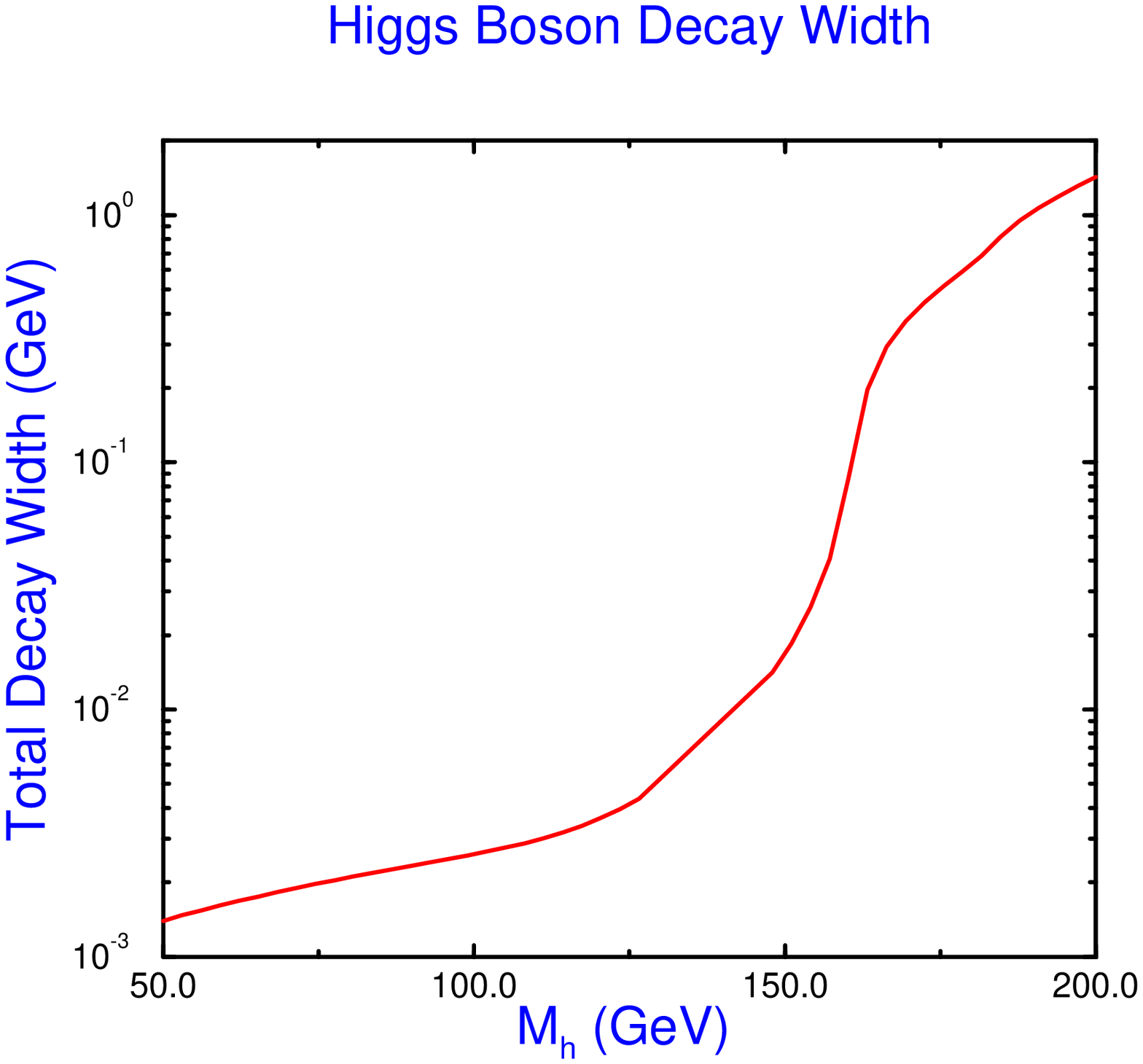}
\caption{Total Higgs boson decay width in the Standard Model,
including QCD radiative corrections. The turn-on of 
the $W^+W^-$ threshold 
at $M_h\sim 160~GeV$ is obvious.}
\label{totwid}
\end{figure}  

In the limit where the particle in the loop is much heavier 
than the Higgs boson, $\tau\rightarrow \infty$, 
\beqn
F_{1/2}&\rightarrow & -{4\over 3}
\nonumber \\
F_W&\rightarrow & ~7
\quad .
\label{fwdef}
\eeqn
The top quark contribution $(F_{1/2})$ is therefore much
smaller than that of the $W$ loops $(F_W)$ and so we expect the QCD
corrections to be 
less important than is the case for the $h\rightarrow gg$
decay. 
In fact the QCD corrections to the total width for $h\rightarrow
\gamma\gamma$ are quite small.\cite{spgg} 
The $h\rightarrow \gamma\gamma$ branching ratio is
the dotted line in Fig. \ref{gagwid}.
For small Higgs masses it rises with increasing $M_h$
and peaks at around $2\times 10^{-3}$ for $M_h\sim 125~GeV$.  
Above this mass, the $W W^*$ and $ZZ^*$ decay modes are
increasing rapidly with increasing Higgs mass and the 
$\gamma \gamma$ mode
becomes further suppressed.

The total  Higgs boson width for  a Higgs boson
with mass less than $M_h\sim 200~GeV$
is shown in
Fig. \ref{totwid}.  As the Higgs boson becomes heavier twice
the $W$ boson mass,
its width becomes extremely large (see Eq. 122).
 Below around $M_h\sim 150~GeV$, the Higgs boson is
quite
narrow with $\Gamma_h< 10~MeV$.  As the $WW^*$ and $ZZ^*$ channels
become accessible, the width increases rapidly with $\Gamma_h\sim 1~GeV$
at $M_h\sim 200~GeV$.  Below the $W^+W^-$
threshold,
the Higgs boson width is too narrow to be resolved experimentally.
The total width for the lightest neutral Higgs boson in
the minimal supersymmetric model is typically much smaller than the 
Standard Model width for the same  Higgs boson 
mass and so a measurement
of the total width could serve to discriminate between the
two models.

\section{Higgs Production at LEP and LEP2}

\begin{figure}[t]
\centering
\begin{picture}(200,100)
\SetScale{1.}
\ArrowLine(0,100)(50,50)
\ArrowLine(50,50)(0,0)
\Photon(50,50)(100,50){3}{6}
\Photon(100,50)(150,100){3}{6}  
\DashLine(100,50)(150,0){5} 
\put(65,27){$Z$} 
\put(0,82){$e^-$}
\put(0,-8){$e^+$}
\put(125,90){$Z$}
\put(125,-8){$h$}
\end{picture}
\vskip .2in  
\caption{Higgs production through $e^+e^- \rightarrow Z h$.}
\label{eezhfig}
\end{figure}
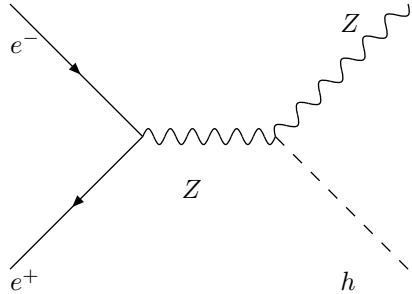

\begin{figure}[t]
\centering   
\epsfxsize=4.in
\leavevmode
\epsffile{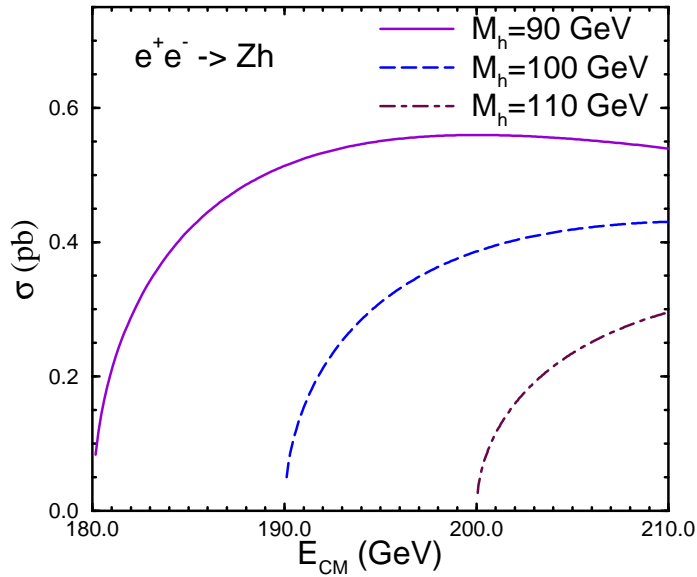}
\caption{Born cross section for $e^+e^- \rightarrow Zh$ as a function
of center of mass energy.}
\label{eezhsug}
\end{figure}

Since the Higgs boson coupling to the electron is very small, $\sim m_e/v$,
the $s-$channel production mechanism,
 $e^+e^-\rightarrow h$, is minute and
the dominant Higgs boson production mechanism
at LEP and LEP2
 is
the associated production with a $Z$, 
$e^+e^-\rightarrow Z^*\rightarrow Zh$, as shown in Fig. \ref{eezhfig}.
At LEP2, a physical
$Z$  boson can be produced  and  the cross section is,\cite{sigee}
\beq
{\hat \sigma}(e^+e^-\rightarrow  Zh )={\pi \alpha^2\lambda_{Zh}^{1/2}
[\lambda_{Zh}+12 {M_Z^2\over s}][1+(1-4 \sin^2\theta_W)^2]
\over 192 s \sin^4\theta_W\cos^4\theta_W(1-M_Z^2/s)^2}
\label{eezhborn}
\eeq
where
\beq
\lambda_{Zh}\equiv (1-{M_h^2+M_Z^2\over s})^2-{4 M_h^2M_Z^2
\over s^2}
\quad .
\label{bjzh}
\eeq
The center of mass momentum of the produced $Z$ is $
\lambda_{Zh}^{1/2}\sqrt{s}/2$ and  the cross section is
shown in Fig. \ref{eezhsug}
  as a function of $\sqrt{s}$ for  different
values of the Higgs boson mass.  The cross section
peaks at $\sqrt{s}\sim M_Z+ 2 M_h$.  From Fig. \ref{eezhsug},
 it is apparent that
the cross section increases rapidly with increasing energy and so the
best limits on the Higgs boson mass
will be obtained at the highest energy. 
  
\begin{figure}[t]
\centering   
\epsfxsize=4.in
\leavevmode
\epsffile{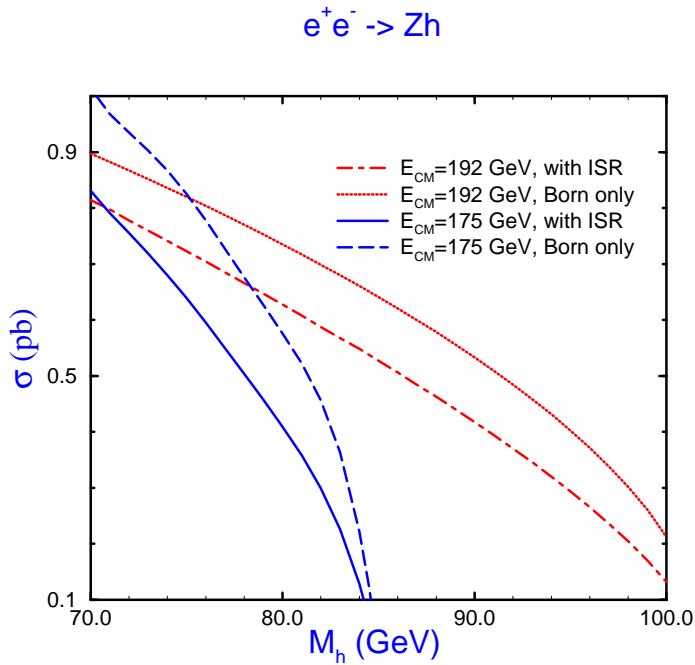}
\caption{Effects of initial state radiation(ISR) on the process $e^+e^-
\rightarrow Zh$. The curves labelled ``Born only'' are the
results of Eq. \ref{eezhborn},
 while those labelled  ``with ISR'' include
the photon radiation as in Eq. \ref{eeisrform}.}
\label{leplimfig}
\end{figure}

The electroweak radiative
corrections are quite small at LEP2 energies.\cite{kniehl}
Photon bremsstrahlung can be important, however, since it
is enhanced by a large logarithm, $\log(s/m_e^2)$.  The photon radiation
can be accounted for by integrating
the Born cross section of Eq. \ref{eezhborn}  with a radiator function
$F$
which includes virtual and soft photon effects, along with
hard photon radiation,\cite{berends} 
\beq
\sigma={1\over s} \int  ds^\prime F(x,s)
{\hat \sigma}(s^\prime)
\label{eeisrform}
\eeq
where $x=1-s^\prime/s$ and
the radiator function $F(x,s)$ is known to ${\cal O}(\alpha^2)$, along
with the exponentiation of the infrared contribution,  
\beqn 
F(x,s)&=& tx^{t-1}\biggl\{1+{3\over 4} t\biggr\}
+\biggl\{{x\over 2}-1\biggr\} t+{\cal O}(t^2) 
\nonumber \\
t&\equiv & {2\alpha\over
\pi}\biggl[ \log\biggl({s\over m_e^2}\biggr) -1\biggr]
\quad .
\eeqn  
Photon radiation significantly reduces
  the $Zh$ production
rate from the Born cross section
as shown in Fig.~\ref{leplimfig}.  

A Higgs boson which can be produced
at LEP or LEP2 will decay mostly to $b {\overline b}$ pairs,
so the final state from $e^+e^-\rightarrow Zh$
 will have four fermions.
The dominant background
 is $Z b {\overline b}$ production, which can
be efficiently eliminated by $b$-tagging
 almost up 
to the kinematic limit for producing the Higgs boson.  
LEP2 studies estimate that with $\sqrt{s}=200~GeV$ and 
${\cal L}=100~ pb^{-1}$ per experiment, a Higgs boson mass of $107~GeV$
could be observed at the $5\sigma$ level.\cite{lepstud}
A higher energy $e^+e^-$ machine
(such as an NLC with $\sqrt{s}\sim 500~GeV$)
 could push the Higgs mass limit
to around $M_h\sim .7 \sqrt{s}$.  

Currently the highest energy
data at LEP2 is $\sqrt{s}=183~GeV$. 
The combined limit from the four  LEP2 detectors is\cite{kar}
\beq
M_h> 89.8~GeV
\quad        \quad {\hbox{at~95\%~c.l.}}
\eeq
This limit includes  both hadronic and leptonic decay modes of
the $Z$.  Note how close the result is
to the kinematic boundary.

The cross section for $e^+e^-\rightarrow Zh$ is $s$-wave and
so has a very steep dependence on energy and on the Higgs boson
mass at threshold, as is clear from Fig. \ref{eezhsug}.
This  makes possible a precision measurement of the Higgs
mass.
By measuring the cross section at threshold and
normalizing to a second measurement above threshold in order
to minimize systematic uncertainties
a high energy $e^+e^-$ collider
with $\sqrt{s}=500~GeV$ could obtain  a $1\sigma$
measurement of the mass~\cite{bbgh}
\beq
\Delta M_h\sim 60~MeV\sqrt{{100 fb^{-1}\over L}}
\quad\quad {\rm for~~}M_h=100~GeV       \quad ,
\eeq
where $L$ is the total integrated luminosity.  
The precision becomes worse for larger
 $M_h$ because of the
decrease in the signal cross section.  (Note that the luminosity
at LEP2 will not be high enough to perform this measurement.) 

 The angular distribution of the Higgs boson from the
$e^+e^-\rightarrow Zh$ process  is
\beq
{1\over\sigma}
{d\sigma\over d\cos\theta}\sim \lambda_{Zh}^2
\sin^2\theta+{8M_Z^2\over s}
\eeq
so that at high energy the distribution is
that of a scalar particle,
\beq
{1\over\sigma}{d\sigma\over d \cos\theta}\sim \sin^2
\theta
\quad .
\eeq  
If the Higgs boson were CP odd, on the other hand, the angular
distribution would be $1+\cos^2\theta$. Hence the angular distribution
is sensitive to the spin-parity assignments of the Higgs boson.\cite{spin}    
The angular distribution of this process is also quite sensitive to
non-Standard Model $ZZh$ couplings.\cite{zhnsm}

\section{Higgs Production in Hadronic Collisions}
We turn now to the production of the Higgs boson in $pp$ and
$p \overline{p}$
collisions. 

\subsection{Gluon Fusion}

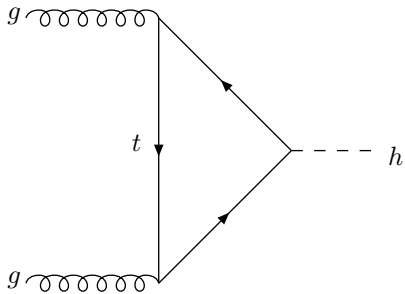
\begin{figure}[t]
\centering 
\begin{picture}(200,100)
\SetScale{1.}
\ArrowLine(50,0)(100,50) 
\ArrowLine(100,50)(50,100) 
\ArrowLine(50,100)(50,0)
\DashLine(100,50)(130,50){5}
\Gluon(0,100)(50,100){3}{6} 
\Gluon(0,0)(50,0){3}{6} 
\put(137,45){$h$} 
\put(-7,0){$g$}
\put(-7,100){$g$}
\put(40,50){$t$}
\SetScale{1}
\end{picture}
\caption{Higgs boson production through gluon fusion with a quark loop.}
\label{gghfig}
\end{figure}

 Since the coupling of a Higgs boson to an up quark
or a down quark is proportional to the quark mass, this coupling
is very small.  The primary production mechanism for a Higgs boson
in hadronic collisions is through gluon fusion, $g g \rightarrow h$,
which is shown in Fig. \ref{gghfig}.\cite{glue}
 The loop contains all quarks in the theory 
and is 
the dominant contribution to Higgs boson production at the
LHC for all $M_h<1~TeV$. 
(In extensions of the standard model, all massive colored particles
run in the loop.)

The lowest order cross section
for $gg\rightarrow h$ is,\cite{hhg,glue}
\beqn
{\hat \sigma}(gg\rightarrow h)&= &
{\alpha_s^2\over 1024 \pi v^2}
\mid \sum_q  F_{1/2}(\tau_q)\mid^2\delta    
\biggl(1-{{\hat s}\over M_h^2}\biggr)
\nonumber \\
&\equiv &{\hat \sigma}_0(gg\rightarrow h)\delta
\biggl(1-{{\hat s}\over M_h^2}\biggr)
\quad ,
\label{sigdef}
\eeqn     
where ${\hat s}$ is the gluon-gluon sub-process center of mass energy,
$v=246~GeV$,
and $F_{1/2}(\tau_q)$ is defined in Eq. \ref{etadef}.
In the heavy quark
limit, $(M_t/M_h \rightarrow \infty)$, the cross
section becomes a constant,
\beq
{\hat \sigma}_0(gg\rightarrow h)\sim {\alpha_s^2\over
576 \pi v^2}      \quad .
\label{siginf}
\eeq
Just like the decay process, $h\rightarrow gg$,
this rate counts the number of heavy quarks and  so could
be a window into a possible fourth generation of quarks.

The Higgs boson production cross section at a hadron collider can be 
found by integrating the parton cross section,
$\sigma_0(pp\rightarrow h)$,
 with the gluon
structure functions, $g(x)$,
\beq
\sigma_0(pp\rightarrow h)={\hat \sigma_0}\tau\int_\tau^1
{dx\over x} g(x)g\biggl({\tau\over x}\biggr),
\label{logh}
\eeq
where $\sigma_0$ is given in Eq. \ref{sigdef},
 $\tau\equiv M_h^2/S$, and $S$ is
the hadronic center of mass energy.   

\begin{figure}[t]
\centering   
\epsfxsize=4.in
\leavevmode
\epsffile{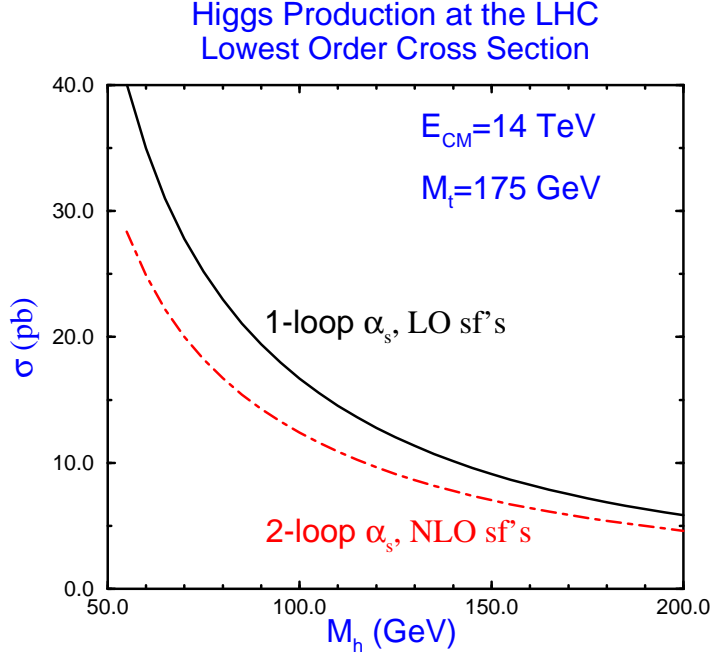} 
\caption{Cross section for gluon fusion, $gg\rightarrow h$, at the
LHC using the lowest order parton cross section of Eq. \ref{sigdef}.  The
solid (dot-dashed) line uses the one-loop (two-loop) expression
for $\alpha_s(\mu)$, along with structure functions fitted
to lowest order (next to lowest order) expressions for the 
data.  The renormalization scale is $\mu=M_h$.}
\label{lhcgg}
\end{figure}

We show the rate obtained using the lowest order parton cross section
of Eq. \ref{sigdef}
in Fig. \ref{lhcgg} for the LHC.  When computing the lowest order result 
from Eq. \ref{sigdef}, it is ambiguous whether to use the one-
 or  two- loop
result for $\alpha_s(\mu)$
 and which structure functions to use;
a set fit to data using only the lowest order  in $\alpha_s$ predictions
or a set which includes some higher order effects.  The difference
between the equations for $\alpha_s$ and the different structure functions
is ${\cal O}(\alpha_s^2)$ and hence  higher order in $\alpha_s$ when one is
computing the ``lowest order'' result.  In Fig. \ref{lhcgg},
 we show two 
different definitions of the lowest order result and see that they
differ significantly from each other.  We will see in the next
section that the result obtained using the $2$-loop $\alpha_s$ and
NLO structure functions, but the lowest order parton  cross section,
is a poor approximation to the radiatively corrected rate.  
Fig. \ref{lhcgg} takes the scale factor $\mu=M_h$ and  the results are
quite sensitive to this choice.  

\subsection{QCD Corrections to $gg\rightarrow h$}

In order to obtain reliable predictions
for the production rate, it is important to compute
the $2-$loop QCD
radiative corrections to $gg\rightarrow h$.  The complete
${\cal O}(\alpha_s^3)$
calculation is available in Ref 39.  
The analytic result is quite complicated, but the 
computer code including all QCD radiative corrections
is readily available.

The result in the $M_h/M_t\rightarrow 0$ limit
turns out to be an excellent approximation to the
exact result for the 2-loop corrected
rate for $gg\rightarrow h$ and can be used in most cases.  The 
heavy quark limit can be obtained from
 the gauge invariant effective Lagrangian,\cite{qcd1,qcd2}
\beq
{\cal L}=-{1\over 4} \biggl[1-{2\beta_s\over g_s(1+\delta)}
{h\over v}\biggr]
G_{\mu\nu}^AG^{A\mu\nu}
-{M_t\over v} {\overline t} t h
\quad ,
\label{lowen}
\eeq
where $\delta=2\alpha_s/\pi$ is the
anomalous mass dimension  arising from the renormalization
of the $t {\overline t}h$ Yukawa coupling constant, $g_s$ is the
QCD coupling constant, and $G_{\mu\nu}^A$ is the color $SU(3)$
field.  This Lagrangian can be derived using low energy
theorems which are valid in the limit $M_h<<M_t$
and  yields momentum dependent $ggh$, $gggh$,
and $ggggh$ vertices which can be used to compute the 
rate for $gg\rightarrow h$ to ${\cal O}(\alpha_s^3)$.\cite{russ}  
  
Since the $hgg$ coupling  in the $M_t\rightarrow\infty $ limit
results from heavy fermion loops, it is
only the heavy fermions which contribute to $\beta_s$ in Eq.
\ref{lowen}.  To 
${\cal O}(\alpha_s^2)$, the heavy fermion contribution to the QCD $\beta$ 
function is,
\beq
{\beta_s\over g_s}\mid{\rm heavy~fermions}=
N_H{\alpha_s\over 6\pi}\biggl[
1+{19\alpha_s\over 4}\biggr]
\quad ,
\eeq
where $N_H$ is the number of heavy fermions.

The parton level cross section for $gg\rightarrow h$ is
found by computing the ${\cal O}(\alpha_s^3)$ virtual
graphs for $gg\rightarrow h$ and combining them with the 
bremsstrahlung process $gg\rightarrow gh$.  The answer
in the heavy top quark limit is,\cite{qcd1,qcd2}
\beq 
{\hat \sigma}_1(gg\rightarrow h X)
={\alpha_s^2(\mu)\over 576 \pi v^2}
\biggl\{ \delta(1-z)+{\alpha_s(\mu)\over\pi}
\biggl[ h(z)+{\bar h}(z)\log\biggl({M_h^2\over \mu^2}\biggr)
\biggr]\biggr\}
\label{qcdsig}
\eeq
where
\beqn
h(z)&=& \delta(1-z)\biggl(\pi^2+{11\over 2}\biggr)-{11\over 2}(1-z)^3
\nonumber \\
&&+6\biggl(1+z^4+(1-z)^4\biggr)\biggl({\log(1-z)\over 1-z}\biggr)_+
-{\overline h}(z) \log(z)
\nonumber \\
{\bar h}(z)&=& 6 \biggl({z^2\over (1-z)_+} 
+(1-z)+z^2(1-z)\biggr)
\label{hdefs}
\eeqn
and $z\equiv M_h^2/{\hat s}$. 
The answer is written in terms of ``+'' distributions,
which are defined by the integrals,
\beq
\int_0^1{f(x)\over (1-x)_+}\equiv\int^1_0{f(x)-f(1)
\over 1-x}
\quad .
\eeq
The factor $\mu$ is an arbitrary renormalization
point.  To $\alpha_s^3$, the physical hadronic cross
section is independent of $\mu$.  
  There are also ${\cal O}
(\alpha_s^3)$ contributions from $q {\overline q}$, $qg$
and ${\overline q}g$
initial states, but these are numerically small.

We can define a $K$ factor as
\beq
K\equiv{{\sigma}_1(pp\rightarrow hX)\over
{ \sigma}_0(pp\rightarrow h)}
\quad ,
\label{kfacdef}
\eeq
where $\sigma_1(pp\rightarrow hX)$ is the ${\cal O}(\alpha_s^3)$
radiatively corrected rate for Higgs production and $\sigma_0$
is the lowest order rate found from  Eq. \ref{sigdef}.
From Eq. \ref{hdefs},
 it is apparent that a significant portion of the corrections
result from the rescaling of the lowest order result,
\beq
K\sim 1+{\alpha_s(\mu)\over \pi}\biggl[ \pi^2+{11\over 2}
+ \cdots\biggr]
\quad .
\eeq
Of  course $K$ is not a constant, but depends on the renormalization
scale $\mu$ as well as $M_h$.  
The radiatively corrected cross section ${\hat \sigma}_1$ should
be convoluted with next-to-leading order structure functions,
while it is ambiguous which structure
functions and definition of $\alpha_s$ to use
in defining the lowest order result, ${\hat \sigma}_0$, as
discussed above. 

The
$K$ factor varies between $2$ and $3$ at the LHC
and so the QCD corrections significantly increase the
rate from the lowest order result.
The heavy top quark limit is an
excellent approximation to the $K$ factor.  
The easiest way to compute the radiatively corrected cross section is
therefore to calculate the lowest order cross section including the
complete mass dependence of Eq. \ref{sigdef}
 and then to multiply by the $K$ factor
computed in the $M_t\rightarrow \infty$ limit.
This result will be extremely accurate.

The other potentially
 important correction to the $hgg$ coupling is the
two-loop electroweak
 contribution involving the top quark, which is of
${\cal O}(\alpha_S G_F M_t^2)$.  In the heavy quark limit,
the function $F_{1/2}(\tau_q)$
 of Eq. \ref{etadef} receives a contribution,~\cite{ewh}
\beq
F_{1/2}(\tau_q)\rightarrow
F_{1/2}(\tau_q)\biggl(1+{G_FM_t^2\over 16\sqrt{2}\pi^2}
\biggr)
\quad .
\eeq
When the total rate for Higgs production is computed, the
${\cal O}(\alpha_s G_F M_t^2)$ contribution is $<.2\%$ and
so can be neglected.  
The ${\cal O}(\alpha_s G_F M_t^2)$ contributions 
therefore do not spoil
the usefulness of the $gg\rightarrow h$ mechanism 
as a means of  counting
heavy quarks.

\begin{figure}[t]
\centering   
\epsfxsize=4.in
\leavevmode
\epsffile{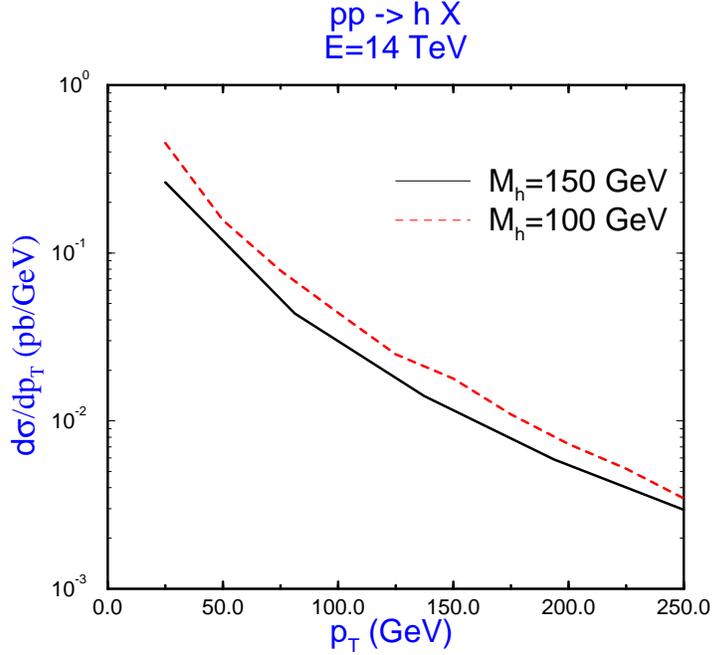} 
\caption{Higgs boson transverse momentum distribution 
 from Eq. \ref{pth}.}
\label{ggpt}
\end{figure}

At lowest order the gluon fusion process yields a Higgs boson with
no transverse momentum. 
At the next order in perturbation theory, gluon fusion
produces a Higgs boson with finite $p_T$, primarily through
the process $gg\rightarrow gh$.  At low $p_T$, 
the parton cross section diverges as $1/p_T^2$,\cite{qcdsum}
\beqn 
{d{\hat\sigma}\over d {\hat t}}(gg\rightarrow g h)
&=&{\hat \sigma}_0{3 \alpha_s\over 2 \pi}\biggl\{
{1\over p_T^2}\biggl[
\biggl(1-{M_h^2\over {\hat s}}\biggr)^4 + 1
+\biggl( {M_h^2\over {\hat s}}\biggr)^4\biggr]
\nonumber \\
&& - {4\over {\hat s}}\biggl(1-{M_h^2\over {\hat s}}
\biggr)^2+{2 p_T^2\over {\hat s}}\biggr\}\quad .
\label{pth}
\eeqn 
 The hadronic cross section can be found by integrating
 Eq. \ref{pth} with
the gluon structure functions.  In Fig. \ref{ggpt}, we show the $p_T$
spectrum of the Higgs boson at ${\cal O}(\alpha_s^3)$.
At the LHC, the event rate even at large $p_T$ is significant.  This figure
clearly demonstrates the singularity at $p_T\rightarrow 0$.

\subsection{Finding the Higgs Boson at the LHC}

We turn now to a discussion of search techniques for the
Higgs boson at the LHC.  For $M_h< 1~TeV$, gluon
fusion is the primary production mechanism.  \begin{figure}[t]
\centering   
\epsfxsize=4.in
\leavevmode
\epsffile{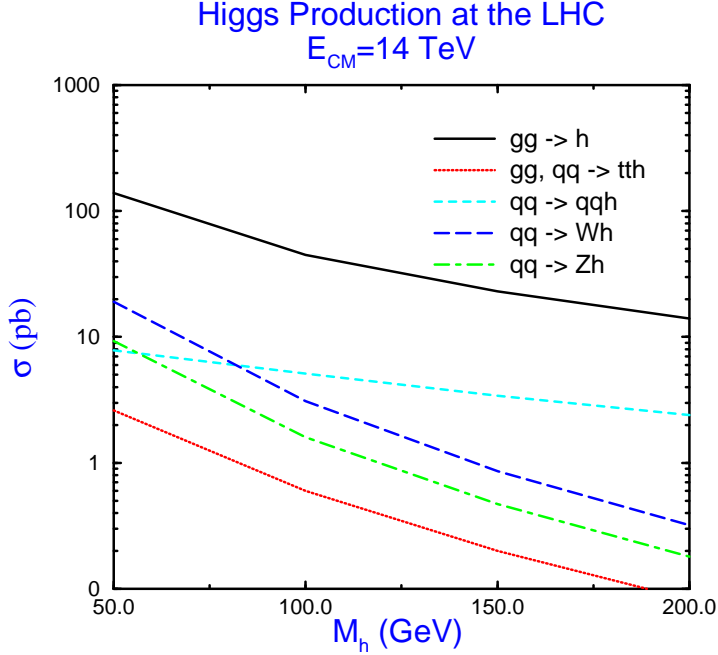}
\caption{Processes contributing to Higgs production at the LHC.}
\label{lhcprod}
\end{figure}
Fig. \ref{lhcprod} shows the Higgs boson production rates from
various processes at the LHC.
  At the present time, there are two
large detectors planned for the LHC; the ATLAS detector\cite{atlas}
and the CMS detector\cite{cms}. A more detailed
discussion of the experimental
issues involved in searching for the Higgs boson at the LHC can be found
in the 1997 TASI lectures of F.~ Paige.\cite{fptasi}

The production rate for the Higgs boson at the LHC
is significant, $\sigma_h\sim  10~pb$ for $ M_h
\sim 200~GeV$.  However, in order to see the Higgs boson it must decay into
some channel where it is not overwhelmed by the background. 
For $M_h < 2 M_W$ the Higgs boson decays predominantly to
$b {\overline b}$ pairs.
Unfortunately, the QCD production
of $b$ quarks is many orders of magnitude larger than Higgs production
and so this channel is thought to be useless.\cite{agel}
  One is led to consider
rare decay modes of the Higgs boson where the backgrounds may be smaller.
The decay channels which have received the most attention 
are $h\rightarrow \gamma\gamma$ and 
$h\rightarrow Z Z^*$.\cite{gg}\footnote{References to the many
studies of  the decays 
$h\rightarrow\gamma\gamma$ and $h\rightarrow Z Z^*$
at the LHC
can be found in Ref. 45.}
The branching ratios for these decays are shown in Fig. \ref{gagwid}
 and can
be seen to be  quite small.  (The rates for
off-shell gauge bosons,
 $h\rightarrow V V^*$, $V=W^\pm,Z$
must be multiplied by the relevant
branching ratios, $V\rightarrow f^{\prime} {\overline f}$.)

The  $h\rightarrow Z Z^*$ decay mode can lead to a final state
with $4$ leptons, $2$ of whose mass reconstructs to $M_Z$ while the
invariant mass of the $4$ lepton system reconstructs to $M_h$.  The
largest background to this decay is $t \overline{t}$ production
with $t\rightarrow W b \rightarrow (l \nu) ( c l \nu)$.  There are also
backgrounds from $Z b \overline{b}$ production, $Z Z^{*}$ production,
etc.
For $M_h=150~GeV$, the  ATLAS collaboration
estimates that there will be 184 signal events and 840 background
events in their detector in one year from $h\rightarrow Z Z^*
\rightarrow (4l)$ with the 4-lepton invariant mass in a 
mass bin within $\pm 2 \sigma$ of $M_h$.\cite{atlas}
  The leptons from Higgs decay            
tend to be more isolated from other particles than those coming
from the backgrounds and a series of isolation cuts
can be used to reduce the rate
to 92 signal and 38 background events. 
 The ATLAS collaboration
claims that they will be able to discover the Higgs boson in the
$h\rightarrow Z Z^*\rightarrow l^+l^-l^+l^-$ mode for
$130~GeV < M_h < 180~GeV$ with an integrated luminosity
of $10^5~ pb^{-1}$ (one year of running at the LHC
with design luminosity) and using both the electron and muon
signatures.  For $M_h< 130~GeV$, there are not
enough events since the branching ratio is too small (see Fig. \ref{gagwid}),
while for $M_h> 180~GeV$ the Higgs search can proceed via the
$h\rightarrow ZZ$ channel, which we discuss in Section 7.2.  

For $M_h< 130~GeV$, the Higgs boson can be searched for through
its decay to two photons.  The branching ratio in this region
is about $10^{-3}$, so for a Higgs boson with $M_h\sim 100~GeV$
there will be about $3000$ events per year.
  The
Higgs boson decay into the $\gamma \gamma$ channel
is an extremely narrow resonance in this region  with
a width around $1~KeV$.  From Fig. \ref{gagwid}  we see that the branching
ratio for $h\rightarrow \gamma\gamma$ falls  rapidly with
increasing $M_h$ and so this decay mode is probably only useful in
the region $80~GeV < M_h < 130~GeV$.

The irreducible background to $h\rightarrow \gamma\gamma$ comes from $q 
\overline{q}\rightarrow \gamma\gamma$ and $gg \rightarrow \gamma
\gamma$.
Extracting the  narrow signal from the immense background poses a formidable
experimental challenge.  The detector must have a mass resolution
on the order of $\delta m/m\sim 1.5 \%$ in order to be able to hope
to observe this signal.
For $M_h=110~GeV$ the ATLAS
collaboration estimates that there
will be $1430$ signal events and $25,000$ background
events in a mass bin equal to the Higgs width.  This leads to a
ratio ,
\beq
 { {\rm Signal} \over \sqrt{{\rm Background}}}\sim 9
.\eeq
  A ratio greater than
$5$ is usually ${\it defined}$ as a discovery.  ATLAS  claims that
they will be able to discover the Higgs boson in this channel for
$100~GeV < M_h < 130~GeV$.  (Below $100~GeV$ the background is too
large and above $130~GeV$ the event rate is too small.)
Because of its finely grained calorimeter, the CMS collaboration 
expects to do slightly better.\cite{cms}

There are many additional difficult experimental problems associated with
the decay channel $h\rightarrow \gamma \gamma$.  The most significant
of these is the confusion of a photon with a jet.  Since the cross
section for producing jets is so much larger than that of $h\rightarrow
\gamma\gamma$ the experiment must not mistake a photon for a jet more
than one time in $10^4$.  It has not yet been demonstrated that this
is experimentally feasible.

One might think that the decay $h\rightarrow \tau^+\tau^-$ 
could be measured
since as shown in Fig. \ref{brferm}
 its branching ratio is considerably larger than $h\rightarrow
Z Z^* $ and $h\rightarrow \gamma \gamma$; $BR(h\rightarrow \tau^+
\tau^-)\sim 3.5\%$ for $M_h$ in the $100~GeV$ region.
The problem is that for the
dominant production mechanism, $gg\rightarrow h$, the Higgs boson
has no transverse momentum and so the $\tau^+\tau^-$ invariant mass
cannot be reconstructed.  If we use the production mechanism
$g g \rightarrow h g$, then the Higgs is produced at large transverse
momentum and it is  possible to reconstruct the $\tau^+\tau^-$ invariant
mass.  Unfortunately, the background from 
$ q\overline{q}
\rightarrow \tau^+\tau^-$ and from $t \overline{t}$ decays 
is much larger than
the signal.\cite{keith}
Recent studies, however, have shown that the $h\rightarrow \tau^+\tau^-$
decay channel may be  useful
at the LHC for $M_h\sim~110-150~GeV$ with $30~fb^{-1}$
of data.\cite{rain}

\subsection{Associated Higgs Boson Production}

At the Tevatron and the LHC
the process $q {\overline q}\rightarrow Wh$ offers the hope 
of being able to tag the Higgs boson by the 
$W$ boson decay products.\cite{stange} 
This process has the rate:
\beq
{\hat \sigma}(q_i {\overline q}_j\rightarrow W^\pm h)
={G_F^2 M_W^6 \mid V_{ij}\mid ^2\over 6\pi {\hat s}^2
(1-M_W^2/{\hat s})^2}\lambda_{Wh}^{1/2}\biggl[
1+{{\hat s}\lambda_{Wh}\over 12 M_W^2}\biggr]
\quad ,
\eeq
where $\lambda_{Wh}
=1-2(M_W^2+M_h^2)/{\hat s}+(M_W^2-M_h^2)^2/{\hat s}^2$
and $V_{ij}$ is the Kobayashi-Maskawa angle associated with the
$q_i {\overline q}_j W$ vertex.  
This process is sensitive to the $W^+W^-h$ coupling and
so will be different in extensions of the Standard Model.

\begin{figure}[t]
\centering   
\epsfxsize=4.in
\leavevmode
\epsffile{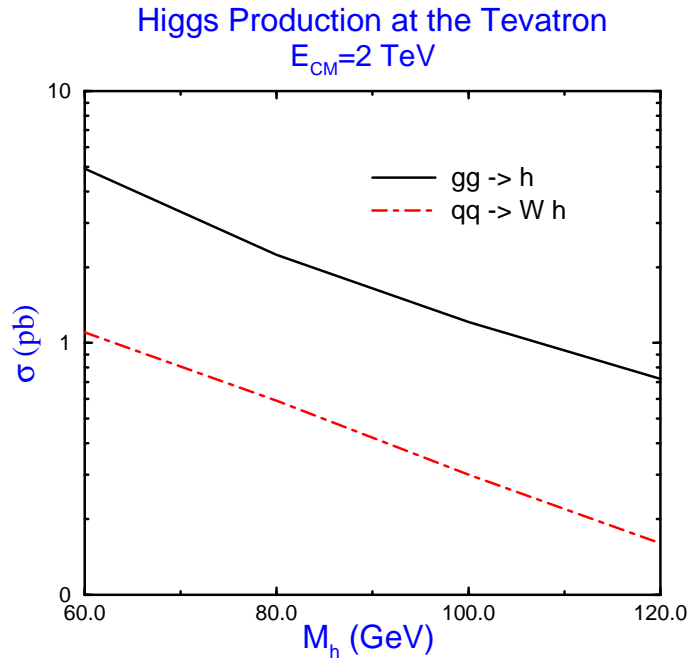}
\caption{Next to leading order QCD  predictions for Higgs
boson production at the Tevatron.
 The dot-dashed line is the $W^\pm h$ production rate (summed
over $W^\pm$ charges),
while the solid line is the rate for
Higgs production from gluon
fusion.}
\label{tevwh}
\end{figure}

\begin{figure}[t]
\centering   
\epsfxsize=4.in
\leavevmode
\epsffile{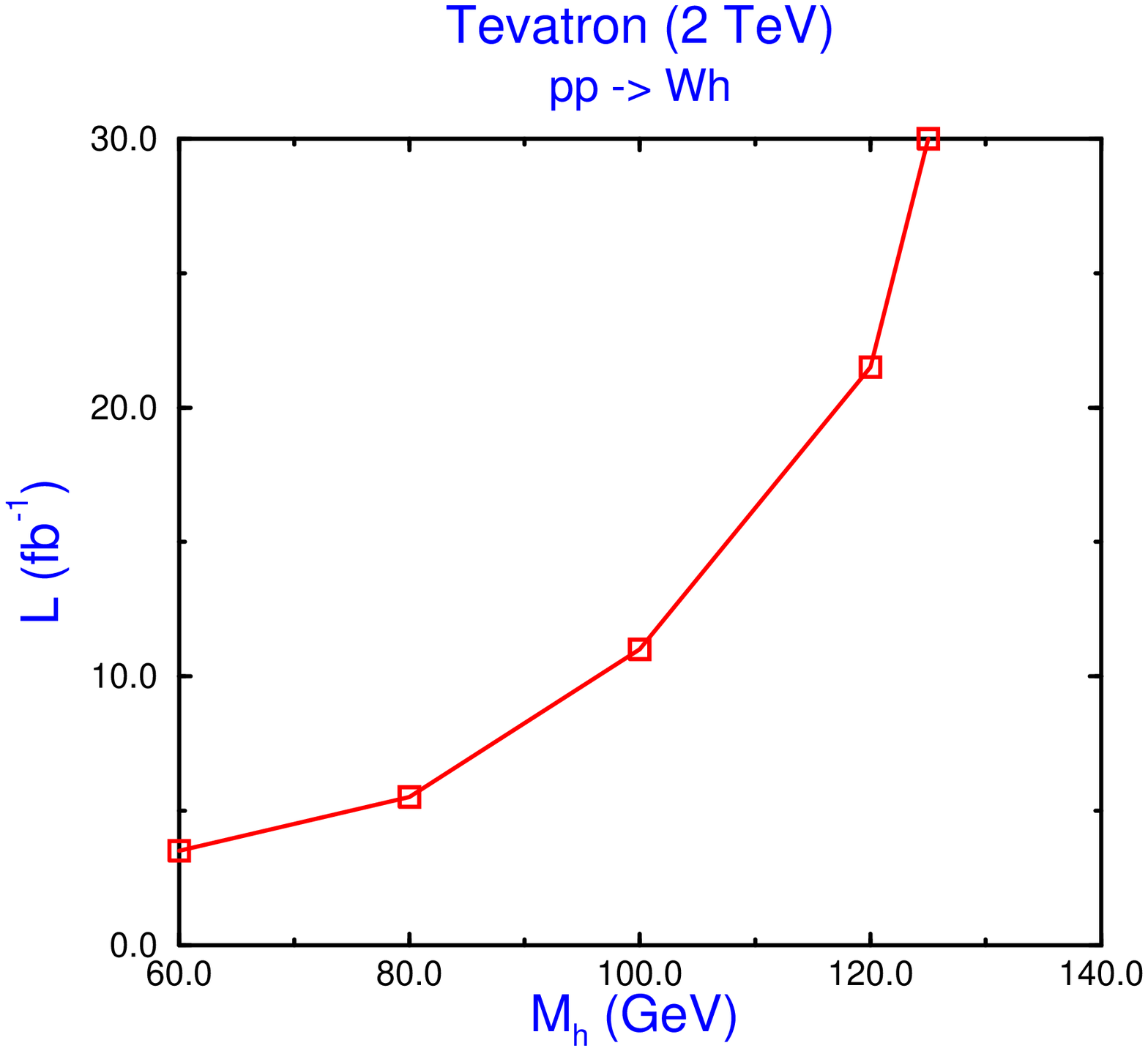}
\caption{Luminosity required to obtain a $5\sigma$ signal in
the process, $p {\overline p} \rightarrow Wh, W\rightarrow
l \nu, h\rightarrow b {\overline b}$ at $\sqrt{s}=2~TeV$.
  From Ref. 54.}
\label{tevlum}
\end{figure}  

Since this mechanism produces a relatively small number of
signal events, (as can be seen clearly in Figs. \ref{lhcprod} 
and \ref{tevwh}),
it is
important to compute the rate as accurately as possible
by including the QCD radiative corrections.  This has been
done in Ref. 51, where  it is shown that
the cross section  can be  written as
\beq
{d\sigma\over d q^2
} (pp\rightarrow W^\pm h)=\sigma (pp\rightarrow W^{\pm *}
) 
{G_F M_W^4\over \sqrt{2}\pi^2 (q^2-M_W^2)^2}
{\mid {\vec p}\mid\over \sqrt{q^2}}\biggl(
1+{\mid {\vec p}\mid^2\over 3 M_W^2}\biggr)
\label{wvh}
\eeq 
to all orders in $\alpha_s$.  In Eq. \ref{wvh}, $W^*$ is a virtual
$W$ with momentum $q$ and $\mid {\vec p}\mid= 
\sqrt{s}\lambda_{Wh}^{1/2}/2$ is
the momentum of the outgoing $W^\pm$ and $h$.
  From Eq. \ref{wvh}, it is clear
that the radiative corrections to $W^\pm h$ production are identical to
those for the Drell-Yan process which have been known
for some time.  Using the DIS factorization scheme, the
cross section at the LHC is increased by roughly $17\%$ 
over the lowest order rate.  The QCD
corrected cross section is relatively insensitive to the choice
of renormalization and factorization scales.  It is, however,
quite sensitive to the choice of structure functions.  The rate for
$pp\rightarrow W^\pm h$ at the LHC is shown in Fig. \ref{lhcprod} 
(the long-dashed curve) and is more than
an order of magnitude smaller than the rate from gluon fusion.
The rate for $pp\rightarrow Zh$ is smaller still.  
   
The $Wh$ events can be tagged by identifying the charged lepton from the
$W$ decay.  Imposing isolation cuts on the lepton significantly reduces the
background.  At the LHC, there are sufficient events that the Higgs
produced in association with a $W$ 
boson can be identified through the $\gamma\gamma$ decay mode.\cite{stange}
ATLAS claims a $4\sigma$ signal in this channel for $80~GeV < M_h < 120~GeV$
with $100~ fb^{-1}$, (this corresponds to about $15$ signal events),
while CMS hopes to find a $6-7\sigma$ effect in this channel.
There  are a large number of $Wh$ events with $W\rightarrow l \nu$
and $h\rightarrow b {\overline b}$, but unfortunately the
backgrounds  to this decay chain
are difficult to reject and observation of this signal
will probably require a high luminosity.\cite{stange}

Associated production of a
Higgs boson with a $W^\pm$ boson can also
potentially be observed at the Tevatron.\cite{tev2000}
For a $100~GeV$ Higgs boson, the lowest order cross section
is roughly $.2~pb$.  Including the next-to-leading order corrections
increases this to $.3~pb$, while summing over the soft
gluon effects increases the NLO result by $2-3\%$.
\cite{mren}
The next to leading order rate 
for $p {\overline p}\rightarrow Wh$
is shown in Fig. \ref{tevwh} and is much
smaller that that from gluon fusion.
  At the
Tevatron, the Higgs boson  from $W^\pm h$
production must be searched for in the $b {\overline b}$
decay mode since the $\gamma\gamma$ decay mode produces
too few events to be observable.
  The largest backgrounds are $W b {\overline b}$ 
and $WZ$, along with top quark production.
The background from top quark production is considerably smaller at
the Tevatron than at the LHC, however. 
  The Tevatron with 
$2-4~fb^{-1}$ will be sensitive to $M_h< M_Z$, a region 
already probed by LEP2.

  An upgraded Tevatron
with higher luminosity (say $25-30~fb^{-1}$) may be able to probe  
 a Higgs boson
mass up to about $130~GeV$ through the $Wh$ production
mechanism.\cite{tev2000,tevwhstud} 
This result
depends critically on the $b$ tagging capabilities of the
detectors, since it requires reconstructing the mass of both
$b-$ jets. Fig. \ref{tevlum} shows the luminosity required to obtain
a $5~\sigma$ signal at the Tevatron as a function of Higgs 
mass.
The $120-130~GeV$ mass region is
particularly important since it is above the kinematic
threshold of LEP2 and is the most challenging region to probe
at the LHC.

\section{Higgs Boson Production from Vector Bosons}
\subsection{The Effective $W$ Approximation}

We turn now to the  study  of
the couplings of the Higgs boson to gauge
bosons.  We begin by studying the diagram in Fig. 18.
\begin{figure}[t]
\centering
\begin{picture}(200,100)
\SetScale{1.}
\ArrowLine(0,100)(50,75)
\ArrowLine(50,75)(100,100) 
\Photon(50,75)(75,50){3}{6}
\Photon(75,50)(50,25){3}{6}
\ArrowLine(0,0)(50,25) 
\ArrowLine(50,25)(100,0)
\DashLine(75,50)(125,50){5}
\put(70,30){$W, Z$} 
\put(140,45){$h$}
\put(70,60){$W,Z$}
\end{picture}
\label{wwhdiag}
\caption{Contribution to heavy Higgs boson production from
vector boson scattering.}
\end{figure}
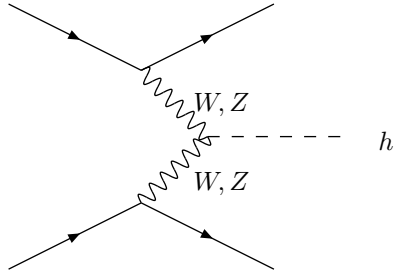 
Naively, one expects this diagram to give a negligible contribution
to Higgs production because of the two $W$ boson propagators.  However,
it  turns out that this production mechanism 
can give an important contribution.
The diagram of Fig.~18
can be interpreted in parton model language
as the resonant scattering of two $W$ bosons to form a Higgs 
boson\cite{effw} and we
can compute the distribution of $W$ bosons in a quark in an
analogous manner to the computation of the distribution of photons
in an electron.\cite{effgam}
By considering the $W$ and $Z$ gauge bosons as partons, calculations
involving gauge bosons in the intermediate states can be considerably
simplified.

In order to treat the $W^\pm$ and $Z$ bosons as partons, we consider them as
on-shell physical bosons.  We make the approximation that the 
partons have zero
transverse momentum, which ensures that the longitudinal and transverse
projections of the $W$ and $Z$ partons are uniquely specified.
We want to be able to write a parton level relationship:
\beq
\sigma(q_1+q_2\rightarrow q_1^\prime+X) =
\int^1_{M_W\over E} dx f_{q/W}(x)\sigma(W+q_2\rightarrow X).
\label{pardef}
\eeq
The function $f_{q/W}(x)$ is  called the distribution of
$W$'s in a quark and it is defined by Eq. \ref{pardef}.
The longitudinal and transverse  $W$ distributions in a quark 
can be found in a manner directly
analogous to the derivation of the effective
photon approximation,\cite{effw}

\beqn
f_{q/W}^L(x)&=
&{g^2\over 16 \pi^2}\biggl({1-x\over x}\biggr)
\nonumber \\
f_{q/W}^T(x)&=&{g^2\over 64\pi^2 x}\log\biggl({4 E^2\over M_W^2}\biggr)
\biggl[1+(1-x)^2\biggr],
\label{wdists}
\eeqn
where we have averaged over the two transverse polarizations
and $E$ is the relevant energy scale of the $WW$ subprocess.
The logarithm in Eq.~\ref{wdists} is the same logarithm which appears in the
effective photon approximation.
The result of Eq. \ref{wdists} violates our intuition that longitudinal gauge
bosons don't couple to massless fermions.
This is because the integral
over the angle between the outgoing quark and the emitted $W$
boson
picks out the small angle region and hence the
subleading term in the $W$ polarization tensor.

It is  straightforward to compute
the rates for
 processes involving $WW$ scattering.
The  hadronic cross section can be written in terms of a luminosity
of $W$'s in the proton and a subprocess cross section for the $WW$
scattering,
\beq
\sigma_{pp\rightarrow WW\rightarrow X}(s)=\int_{\tau_{min}}^1
d \tau {d {\cal L}\over d \tau}\mid_{pp/WW}\sigma_{WW\rightarrow X}
(\tau s)
\label{effwres}
\eeq
where the luminosities are defined: 
\beqn
 {d {\cal L}\over d \tau}\mid_{pp/WW}&=&\sum_{ij}
\int_\tau^1{d \tau^\prime \over \tau^\prime}
\int^1_{\tau^\prime}{dx\over x}
f_i(x)f_j\biggl({\tau^\prime\over x}\biggr)
 {d {\cal L}\over d \zeta}\mid_{q_i q_j/WW}
\nonumber \\
 {d {\cal L}\over d \tau}\mid_{qq/WW}&=&
\int_\tau^1{dx\over x}
f_{q/W}(x)f_{q/W}\biggl({\tau\over x}\biggr).
\label{lumo}
\eeqn
$f_i(x)$ are the quark distribution functions in the proton and
$\zeta\equiv \tau/\tau^{\prime}$.
The distributions of $Z$ bosons are found in an identical manner
and are roughly a factor of three smaller than the $W$
distributions due to the smaller couplings of the $Z$ to 
the fermions.   The approximation of Eq. \ref{effwres} has
been shown to be extremely accurate for heavy Higgs boson
production.

The effective $W$ approximation is particularly useful in models
where the electroweak symmetry breaking is  due not to the Higgs
mechanism, but rather to some strong interaction dynamics (such
as in a  technicolor model).  In these models one typically estimates
the strengths of the three and 
four gauge boson couplings due to the
new physics.  These interactions can then be folded into the 
luminosity of gauge bosons to 
get estimates of the size of the new physics effects.

\subsection{Searching for a Heavy Higgs Boson at the LHC}

We now have the tools necessary to discuss the search for a
very massive Higgs boson.   The rates
for the various production mechanisms
contributing to Higgs production  at the LHC 
are shown in Fig. \ref{lhcprod}.
For $M_h< 1~TeV$ the dominant production mechanism at the LHC is gluon
fusion.  For heavy Higgs boson masses,
the $W^+W^-$ fusion mechanism  is also an important
mechanism.  
  Searching for a Higgs boson on the TeV mass scale
will be extremely difficult due to the small rate.
   For example, a $700~GeV$ Higgs boson
has a cross section near $1~pb$ leading to around $10^5$ events/LHC
year.  The cleanest way to see these events is the so-called
``gold-plated" decay channel,
\beq
h\rightarrow ZZ\rightarrow (l^+l^-) (l^{\prime~+} l^{\prime~-}) .
\eeq
The lepton pairs will reconstruct to the $Z$ mass and the $4$
lepton invariant mass will give the Higgs boson  mass.
Since the branching ratio,
 $Z\rightarrow (4l)$, $l=e, \mu$ is $\sim .36\%$
the number of events for a $700~ GeV$ Higgs is reduced to around
$360$ four lepton events per year.  Since this number will be further
reduced by cuts to separate the signal from the background, it 
is clear that this channel will run out of events as the Higgs
mass becomes heavier.\cite{zzgold}

In order to look for still heavier Higgs bosons, one can look in the
decay channel,
\beq
h\rightarrow ZZ\rightarrow (l^+l^-)( \nu \overline{\nu}).
\eeq
Because  the branching ratio, $Z\rightarrow \nu \overline{\nu}$, 
is  approximately $ 20~\%$,
this decay channel has  a larger rate than the
four lepton channel.  However, the price is that
because of the neutrinos, events of this type cannot be fully
reconstructed.
This channel extends
the Higgs mass reach of the LHC slightly.

Another idea which has been proposed is to use the fact that events
coming from $WW$ scattering have outgoing jets at small angles,
whereas the $WW$ background coming from $q \overline{q}\rightarrow
W^+W^-$ does not have such jets.\cite{barger} Additional sources of 
background to Higgs detection
such as $W$ plus  jet production have jets at all angles.

The LHC will have the capability
to observe the Higgs boson between around $100~GeV <M_h
< 800~GeV$.
Since LEP2 will cover the region up to around $M_h\sim 100~GeV$
there will be no holes in the experimental
 coverage of the Higgs boson mass regions.

\section{Higgs Production at a High Energy $e^+e^- $ Collider}
\subsection{$e^+e^-\rightarrow {\overline l} l h$} 

\begin{figure}[t]
\centering   
\epsfxsize=4.in
\leavevmode
\epsffile{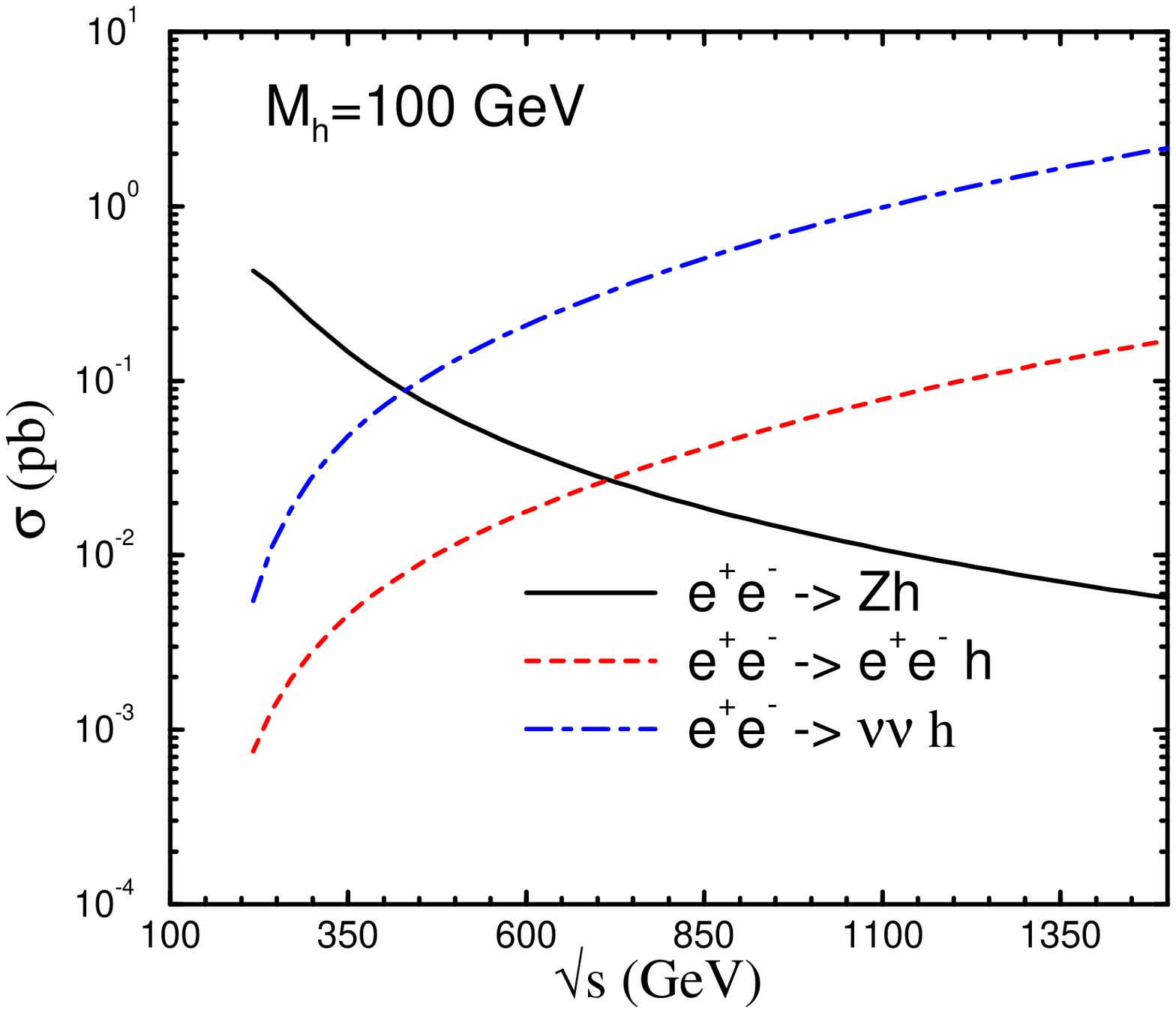} 
\caption{Higgs boson production in $e^+e^-$ collisions
for $M_h=100$.}
\label{eenlcprod}
\end{figure}

\begin{figure}[t]
\centering   
\epsfxsize=4.in
\leavevmode
\epsffile{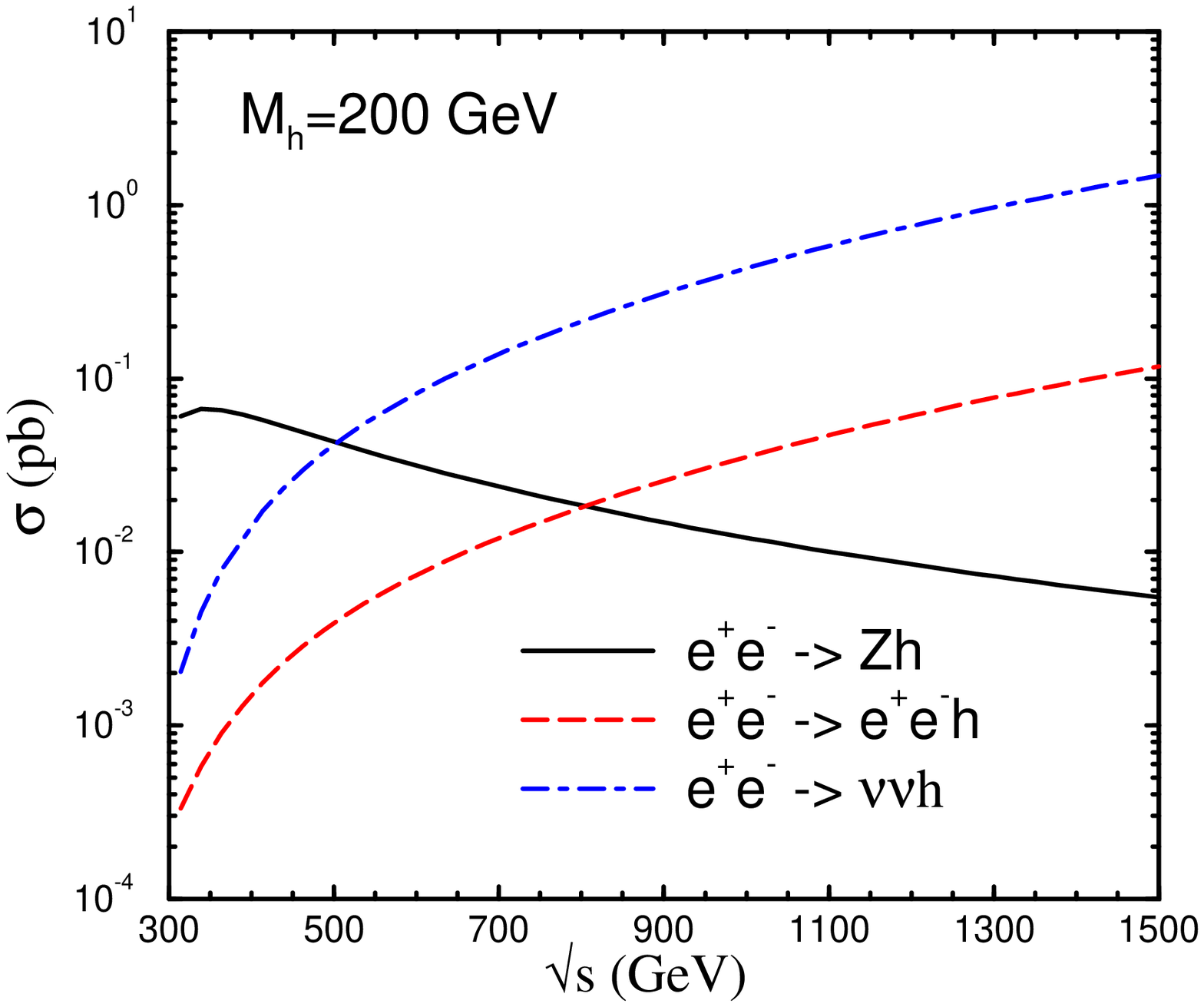} 
\caption{Higgs boson production in $e^+e^-$ collisions
for $M_h=200$.}
\label{eenlcprod2}
\end{figure}

In $e^+e^-$ collisions
the Higgs boson can be produced by
$e^+e^-\rightarrow Zh$, as discussed in Sec. 5.
This mechanism will probe close to the kinematic bound at LEP2. 
At higher energies the  $W^+W^-$ and $ZZ$ fusion processes become 
important,\cite{hhg,dr,dawros}
\beqn 
e^+e^-&\rightarrow& W^+W^- \nu {\overline \nu}
\rightarrow h \nu {\overline \nu},
\nonumber \\ 
e^+e^-
& \rightarrow  &ZZe^+e^-\rightarrow he^+e^-
.
\eeqn 
The fusion cross sections are easily found,~\cite{dr} 
\beqn
\sigma(e^+e^-\rightarrow VV\rightarrow {\overline l}
l h)=&{G_F^3M_V^4\over 64\sqrt{2}\pi^3}
\int^1_{{M_h^2\over s} } dx \int^1_x {dy\over
(1+s(y-x)/M_V^2)^2} \nonumber \\ &
\cdot \biggl[(v_e^2+a_e^2)^2f(x,y)+4v_e^2a_e^2g(x,y)\biggr]
\eeqn
where,
\beqn  
f(x,y)&=&\biggl({2x\over y^3}-{1+2x\over y^2}
+{2+x\over 2 y}-{1\over 2}\biggr)\biggl(
{w\over 1+w}-\log(1+w)\biggr)
\nonumber \\
&&~~+{x\over y^2}{w^2(1-y)\over 1+w}
\nonumber \\
g(x,y)&=& \biggl( -{x\over y^2}+{2+x\over 2 y}-{1\over 2}
\biggr)\biggl({w\over 1+w}-\log(1+w)\biggr)
\nonumber \\
w&\equiv & {y(sx-M_h^2)\over M_V^2  x}
\eeqn 
 and $v_e=a_e=\sqrt{2}$ for $e^+e^-\rightarrow W^+W^-\nu
\overline{\nu}\rightarrow \nu{\overline\nu} h$  
 and $v_e=-1+4 \sin^2\theta_W,~a_e=-1$
for $e^+e^-\rightarrow ZZe^+e^-\rightarrow e^+e^-h$.
The vector boson fusion cross sections are shown in Figs. \ref{eenlcprod} 
and \ref{eenlcprod2} as
a function of $\sqrt{s}$.  
                               
The $ZZ$ fusion cross section, $e^+
e^-\rightarrow e^+e^- h$,
 is an order of magnitude smaller than the
$W^+W^-$ fusion process due to the smaller neutral current couplings.
This suppression
is partially compensated for experimentally
 by the fact that the $e^+e^-h$ final
state permits a missing mass analysis to determine the Higgs
boson mass.

At an $e^+e^-$ collider with  $\sqrt{s}\sim 350~GeV$,
 the cross section for vector boson fusion,
 $e^+e^-\rightarrow W^+W^-\nu {\overline \nu}
\rightarrow h \nu {\overline \nu}$, and  that for
$e^+ e^-\rightarrow Zh$ are
of similar  size for a $100~GeV$ Higgs boson.   
The fusion processes grow as $(1/M_W^2)\log(s/M_h^2)$, while
the $s-$ channel process, $e^+e^-\rightarrow Zh$, falls as
$1/s$ and so at high enough energy the fusion process will
dominate, as can be seen in Figs. \ref{eenlcprod} and \ref{eenlcprod2}.

\subsection{$e^+e^-\rightarrow t {\overline t}h$}  
Higgs boson
 production in association with a $t {\overline t}$ pair is small at
an $e^+e^-$ collider.  At $\sqrt{s}=500~GeV$, $20~fb^{-1}$ of
luminosity would produce only $20$ events for $M_h=100~GeV$.  
The signature for this
final state is  spectacular, however, since the final
state  is
predominantly $W^+W^- b {\overline b} b {\overline b}$, which
has  a very small background.

The process $e^+e^-\rightarrow t {\overline t} h$ provides a direct
mechanism for measuring the $t {\overline t} h$ Yukawa coupling.
Since this coupling can be significantly different in a supersymmetric
model from that in the Standard Model, the measurement would provide a
means of discriminating between models.  The $ t {\overline t}h$
Yukawa coupling also enters into the rates for $gg\rightarrow h$ and
$h\rightarrow \gamma\gamma$ as these processes have large
contributions from top quark loops.  However, in these cases it is
possible that there is unknown new physics which also enters into the
calculation of the
rate and dilutes the interpretation of the signal as the measurement
of the $t {\overline t } h$ coupling.

The rate for $e^+e^-\rightarrow b {\overline b}h$ in the Standard
Model is probably not large enough to be measured due to the smallness
of the $b {\overline b}h$ Yukawa coupling.  In supersymmetric models
with large $\tan \beta$, however, the rate can be enhanced.

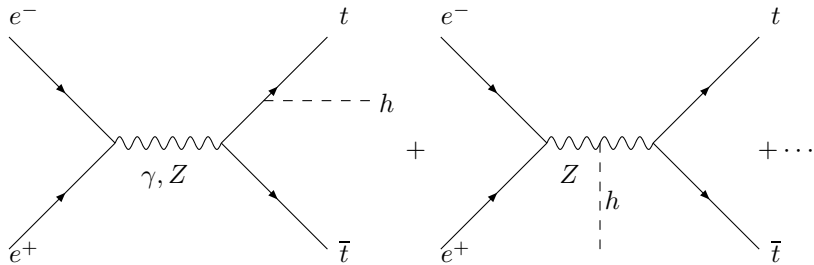
\begin{figure}[t]
\centering
\begin{picture}(100,100)(60,-5)
\SetScale{0.8}
\ArrowLine(0,100)(50,50)
\ArrowLine(0,0)(50,50)
\Photon(50,50)(100,50){3}{6}
\ArrowLine(100,50)(150,100) 
\ArrowLine(100,50)(150,0)
\DashLine(120,70)(170,70){5}
\put(150,35){$+$} 
\put(50,25){$\gamma, Z$} 
\put(0,85){$e^-$}
\put(0,-5){$e^+$}
\put(140,52){$h$}
\put(125,85){$t$}
\put(125,-5){${\overline t}$}
\SetScale{1}
\end{picture}
\begin{picture}(100,100)(0,-5)
\SetScale{0.8}
\ArrowLine(0,100)(50,50)
\ArrowLine(0,0)(50,50)
\Photon(50,50)(100,50){3}{6}
\DashLine(75,50)(75,0){5}
\ArrowLine(100,50)(150,100) 
\ArrowLine(100,50)(150,0)
\put(120,35){$+ \cdots$} 
\put(45,25){$Z$} 
\put(0,85){$e^-$}
\put(0,-5){$e^+$}
\put(62,15){$h$}
\put(125,85){$t$}
\put(125,-5){${\overline t}$}
\SetScale{1}
\end{picture} 
\caption[]{Feynman diagrams contributing to the lowest
order process, $e^+e^-\rightarrow t {\overline t} h$.}
\label{lofeyndiag}
\end{figure}

The cross section for $e^+e^-\rightarrow t\bar t h$ occurs through the
Feynman diagrams of Fig. \ref{lofeyndiag}.\cite{tth}

\begin{figure}[t]
\centering
\epsfxsize=4.in
\leavevmode\epsffile{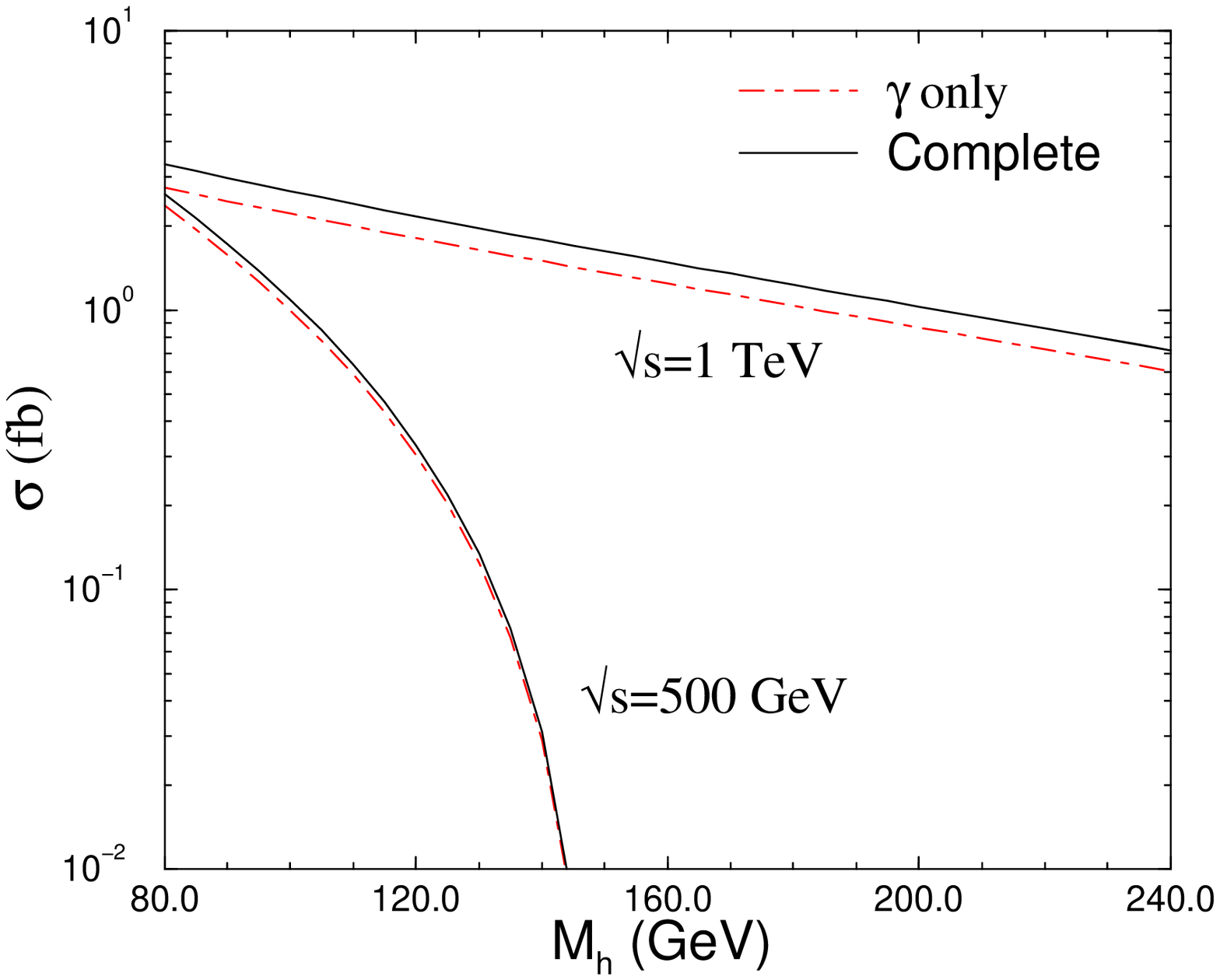}
\caption[]{Lowest order cross section for $e^+e^-\rightarrow t
{\overline t}h$ at $\sqrt{s}=500$~GeV and $\sqrt{s}=1~\mbox{TeV}$.
The curve labelled {\it complete} includes $\gamma,Z$ exchange,
along with bremsstrahlung from the $Z$ boson. We take $M_t=175~GeV$.}
\label{tthlofig}
\end{figure}

\beqn
\frac{d\sigma(e^+e^-\rightarrow t\bar t h^0)}{dx_h} &=& 
N_c\frac{\sigma_0}{(4\pi)^2}\left\{
\left[Q_e^2Q_t^2+\frac{2Q_eQ_tV_eV_t}{1-M_Z^2/s}
\right. \right. 
\nonumber \\ &&\left. +
\frac{(V_e^2+A_e^2)(V_t^2+A_t^2)}{(1-M_Z^2/s)^2}\right]G_1
\nonumber\\
&&+ \left. \frac{V_e^2+A_e^2}{(1-M_Z^2/s)^2}\left[A_t^2\sum_{i=2}^6G_i+
V_t^2(G_4+G_6)\right]
\right. \nonumber \\ &&\left. 
+\frac{Q_eQ_tV_eV_t}{1-M_Z^2/s}G_6 \right\}
\,\,\,\,,
\label{dsig0}
\eeqn

\noindent where $\sigma_0\!=\!4\pi\alpha^2/3s$, $\alpha$ is the
QED fine structure constant, $N_c=3$ is the number of colors, $x_h=2
E_h/\sqrt{s}$ with $E_h$ the Higgs boson energy, and $Q_i$, $V_i$ and
$A_i$ ($i\!=\!e$, $t$) denote the electromagnetic and weak couplings
of the electron and of the top quark respectively,

\beq
V_i =\frac{2I_{3L}^i-4Q_is_W^2}{4s_Wc_W}\,\,\,\,\,\,\,\,,\,\,\,\,\,\,\
A_i = \frac{2 I_{3L}^i}{4s_Wc_W}\,\,\,\,,
\eeq

\noindent with $I_{3L}^i\!=\!\pm 1/2$ being the weak isospin of the
left-handed fermions and $s_W^2\!=\!1-c_W^2=0.23$.  

The coefficients $G_1$ and $G_2$ describe the radiation of the Higgs
boson from the top quark.   
The other four coefficients, $G_3,\ldots,G_6$ describe the emission of
a Higgs boson from the $Z$-boson.
Analytic expressions for $G_i$ can be found in the first paper of
Ref. 62.  
 The most relevant
contributions are those in which the Higgs boson is emitted from a top
quark leg, i.e. those proportional to $G_1$ and $G_2$ in
Eq.~(\ref{dsig0}).  The contribution from the Higgs boson coupling to
the $Z$ boson is always less than a few per cent at $\sqrt{s}=500$~GeV
and $1$~\mbox{TeV} and can safely be neglected.  In
Fig.~\ref{tthlofig}, we show the complete cross section for
$e^+e^-\rightarrow t {\overline t}h$ production and also the
contribution from the photon exchange contribution only.  We see that at
both $\sqrt{s}=500$~GeV and $1$~TeV, the cross section is well
approximated by the photon exchange only. 

\begin{figure}[t]
\centering
\epsfxsize=4.in
\leavevmode\epsffile{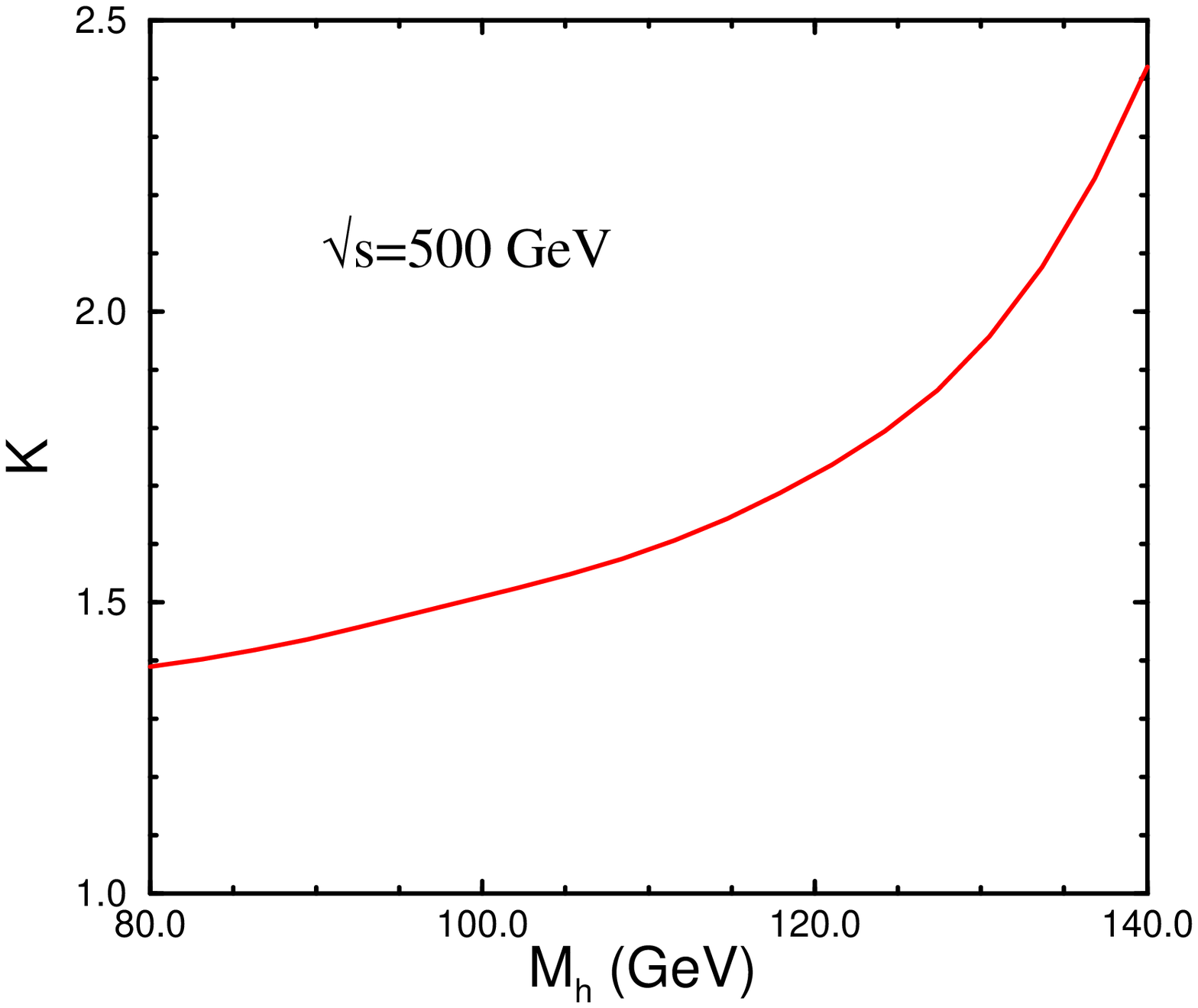}
\caption[]{Ratio of the ${\cal O}(\alpha_s)$ corrected rate to
the lowest order cross section for $e^+e^-\rightarrow t 
{\overline t}h$ at $\sqrt{s}=500$~GeV.
We take $M_t=175~GeV$
and $\alpha_s(M_t^2)=.1116$.}
\label{qcdeetth}
\end{figure}

The ${\cal O}(\alpha_s)$ corrections to the rate
for $e^+e^-\rightarrow t 
{\overline t}h$ increase the rate significantly at $\sqrt{s}=500$
and are shown in Fig. \ref{qcdeetth}.\cite{ttqcd}
The factor $K$ is defined to be the ratio of the ${\cal O}(\alpha_s)$
corrected rate to the lowest order rate.
At $\sqrt{s}=1~TeV$, the corrections are small.
Note that 
the only $\mu$ dependence occurs in $\alpha_s(\mu)$.  If $\mu=\sqrt{s}$,
then $K(M_h=100\,\mbox{GeV})$ is reduced to $1.4$ from the value
$K=1.5$ obtained with $\mu=M_t$ for $\sqrt{s}=500~GeV$. 
Further study is needed of the viability of this process as a means
of measuring the $t {\overline t} h$ Yukawa coupling.

\section{Strongly Interacting Higgs Bosons}

We can see from the scalar potential
of Eq.~43 that as the Higgs boson
grows heavy, its self interactions become large and 
it becomes
strongly interacting. The study of this regime
will therefore 
require new techniques.  For $M_h>1.4~TeV$, the total Higgs boson
decay  width is 
larger than its mass and it no longer makes sense to think of
the Higgs boson as a particle.\cite{marwil} A handy rule of thumb 
for the heavy Higgs boson is
\beq
\Gamma(h\rightarrow W^+W^-+ZZ)\sim 500~GeV
 \biggl({M_h\over 1~ TeV}\biggr)^3
\,\,.
\eeq

The $TeV$ Higgs boson
 regime can most easily be studied by going to the Goldstone
boson sector of the theory.  In Feynman gauge, the three Goldstone
bosons, $\omega^\pm,z$, have mass $M_{\omega,z}=M_{W,Z}$ and have
the interactions,
\beq
V={M_h^2\over 2v}h\biggl(h^2+Z^2+2 \omega^+\omega^-\biggr)
+{M_h^2\over 8 v^2}\biggl(h^2+z^2+2\omega^+\omega^-\biggr)^2.
\label{vgold}
\eeq

Calculations involving only the Higgs boson and the  Goldstone
bosons are easy since the scalar contribution is enhanced
by $M_h^2/v^2$, while contributions involving the gauge
bosons are proportional to $g^2$ and so can
be neglected for heavy Higgs bosons. For example
the amplitude for $\omega^+\omega^-\rightarrow \omega^+\omega^-$
is, \cite{lqt} 

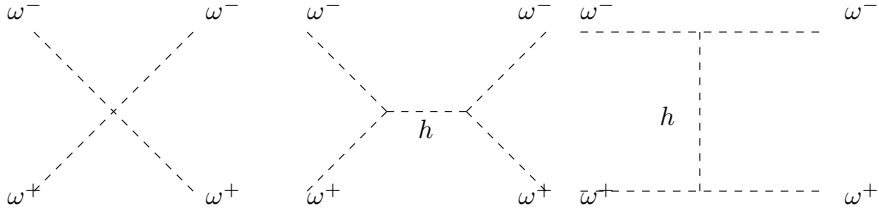
\begin{figure}[t]
\centering
\begin{picture}(100,100)(0,-5)
\SetScale{0.6}
\DashLine(0,100)(100,0){5}
\DashLine(0,0)(100,100){5} 
\put(65,-5){$\omega^+$} 
\put(65,65){$\omega^-$} 
\put(-10,-5){$\omega^+$}
\put(-10,65){$\omega^-$} 
\SetScale{1}
\end{picture}
\begin{picture}(100,100)(0,-5)
\SetScale{0.6}
\DashLine(0,0)(50,50){5} 
\DashLine(0,100)(50,50){5} 
\DashLine(50,50)(100,50){5}
\DashLine(100,50)(150,100){5}  
\DashLine(100,50)(150,0){5} 
\put(42,20){$h$} 
\put(0,65){$\omega^-$}
\put(0,-5){$\omega^+$}
\put(80,65){$\omega^-$}
\put(80,-5){$\omega^+$}
\SetScale{1}
\end{picture} 
\begin{picture}(100,100)(0,-5)
\SetScale{0.6}
\DashLine(0,0)(150,0){5}
\DashLine(0,100)(150,100){5}
\DashLine(75,0)(75,100){5}
\put(0,65){$\omega^-$}
\put(0,-5){$\omega^+$}
\put(100,65){$\omega^-$}
\put(100,-5){$\omega^+$}
\put(30,25){$h$}
\SetScale{1} 
\end{picture}
\caption{Goldstone boson scattering, $\omega^+\omega^-\rightarrow
\omega^+\omega^-$.}
\label{wwwwfig}
\end{figure}

\beq
{\cal A}( \omega^+\omega^-\rightarrow
\omega^+\omega^-)=-{M_h^2\over v^2}
\biggl({s\over s-M_h^2}+{t\over t-M_h^2}\biggr),
\label{wwwwgold}
\eeq
where $s,t,u$ are the Mandelstam variables in the $\omega^+\omega^-$ center
of mass frame.
It is instructive to compare Eq. \ref{wwwwgold} 
with what
is obtained
 by computing $W^+_LW^-_L\rightarrow W^+_L W^-_L$ 
using longitudinally polarized
gauge bosons and extracting the leading
power of $s$ from each diagram:\cite{duncan}
\beq
 {\cal A}(W^+_LW^-_L\rightarrow W^+_LW^-_L)\sim 
-{1\over v^2}\biggl\{
-s-t
+{s^2\over s-M_h^2}+{t^2\over t-M_h^2}\biggr\}
\quad .
\label{wwwwreal}
\eeq
From Eqs. ~\ref{wwwwgold} and ~\ref{wwwwreal} we find an
amazing result,
\beq
 {\cal A}(W^+_LW^-_L\rightarrow W^+_LW^-_L)= 
{\cal A}( \omega^+\omega^-\rightarrow
\omega^+\omega^-)+{\cal O}\biggl({M_W^2\over s}\biggr).
\eeq
This result means that instead of doing the complicated 
calculation with real gauge bosons, we can instead do
the easy calculation with only scalars if we are at an
energy far above the $W$ mass and are interested only 
in those effects which are enhanced by $M_h^2/v^2$.
  This is a general
result and has been given the name of the electroweak
equivalence theorem.\cite{et} The electroweak equivalence theorem
holds for $S-$ matrix elements; it is not true for individual
Feynman diagrams.    

The formal statement of the electroweak equivalence 
theorem is that 
\beqn 
{\cal A}(V_L^1 V_L^2....V_L^N\rightarrow
V_L^1 V_L^2....V_L^{N^\prime})&=& (i)^N(-i)^{N^\prime}
{\cal A}(\omega_1\omega_2...\omega_N\rightarrow
\omega_1\omega_2...\omega_{N^\prime})
\nonumber \\ && ~~ 
+{\cal O}\biggl({M_V^2\over s}\biggr),
\label{equivth}
\eeqn 
where $\omega_i$ is the Goldstone boson corresponding to
the longitudinal gauge boson, $V_L^i$.  In other words, when
calculating scattering amplitudes of longitudinal
gauge bosons  at high energy, we can
replace the ${\it external}$ longitudinal gauge bosons
by Goldstone bosons.  A formal proof of this theorem
can be found in Ref. 66.

The electroweak equivalence theorem is extremely useful in
a number of  applications.  For example, to compute the
radiative corrections to $h\rightarrow W^+W^-$
or to $W^+_LW^-_L\rightarrow W^+_LW^-_L$, the dominant
contributions which are enhanced by $M_h^2/v^2$ can be 
found by computing the one loop corrections to
$h\rightarrow \omega^+\omega^-$ and to $\omega^+
\omega^-\rightarrow \omega^+\omega^-$ considering  only  scalar
particles.\cite{marwil,sdsw}.
Probably the most powerful application
of the electroweak equivalence
theorem is, however, in the search for the physical effects of strongly
interacting gauge bosons which we turn to now.

\subsection{Unitarity}

\begin{figure}[t]
\centering   
\epsfxsize=3.in
\leavevmode
\epsffile{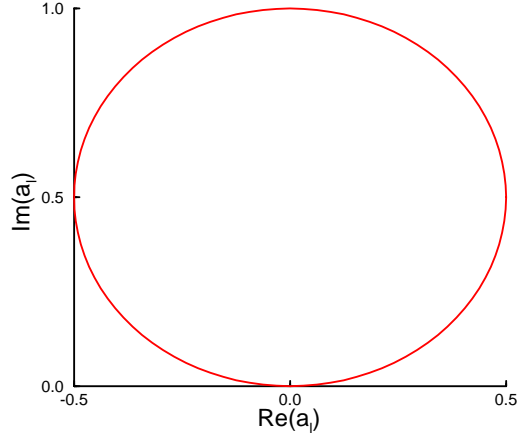} 
\caption{Argand diagram showing unitarity condition on scattering 
amplitudes.}
\label{arg}
\end{figure}

In the previous section we saw that the Goldstone
bosons have interactions which grow with energy.
However,
models which have cross sections rising with $s$ will eventually
violate perturbative unitarity.  To see this we consider 
$2\rightarrow 2$ elastic scattering.  The differential cross section is
\beq
{d \sigma \over d \Omega}={1\over 64 \pi^2 s} \mid {\cal A}\mid^2.
\eeq
Using a partial wave decomposition, the amplitude can be written as
\beq
{\cal A}=16 \pi \sum_{l=0}^{\infty} ( 2 l + 1)
P_l(\cos\theta)a_l   \quad ,
\eeq
where $a_l$ is the spin $l$ partial wave and $P_l(\cos\theta)$ are
the Legendre polynomials.  The cross section
  becomes,
\beqn
\sigma &=& {8 \pi\over s}\sum_{l=0}^{\infty}
\sum_{l^\prime=0}^{\infty}
(2 l+1) (2 l^\prime + 1) a_l a^*_l\nonumber \\
&&\qquad\cdot
\int_{-1}^1 d \cos\theta P_l(\cos \theta) P_{l^\prime}(\cos \theta)
\nonumber \\
&=& {16 \pi \over s}\sum_{l=0}^{\infty}(2l+1)\mid a_l\mid^2 \quad ,
\eeqn
where we have used the fact that the $P_l$'s are orthogonal.
The optical theorem  gives,
\beq
\sigma={1\over s}Im\biggl[{\cal A}(\theta=0)\biggr]
={16 \pi \over s}\sum_{l=0}^{\infty}(2l+1)\mid a_l\mid^2.
\eeq
This  immediately yields
 the unitarity requirement which is illustrated in
Fig. \ref{arg}.
\beq
\mid a_l\mid^2=Im(a_l).
\eeq
From Fig. \ref{arg}
 we see that one statement of unitarity is the requirement
that 
\beq
\mid Re(a_l)\mid < {1\over 2}.
\eeq

As a demonstration of unitarity restrictions, we consider the scattering
of longitudinal gauge bosons, $W^+_LW^-_L\rightarrow W^+_LW^-_L$,
which can be found to ${\cal O}(M_W^2/s)$ from the Goldstone boson
scattering of Fig. \ref{wwwwfig}.
We begin by constructing the $J=0$ partial wave, $a_0^0$,  in the limit
$M_W^2<<s$ from Eq. \ref{wwwwgold},
\beqn
a_0^0(\omega^+\omega^-\rightarrow
\omega^+\omega^-)& \equiv &{1\over 16 \pi s}
\int^0_{-s}\mid {\cal A}\mid dt
\nonumber \\
&=&- {M_h^2\over 16 \pi v^2 }
\biggl[2 + {M_h^2\over s-M_h^2}-{M_h^2\over s}\log
\biggl(1+{s\over M_h^2}\biggr)\biggr].
\label{wwscat}
\eeqn

If we go to very high energy, $s >>M_h^2$, then Eq. ~\ref{wwscat}
has the limit
\beq
 a_0^0(\omega^+\omega^-\rightarrow
\omega^+\omega^-)\longrightarrow_{s>>M_h^2}
- {M_h^2\over 8 \pi v^2}.
\eeq
Applying the unitarity condition, $\mid Re (a_0^0)\mid< {1\over 2}$ gives
the restriction\cite{lqt}
\beq
M_h< 870~GeV.
\label{hbound}
\eeq
It is important to understand that this does not mean that the
Higgs boson cannot be heavier than $870~GeV$, it simply means that
for heavier masses perturbation theory is  not valid.
By considering coupled  channels, a slightly tighter bound than
Eq. \ref{hbound} can be obtained.
The Higgs boson therefore plays a fundamental role in the
theory since it cuts off the growth of the partial wave
amplitudes and makes the theory obey perturbative unitarity.

We can apply the alternate limit to Eq.~\ref{wwscat} and take the
Higgs boson much heavier than the energy scale.  In this limit\cite{chan}
\beq
a_0^0(\omega^+\omega^-\rightarrow
\omega^+\omega^-)\longrightarrow_{s<<M_h^2}
-{s\over 32 \pi v^2}  \quad .
\eeq
Again applying the unitarity condition we find,
\beq
\sqrt{s_c}< 1.7~TeV
\quad .
\label{units}
\eeq
We have used the notation $s_c$ to denote $s$(critical), the scale
at which perturbative unitarity is violated.
Eq. ~\ref{units} is the basis for the oft-repeated statement,
{\it There must be new physics on the TeV scale}.
Eq. \ref{units} tells us that without a Higgs boson, there
must be new physics which restores perturbative unitarity somewhere
below  an energy scale of $1.7~TeV$.

\subsection{$M_h\rightarrow\infty$, The Non-Linear Theory}

So far we have considered searching for the Higgs boson in
various mass regimes.  In this section we will consider the consequences
of taking the Higgs boson mass much
heavier than the energy scale being probed.\cite{abl}
In fact, we will take the limit $M_h\rightarrow \infty$ and
assume that the effective Lagrangian for electroweak
symmetry breaking is determined by new physics outside the reach
of future accelerators such as the LHC.
It is an important question as to whether we can learn something 
about the nature of the electroweak symmetry breaking in this scenario.

  Since we do not know
the full theory, we build the effective Lagrangian out of
all operators consistent with the unbroken symmetries.  In particular,
we must include operators of all dimensions, whether or not
they are renormalizable.  In this way we construct the most general 
effective Lagrangian that describes electroweak symmetry breaking.

To specify the effective Lagrangian, we must first fix the pattern
of symmetry breaking.  We will assume that the global symmetry in
the scalar sector of the model is
 $SU(2)_L \times SU(2)_R$ as in the minimal Standard Model.
We will also assume a custodial $SU(2)_C$ symmetry.\cite{cgg}  This is the
symmetry which forces $\rho=M_W^2/(M_Z^2\cos\theta_W)=1$.
The pattern of global symmetry breaking is then 
\beq
SU(2)_L\times
SU(2)_R \rightarrow SU(2)_C.
\label{symtrans}
\eeq 
In this case, the Goldstone bosons can be described in terms
of the unitary unimodular field $U$\cite{leut}, 
\beq
U  \equiv  
\exp \bigg( \frac{i\,\omega_i(x)\cdot \tau_i}{2v}\bigg)
\quad .
\eeq
This is analogous to the Abelian Higgs model
where we took the Higgs field to be,
\beq \Phi={1\over \sqrt{2}}e^{{i\chi\over v}}(h+v).
\eeq
Under the global $SU(2)_L\times SU(2)_R$ symmetry the $U$ field
transforms as,
\beq
U \rightarrow L^\dagger U R.
\label{trans}
\eeq

It is straightforward to write down the most general $SU(2)_L\times
U(1)_Y$ gauge invariant Lagrangian which 
respects the global symmetry of Eq. \ref{symtrans}.  
The Lagrangian can be written in terms of a derivative expansion, where
each additional set of derivatives corresponds to an additional power
of $s/\Lambda^2$, with $\Lambda$ some high scale corresponding to 
unknown new physics.  
  The lowest order effective Lagrangian (with two derivatives
acting on the $U$ field) for the
symmetry breaking sector of the theory is then
\beq
{\cal L}_{SM}^{nlr}={v^2\over 4} Tr\biggl[
D_\mu U^\dagger D^\mu U\biggr]
-{1\over 2}Tr \biggl( W^{\mu\nu}W_{\mu\nu}\biggr)
-{1\over 2}Tr \biggl( B_{\mu\nu}  B^{\mu\nu}\biggr),
\label{chiral}
\eeq
where the covariant derivative is given by,
\beq
D_\mu U =\partial_\mu U +{i\over 2} g
 W_\mu^i\tau^i U-{i\over 2} g ^{\prime}
  B_\mu U \tau_3.
\eeq
We use the superscript `nlr' to denote `nonlinear realization', since
this expansion contains derivative interactions.  

The gauge field kinetic energies are  matrices in $SU(2)$ space:
\beqn
 W_{\mu\nu}& \equiv &{1\over 2}\biggl(
\partial_\nu W_\mu-\partial_\mu W_\nu
-{i\over 2} g [ W_\mu,W_\nu]\biggr)\nonumber \\
B_{\mu\nu}&=& {1\over 2}
 \biggl(\partial_\nu B_\mu-\partial_\mu B_\nu
\biggr)\tau_3 
\eeqn
with $W_\nu\equiv W_\nu^i \cdot \tau_i$.
In unitary gauge, $U=1$ and it is easy to see that Eq. \ref{chiral}
generates mass terms for the $W$ and $Z$ gauge bosons. 
The Lagrangian of 
 Eq. \ref{chiral}
{\bf is~exactly } the Standard Model Lagrangian with $M_h\rightarrow \infty$.
 Since no physical Higgs boson is included, ${\cal L}^{nlr}_{SM}$ is
non-renormalizable. 

Using the Lagrangian of Eq. \ref{chiral}
 it is straightforward to compute
Goldstone boson scattering
 amplitudes such
as\cite{wb}
\beq
{\cal A}(\omega^+\omega^-\rightarrow zz)={s\over v^2}\equiv A(s,t,u)
\quad ,
\label{pipi}
\eeq
which of course agree with those found in the Standard Model
when we take $M_h^2 >> s$.
Amplitudes which grow with $s$ are
a disaster 
for perturbation theory
since eventually they will violate perturbative
unitarity as discussed in Sec. 9.2.  Of course, this simply tells
us that there must be some new physics at high energy.

  Because of the custodial $SU(2)_C$ symmetry,
the various scattering amplitudes are related:
\beqn
{\cal A}(\omega^+ z\rightarrow \omega^+ z)&=& A(t,s,u)\nonumber \\
{\cal A}(\omega^+ \omega^-\rightarrow \omega^+ 
\omega^-)&=& A(s,t,u)+A(t,s,u)\nonumber \\
{\cal A}(\omega^+ \omega^+\rightarrow \omega^+ 
\omega^+)&=& A(t,s,u)+A(u,t,s)\nonumber \\
{\cal A}( z z\rightarrow z z)&=& A(s,t,u)+A(t,s,u)+A(u,s,t).
\label{isorel}
\eeqn
The relationships of Eq. \ref{isorel}
 were discovered by Weinberg\cite{wb}
 over 30 years ago
for the case of $\pi\pi$ scattering,  which has the same
global $SU)2)_L\times SU(2)_R$ global symmetry.\footnote{In
the limit $M_h>>\sqrt{s}$ there is an exact
analogy between $\pi\pi$ scattering and $W^+_LW^-_L$ scattering
with the replacement $f_\pi\rightarrow v$.}

Using the electroweak equivalence theorem, the Goldstone boson scattering 
amplitudes can be related to the amplitudes for longitudinal gauge boson
scattering.  The effective $W$ approximation can then be
used to find the physical scattering cross sections for hadronic
and $e^+e^-$ interactions in the scenario where there is an 
infinitely massive (or no) Higgs boson.

Eq. ~\ref{chiral} is a non-renormalizable effective
 Lagrangian which
must be interpreted as an expansion in powers of $s/\Lambda^2$,
where $\Lambda$ can be taken to be the scale of new physics 
which restores unitarity (say
$M_h$ in a theory with a Higgs boson).
The effective Lagrangian can be written as,
\beq
{\cal L}_{\rm eff}^{\rm nlr} = {\cal L}_{\rm SM}^{\rm nlr} + 
\sum_i \alpha_i {\cal O}_i + \cdots \;.
\eeq
To ${\cal O}(s^2)$ we have
the non-Standard Model
operators with four or fewer derivatives which conserve 
CP and preserve the custodial $SU(2)_C$ symmetry,  
 \cite{abl}
\beqn
{\cal L}_1 & = & \frac{1}{2}\alpha_1 gg^\prime Tr 
                 \biggl(B_{\mu\nu} T W^{\mu\nu} \biggr) \;,
\nonumber \\
{\cal L}_2 & = & \frac{i}{2} \alpha_2 g^\prime
        B_{\mu\nu} Tr \biggl( T [ V^\mu , V^\nu ] \biggr) \;,
\nonumber \\
{\cal L}_3 & = & i \alpha_3 g Tr \biggl( W_{\mu\nu} [V^\mu , V^\nu ]
 \biggr) \;,
\nonumber \\
{\cal L}_4 & = & \alpha_4 \biggl[ Tr ( V_\mu V_\nu ) \biggr]^2 \;,
\nonumber \\
{\cal L}_5 & = & \alpha_5 \biggl[ Tr ( V_\mu  V^\mu ) \biggr]^2 \;,
\label{abops}
\eeqn
where
\beqn
T & \equiv & 2 U T^3 U^\dagger \;, \nonumber \\
V_\mu & \equiv & ( D_\mu U ) U^\dagger 
 \;,
\eeqn
and $T_i = \tau_i/2$ with the normalization 
Tr$(T_iT_j) = \frac{1}{2}
\delta_{ij}$. 
This is the most general 
$SU(2)_L\times U(1)_Y$ gauge invariant
set of interactions of ${\cal O}(s^2)$
which preserves the custodial $SU(2)_C$.  The coefficients, 
$\alpha_i$,  have information about the underlying dynamics of the
theory.  By measuring the various coefficients one might hope
to learn something about the mechanism of electroweak symmetry
breaking even if the energy of an experiment is below the scale at
which the new physics occurs.

If the assumption that there is a custodial $SU(2)_C$ is relaxed
then there is an additional
term with two derivatives and six additional 
terms with four derivatives,
\beqn
{\cal L}_1^\prime &=& {\beta_1\over 4}v^2 \biggl[Tr(TV_\mu)\biggr]^2
\;,\nonumber \\
{\cal L}_6 & = & \alpha_6 Tr \biggl( V_\mu V_\nu \biggr) 
        Tr \biggl( T V^\mu \biggr) Tr \biggl( T V^\nu \biggr) \;,
\nonumber\\
{\cal L}_7 & = & \alpha_7 Tr \biggl( V_\mu V^\mu \biggr) 
        Tr \biggl( T V_\nu \biggr) Tr \biggl( T V^\nu \biggr) \;,
\nonumber\\
{\cal L}_8 & = & \frac{1}{4}\alpha_8  g^2  \biggl[
 Tr(TW_{\mu\nu})\biggr]^2 \;,
\nonumber\\
{\cal L}_9 & = & \frac{i}{2}\alpha_9 g Tr \biggl( T W_{\mu\nu} \biggr) 
        Tr \biggl( T [V^\mu , V^\nu ] \biggr) \;,
\nonumber \\
{\cal L}_{10} & = & \frac{1}{2} \alpha_{10}
        \biggl[ Tr ( T V_\mu )  Tr ( T V_\nu ) \biggr]^2 \;,
\nonumber \\
{\cal L}_{11} & = & \alpha_{11} g\, \epsilon^{\mu\nu\rho\sigma}
Tr \biggl( T V_\mu \biggr) Tr \biggl( V_\nu W_{\rho\sigma} \biggr) \;.
\label{e4terms} 
\eeqn
The first term ${\cal L}^\prime$  corresponds
to a non-Standard Model contribution to the $\rho$ parameter.  
 
Since the theory contains no Higgs boson, it is non-renormalizable
and so loop corrections will generate singularities. 
  At each order in the
energy expansion new effective
operators  arise whose effects  will cancel
the singularities generated by loops computed using
the Lagrangian corresponding to the order below in $s/\Lambda^2$.
To ${\cal O}(s^2)$, the infinities which arise at one loop can
all be absorbed by defining renormalized parameters, $\alpha_i(\mu)$.
The coefficients thus depend on the renormalization scale $\mu$.

As an 
example, we 
calculate the loop corrections to the gauge boson two point functions
from the new operators of Eqs. \ref{abops} and \ref{e4terms}.\cite{dvlep}
We compute only the divergent contributions and 
make the identification   
\beq
{1\over\epsilon}(4\pi)^\epsilon\Gamma(1+\epsilon) \rightarrow 
\log\biggl({\Lambda^2\over \mu^2}\biggr)
\eeq
 and drop all nonlogarithmic terms.
Furthermore,  we  chose $\mu^2=M_Z^2$. 
This gives an estimate of the size of the new physics corrections. 

The contributions of the two point functions to four fermion amplitudes is
generally summarized by a set of parameters such as the $S$, $T$ and $U$
parameters of Ref.~75.
Three of the operators, ${\cal L}_{1}^\prime$, ${\cal L}_{1}$ and 
${\cal L}_{8}$, contribute at the tree-level to the gauge-boson 
two-point-functions.
Both ${\cal L}_{1}^\prime$ and 
${\cal L}_{8}$ violate the custodial $SU(2)_C$ symmetry and so are
expected to be small. 
The two point functions arising from
the Lagrangian of Eqs. \ref{abops} and \ref{e4terms} 
give contributions to $S$, $T$, and $U$  to
one-loop, \cite{dawsz}
\begin{eqnarray}
\nonumber
\alpha\Delta S & = & 
\frac{\alpha}{12\pi}\log\biggl({\Lambda^2\over M_h^2}\biggr)
- 4e^2 \alpha_1
\\
 &&\mbox{}
-\frac{g^2 e^2 }{16\pi^2}
\biggl[ {1+30 c_W^2\over 3 c_W^2} 
       +\frac{1-22 c_W^2 }{3c_W^2} \alpha_3
       +\frac{1+6 c_W^2}{3c_W^2} \alpha_9
        \bigg]\loglammz
\;, \nonumber \\
\alpha\Delta T & = & 
\frac{3\alpha}{16\pi c_W^2 }
\log\biggl({\Lambda^2\over M_h^2}\biggr)  + 2\beta_1 
- \frac{g^4}{16\pi^2 c_W^2}
\bigg[ \frac{3 s_W^2 (3c_W^2-1)}{2c_W^2 }\alpha_2 
\nonumber \\  && 
      + 3s_W^2  \alpha_3 
      + \frac{15 s_W^2(c_W^2 +1)}{4c_W^2 }\alpha_4
      + \frac{3 s_W^2 (c_W^2+1)}{2c_W^2 }\alpha_5
\nonumber
\\  && \mbox{}
      + \frac{3(2 c_W^4+11)}{4 c_W^2 }\alpha_6
      + \frac{6(c_W^4+1)}{c_W^2}\alpha_7
      + \frac{3s_W^2}{2}\alpha_9
      + \frac{9}{c_W^2}\alpha_{10}
       \bigg]\loglammz
\;,\nonumber  \\
\alpha\Delta U & = & -4e^2 \alpha_8
        + \frac{g^4}{16\pi^2} \frac{2s_W^2}{3c_W^2}\bigg[
        -s_W^2(2c_W^2+3)\alpha_2 
\nonumber
\\
 && \mbox{}
        + 2 s_W^2  (2 s_W^2 + c_W^2 )\alpha_3
        + (2 c_W^4 - 15 c_W^2  + 1) \alpha_9 \bigg]
        \loglammz
\;,
\eeqn
where $s_W^2=\sin^2\theta_W=.23$ and $c_W^2=\cos\theta_W^2$.
Even
when all the $\alpha_i$ are zero, the expressions for $\Delta S$ and
$\Delta T$ are nonzero.  This is because the nonlinear Lagrangian contains 
singularities which in the Standard
Model  would be cancelled by the contributions of the 
Higgs boson.  In these terms the renormalization scale, ${\hat\mu}$, 
is appropriately taken to be the
 same Higgs-boson mass we use to evaluate the Standard Model  
contributions.

Due to the extraordinary precision of electroweak data at low energy and on the 
$Z$ pole, it is possible to place constraints on models for physics beyond the 
Standard Model  by studying these loop-level contributions of  new physics
to electroweak observables.\cite{dvlep} 
First we analyze the numerical constraints on $\alpha_1$, $\beta_1$ and 
$\alpha_8$ (which arise
from  $\Delta S$, $\Delta T$ and $\Delta U$, 
respectively), and we present the best-fit central values with one-sigma errors,
\begin{equation}
\label{fit-now-nl}
\left.\begin{array}{lll}
\alpha_1 & = & (4.7 \pm 2.6)\times 10^{-3} \;,\\
\beta_1  & = & (0.30 \pm 0.57)\times 10^{-3}\;,\\
\alpha_8 & = & (-0.9 \pm 7.6)\times 10^{-3} \;.
\end{array}\right.
\end{equation}
We use $\Lambda = 2$~TeV everywhere in this section.
These constraints are sufficiently strong that there is no sensitivity to 
these three parameters at LEP~2.\cite{sz}  Observe that a positive 
value for $\alpha_1$ is favored.

Two of the operators, $\call_{1}$ and 
$\call_{8}$, contribute at the tree-level to the gauge-boson 
two-point-functions and 
also to 
nonstandard $WW\gamma$ and $WWZ$ couplings.
These are sufficiently constrained by the limits on the two-point
functions that we will not consider their contributions to the three
point functions.
  Three operators, $\call_{2}$, 
$\call_{3}$ and $\call_{9}$, contribute to $WW\gamma$ and $WWZ$ couplings 
without making a tree-level contribution to the gauge-boson propagators.
Much of the literature describes nonstandard $WW\gamma$ and $WWZ$ vertices 
via the phenomenological effective Lagrangian\cite{hpzh87}, 
\begin{eqnarray}
\nonumber
\lefteqn{\call_{WWV} =}\\&&  - i g_{WWV} \Bigg\{ 
g_1^V \Big( {\tilde W}^+_{\mu\nu} W^{- \, \mu} V^{\nu} 
  - W^+_{\mu} V_{\nu} {\tilde W}^{- \, \mu\nu} \Big)
\nonumber \\ && 
+ \kappa_V W_\mu^+ W_\nu^- {\tilde V}^{\mu\nu}
+ \frac{\lambda_V}{\mwsq} {\tilde W}^+_{\mu\nu}
{\tilde W}^{- \, \nu\rho} {\tilde V}_\rho^{\; \mu}
 \Bigg\}
\;,\label{lagr-phenom}
\end{eqnarray}
where $V=Z,\gamma$, the overall coupling constants are 
$g_{WW\gamma} = e$ and 
$g_{WWZ} = g\cos\theta_W$.  The field-strength tensors include only
the Abelian parts, {\em i.e.}
${\tilde W}^{\mu\nu} = \partial^\mu W^\nu - \partial^\nu W^\mu$
and ${\tilde V}^{\mu\nu} = \partial^\mu V^\nu - \partial^\nu V^\mu$.  
The coefficients of the electroweak chiral 
Lagrangian to ${\cal O}(s^2)$ in the energy expansion can be
matched to the parameterization of Eq. \ref{lagr-phenom}, 
demonstrating that the two approaches are equivalent: 
\cite{dvlep,fls}
\begin{eqnarray}
\gonez(\qsq) & = & 1 + {g^2\over \cos\theta_W^2}\alpha_3 \;,
\\
\kapgam(\qsq) & = & 1 + g^2 \Big( \alpha_2 + \alpha_3 + \alpha_9 \Big)\;, 
\\
\kapz(\qsq) & = & 1 + {g^2\over\cos^2\theta_W}\Big( 
-\sin^2\theta_W\alpha_2  + \cos^2\theta_W(\alpha_3 + \alpha_9 )\Big)  \;, \\
\lamgam(\qsq) & =& \lamz(\qsq)   \approx  0 \;.
\label{nl-lambda-hisz}
\end{eqnarray}
If we impose the custodial SU(2)$_{\rm C}$ symmetry on the new physics, then we 
may neglect the $\alpha_9$ terms.  Eqn.~(\ref{nl-lambda-hisz}) reflects
our prejudice that the $\lambda_V$ couplings, being generated by 
${\cal O}(s^3)$ operators while the other couplings are generated by 
${\cal O}(s^2)$  operators, should be relatively small. 

We place constraints on the operators contributing to the 3-point
function by considering the effects of only one operators at a time.
\begin{table}[t]
\center{
{Table 2:  $95\%$ confidence level constraints for $\Lambda=2~TeV$.
In the first row, all other coefficients are set to zero.  In the
second row, $\alpha_1=5.5\times 10^{-3}$ is chosen according to
Eq. 154.} 
\begin{tabular}{||l||c|c|c||}
 & $\alpha_2$ & $\alpha_3$ & $\alpha_9$ 
\\ \hline\hline
$\alpha_1 = 0$ & 0.21$\pm$0.15 & -0.17$\pm$0.11 & 0.16$\pm$0.67 
 \\\hline
$\alpha_1 = 5.5\times 10^{-3}$ & 0.05$\pm$0.15 & -0.04$\pm$0.11 & -0.02$\pm$0.67 
 \\
\end{tabular}}
\label{table-nl-singly-loop}
\vskip .75in 
\end{table}
Note that in the first row of  Table 2, when $\alpha_1 = 0$,
only $\alpha_9$ is consistent with 
the Standard Model at the 95\% confidence level.  However,
in the second row where
 we have chosen the central value of $\alpha_1$ according 
to the best-fit value of Eqn.~(\ref{fit-now-nl}), all of the central values 
are easily consistent with the Standard Model.
  While the central values easily move 
around as we include additional operators in the analysis, the errors are 
much more robust.

The effects of the non-Standard Model  gauge boson couplings  
can also be searched for in 
$e^+e^-$ and hadron machines which are 
sensitive to both the three and
 four gauge boson vertices
through vector boson scattering.\cite{fls,bdv}
Since the effects grow with energy, the LHC will be much
more sensitive than the Tevatron to non-Standard Model
gauge boson couplings.
To study strong interactions with the Lagrangian of Eqs. ~\ref{abops}
and \ref{e4terms},  
 the effective $W$ approximation  can be used to get results for
hadronic interactions or $e^+e^-$ scattering.

 \subsection{Coefficients of New Interactions in  a Strongly
Interacting Symmetry Breaking Sector}
It is instructive to estimate the size of the  $\alpha_i$ coefficients in 
typical theories.  Using the effective Lagrangian approach this can be
done in a consistent way. 
In any realistic scenario there will be a set of nonzero $\alpha_i$, and it is 
possible (indeed likely) that there will be large interferences between the 
effects of the various coefficients.  In order to see the types of limits which 
might arise in various scenarios of
spontaneous symmetry breaking, we consider a strongly interacting 
scalar in order to obtain an indication of 
the sensitivity  of our results to the underlying dynamics.  Using the 
effective-Lagrangian approach, we can estimate the coefficients in a consistent 
way.

We first consider a model with three Goldstone bosons corresponding to the 
longitudinal components of the $W^\pm$ and $Z$  bosons coupled to a scalar 
isoscalar resonance like the Higgs boson.  We assume that the $\alpha_i(\mu^2)$ 
are dominated by tree-level exchange of the scalar boson.  Integrating out the 
scalar and matching the coefficients at the scale $M_h$ gives the 
predictions, \cite{drv}

\begin{eqnarray}
\alpha_4(\mu^2) & = & 
   \frac{1}{16\pi^2}\frac{1}{12} \log\bigg(\frac{\mhsq}{\mu^2}\bigg)
\\ & = & 2\alpha_2(\mu^2)=2\alpha_3(\mu^2)=-\alpha_1(\mu^2)
\\
\alpha_5(\mu^2) & = & \frac{1}{16\pi^2}\Bigg[\frac{1}{24} 
\log\bigg(\frac{\mhsq}{\mu^2}\bigg)
+\frac{64\pi^3}{3}\frac{\Gamma_h v^4}{m_H^5}\Bigg],
\label{scalform}
\end{eqnarray}
where $\Gamma_h$ is the width of the scalar into Goldstone bosons.  All of the 
other $\alpha_i$ are zero in this scenario.  It is important to note that only 
the logarithmic terms are uniquely specified.  The constant terms depend on the 
renormalization scheme. 

In Fig.~\ref{fig-lin-nl}
\begin{figure}[t]
\centering   
\epsfxsize=4.in
\leavevmode
\epsffile{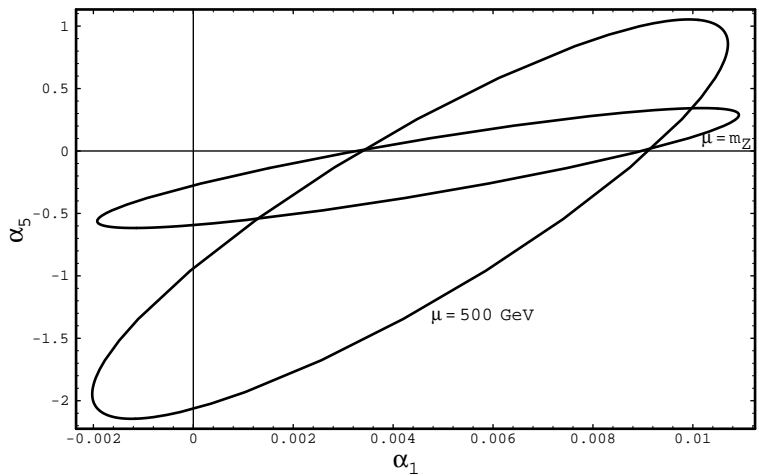}
\caption{95\% confidence level contours for $\Lambda = $~2~TeV and 
$M_t = $~175~GeV for two values of $\mu$ in a model with a strongly
interacting scalar, ({\it cf} Eqs. 160-162 ).}
\label{fig-lin-nl} 
\end{figure}
we plot $\alpha_5(\mu^2)$ 
{\em vs.}\/ $\alpha_1(\mu^2)$ with the pattern typical of a theory dominated by 
a heavy scalar given in Eqn.~\ref{scalform}, 
$-\alpha_1(\mu^2) = 2\alpha_2(\mu^2) = 2\alpha_3(\mu^2) = \alpha_4(\mu^2)$.  
First of all, notice that the contour obtained depends rather strongly upon 
our choice of the renormalization scale, $\mu$, especially with regard to the 
$\alpha_5$ axis. Everything to the right of $\alpha_1 = 0$ corresponds to 
$m_H < \mu$.  Furthermore, since we require that $\Gamma_h$ be non-negative,
we may approximately exclude everything below the $\alpha_5 = 0$ axis.  The 
allowed region to the upper right of the figure corresponds to a Higgs-boson 
with a mass in the MeV range and an extremely narrow width; this portion of 
the figure is already excluded by experiment. 
The positive central value of $\alpha_1$ indicates that a heavy scalar
resonance is disfavored,  in agreement with the indirect limits from LEP2
discussed in Section 3.3.

\section{Problems with the Higgs Mechanism}

In the preceeding sections we have discussed many features of the
Higgs mechanism.  However, most theorists firmly believe
that the Higgs mechanism cannot be the entire story behind electroweak
symmetry breaking.  The primary reasons are:
\begin{itemize}
\item  The Higgs sector of the theory is trivial
	($\lambda\rightarrow 0$ as the energy scale $\rightarrow
	\infty$)  unless the Higgs mass is
	in a very restricted range.

\item  The Higgs mechanism doesn't explain why $v=246~GeV$.

\item  The Higgs mechanism doesn't explain why fermions have
the masses they do.

\item  Loop corrections involving the Higgs boson are quadratically
divergent and counterterms must be adjusted order by order in
perturbation theory to cancel these divergences.  This fine tuning
is  considered by most theorists to be unnatural.

\end{itemize}

\subsection{Quadratic Divergences} 
The most compelling
argument against the simplest  version of the Standard Model
is the quadratically divergent contributions to the Higgs  boson
mass which arise when loop corrections are computed.
 At one loop, the quartic self- interactions
of the Higgs boson  generate a quadratically
divergent contribution
 to the Higgs boson mass which must be cancelled
by a  mass  counterterm.  This counterterm must be fine tuned
at each order in perturbation theory.  
 We begin by considering  a theory with a single  fermion, $\psi$,
coupled to a massive  Higgs scalar,
\beq
{\cal L}_\phi=
{\overline \psi}(i \partial)\psi
+\mid \partial_\mu \phi\mid^2-m_S^2 \mid \phi\mid^2
-\biggl( \lambda_F
{\overline \psi}\psi \phi +{\rm h.c.} \biggr)  
\quad .
\label{ephilag}
\eeq
\begin{figure}[t]
\vskip -3.5in 
\centering   
\hskip -1in 
\epsfxsize=6.in
\leavevmode
\centerline{\epsffile{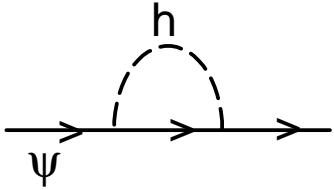}}
\vskip -.2in 
\caption{Scalar contribution to fermion mass renormalization.}
\label{figferm} 
\end{figure}  
We will assume that this Lagrangian leads to spontaneous symmetry 
breaking and so take $\phi=(h+v)/\sqrt{2}$, with $h$ the physical Higgs
boson. 
 After spontaneous symmetry breaking,
the fermion acquires a mass, $m_F=\lambda_Fv/\sqrt{2}$. 
First, let us consider the fermion self-energy  arising from
the  scalar loop corresponding to Fig. \ref{figferm}.
\beq
-i\Sigma_F(p)=\biggl({-i\lambda_F\over \sqrt{2}}\biggr)^2
(i)^2
\int{d^4k\over (2\pi)^4}{(k+m_F)\over [k^2-m_F^2][(k-p)^2-m_S^2]}
\quad .
\eeq
The renormalized fermion mass is $m_F^r=m_F+\delta m_F$, with
\beqn
\delta m_F &=& \Sigma_F(p)\mid_{ p=m_F}
\nonumber \\
&=& i {\lambda_F^2\over 32\pi^4}\int_0^1 dx \int  d^4 k^\prime
{m_F (1+x)\over [k^{\prime 2} -m_F^2 x^2-m_S^2(1-x)]^2}
\quad .
\label{meren}  
\eeqn

  The integral can be performed in Euclidean space, which
amounts to making the following transformations,
\beqn
k_0  &\rightarrow & i k_4
\nonumber \\
d^4k^\prime & \rightarrow & i d^4 k_E
\nonumber \\
k^{\prime 2} & \rightarrow & - k_E^2
\quad .
\eeqn
Since the integral of Eq. \ref{meren} depends only on
$k_E^2$, it can easily be performed using the
result (valid for symmetric integrands),
\beq \int d^4 k_E f(k_E^2)=\pi^2
\int^{\Lambda^2}_0 y dy f(y)
\quad .
\label{intfacs}  
\eeq 
In Eq. \ref{intfacs}, $\Lambda$ is a high energy cut-off, presumably of the
order of the Planck scale or a grand unified scale.
The renormalization of the fermion   mass is then, 
\beqn
\delta m_F&=& -{\lambda_F^2 m_F\over 32\pi^2}\int_0^1~dx (1+x)
\int^{\Lambda^2} _0 {y dy\over [y+m_F^2x^2+m_S^2(1-x)]^2}
\nonumber \\
&=& -{3 \lambda_F^2 m_F\over 64\pi^2} \log\biggl({\Lambda^2
\over m_F^2}\biggr) + ....
\eeqn
where the $....$  indicates terms independent of the cutoff
or which vanish when $\Lambda\rightarrow\infty$.
This correction clearly corresponds to a well-defined expansion for $m_F$.
 The corrections to
fermion masses are said to be {\it natural}.  In the limit
in which the fermion mass vanishes,
Eq. \ref{ephilag} is invariant under the chiral transformations,
\beqn
\psi_L & \rightarrow  & e^{i\theta_L}\psi_L
\nonumber \\
\psi_R &\rightarrow & e^{i\theta_R}\psi_R,
\eeqn
and so 
 setting the fermion mass to zero increases the symmetry
of the theory.  
Since the Yukawa coupling (proportional to the mass term) breaks
this symmetry, the corrections to the mass must be proportional to $m_F$.

The situation is quite different, however, when we consider the
renormalization of the scalar  mass  from
a fermion loop (Fig. \ref{fig2fig})
 using the same Lagrangian (Eq. \ref{ephilag}),

\begin{figure}[t,b]
\vskip -1.5in 
\centering
\epsfxsize=5.in 
\leavevmode
\centerline{\epsffile{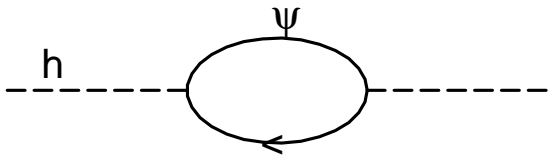} }
\vskip -2.2in 
\caption{Fermion contribution to the renormalization of 
a scalar mass.}
\label{fig2fig}
\end{figure}  

\beq
-i\Sigma_S(p^2)=\biggl({-i\lambda_F\over\sqrt{2}}\biggr)
(i)^2 (-1)\int {d^4 k \over (2 \pi)^4}
{Tr[(k+m_F)((k-p)+m_F)]\over
(k^2-m_F^2)[(k-p)^2-m_F^2]}
\quad .
\eeq 
The minus sign is the consequence of Fermi statistics
and will be quite important later.  Integrating with a momentum
space cutoff as above we find the contribution to the Higgs mass,
\beqn
(\delta M_h^2)_a\equiv\Sigma_S(m_S^2)&
=&-{\lambda_F^2\over 8\pi^2}\biggl[
\Lambda^2+(m_S^2-6m_F^2)\log\biggl({\Lambda\over m_F}\biggr)
\nonumber \\ &&~~+  
(2m_F^2-{m_S^2\over 2})
\biggl(1+I_1\biggl({m_S^2\over m_F^2}\biggr)\biggr)\biggr]~
\nonumber \\ && ~~ 
+{\cal O}\biggl({1\over \Lambda^2}\biggr),
\label{quads}
\eeqn
where $I_1(a)\equiv \int^1_0 dx \log(1-ax(1-x))$.
The Higgs boson mass diverges {\it quadratically}! 
The Higgs boson thus does not obey the decoupling theorem\cite{apc}
and  this quadratic divergence appears independent of
the mass of the Higgs boson. 
Note that the correction is {\it not} proportional to $M_h$.  This 
is because setting the Higgs mass equal to zero does not 
increase the symmetry of the Lagrangian.  There is nothing
that protects the Higgs mass from these large corrections
and, in fact, the Higgs mass wants to be close to the largest mass
scale in the theory. 

Since we know that the physical Higgs boson
 mass, $M_h$, must be less than around
$1~TeV$ (in order to keep the $WW$ scattering cross section from
violating unitarity), we have the unpleasant result,
\beq
M_h^2=M_{h,0}^2+\delta M_h^2+{\hbox{counterterm}}
,
\eeq
where the counterterm must be adjusted to a precision of
roughly $1$ part
in $10^{15}$ in order to cancel the 
quadratically divergent contributions to $\delta M_h$.
This is known as the ``{\it hierarchy problem}''.  
Of course, the quadratic divergence can be renormalized
 away in exactly the same manner  
as for logarithmic divergences by adjusting the
cut-off. There is nothing formally wrong with this fine tuning.  
 Most theorists, however,
regard this solution as unattractive.
  
The effects of scalar particles on the Higgs mass renormalization are
quite different from those  of fermions.
We
introduce  two  complex scalar fields, $\phi_1$ and $\phi_2$,
interacting with the Standard Model Higgs boson, $h$,
(the reason for introducing $2$ scalars is that with foresight we
know that a supersymmetric theory associates $2$ complex scalars with
each massive fermion -- we could just as easily make the argument given
below with one additional scalar and slightly different couplings), 
\beqn 
{\cal L}&=& 
\mid \partial_\mu \phi_1\mid^2+
\mid \partial_\mu \phi_2\mid^2
-m_{s_1}^2\mid \phi_1\mid^2-m_{s_2}^2 \mid\phi_2\mid^2
\nonumber \\ &&~~  
+\lambda_S\phi\biggr(\mid \phi_1\mid^2 +\mid \phi_1\mid^2\biggr)
+{\cal L}_\phi
\quad .
\label{lag2h}
\eeqn

\begin{figure}[t]
\vskip -1.5in 
\epsfxsize=3.in
\leavevmode
\centerline{\epsffile{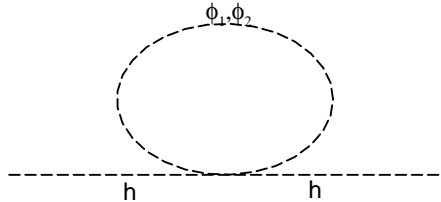}}
\caption{Scalar contributions to Higgs mass renormalization. }
\label{fig3fig}
\end{figure}  
From the diagrams of Fig. \ref{fig3fig},
 we find the contribution to the Higgs mass
renormalization,
\beqn
(\delta M_h^2)_b& =& -\lambda_S
\int {d^4k\over (2\pi)^4}\biggl[
{i\over k^2-m_{s_1}^2}
+{i\over k^2-m_{s_2}^2}\biggr]
\nonumber \\
&=& {\lambda_S\over 16\pi^2}\biggl\{
2 \Lambda^2 -2m_{s_1}^2\log\biggl({\Lambda\over m_{s_1}}\biggr)
-2m_{s_2}^2\log\biggl({\Lambda\over m_{s_2}}\biggr) \biggr\}
\nonumber \\ &&~~
+{\cal O}\biggl({1\over \Lambda^2}\biggr).
\label{quad2}
\eeqn       
From Eqs. \ref{quads} and \ref{quad2}, we see that if 
\beq 
\lambda_S= \lambda_F^2
,
\label{susyreq}
\eeq
the quadratic divergences coming from these two terms exactly cancel
each other.  Notice that the cancellation occurs independent
of the masses, $m_F$ and $m_{s_i}$, and of the magnitude of
the couplings $\lambda_S$ and $\lambda_F$.

In the Standard Model, one could attempt to cancel the quadratic
divergences in the Higgs boson mass by balancing the contribution from
the Standard Model
Higgs quartic coupling with that from the top quark loop
in exactly the same manner as above.  This approach
gives a prediction for the Higgs boson mass in terms of the top quark mass. 
However, since there is no symmetry to enforce this relationship,
 this cancellation of quadratic
divergences fails at $2-$ loops. 

\begin{figure}[t]
\vskip -2.5in 
\centering   
\epsfxsize=3.in
\leavevmode
\epsffile{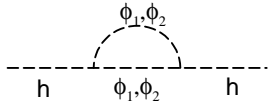}
\vskip -.2in 
\caption{Scalar contributions to Higgs mass renormalization.}
\label{fig4fig}
\end{figure}

After the spontaneous symmetry breaking, Eq. \ref{lag2h} also
leads to a cubic interaction shown in Fig. \ref{fig4fig}.  These   graphs
also 
give a contribution to the Higgs mass renormalization,
although they are not quadratically divergent.
\beqn 
(\delta M_h^2)_c&=& {\lambda_S^2 v^2\over 16\pi^2}\biggl\{ -1+2 \log\biggl(
{\Lambda\over m_{s_1}}\biggr)-I_1\biggl({M_h^2\over m_{s_1}^2}
\biggr)\biggr\} +(m_{s_1}\rightarrow m_{s_2})
\nonumber \\  && ~~ 
+{\cal O}\biggl({1\over \Lambda^2}\biggr).
\eeqn 
Combining the three contributions to the Higgs mass and assuming
$\Lambda_S=\Lambda_F^2$ and $m_{s_1}=m_{s_1}$ we find
no quadratic divergences,
\beq
(\delta M_h^2)_{tot}={\lambda_F^2\over 4\pi^2}
\biggl\{ m_{s_1}^2\log\biggl({\Lambda\over m_{s_1}}\biggr)
-m_F^2\log\biggl({\Lambda\over m_F}\biggr)
\biggr\}+{\cal O}\biggl({1\over\Lambda^2}\biggr).
\eeq
If the mass splitting between the fermion and scalar is small,
$\delta m^2\equiv m_F^2-m_{s_1}^2$, then we have
the approximate result,
\beq
(\delta M_h^2)_{tot}\sim {\lambda_F^2\over 8\pi^2}\delta m^2
\quad .
\eeq
Therefore, as long as the mass splitting between scalars and fermions is
``small'', no unnatural cancellations will be required and
the theory can be considered ``natural''.  

One alternative to
the Standard Model Higgs mechanism is that the Standard
Model becomes supersymmetric.  In these
theories, there is a scalar associated with each fermion and
the couplings are just those of Eq. \ref{susyreq}.  The
electroweak symmetry is still broken by the Higgs mechanism, but the quadratic
divergences in the scalar sector
are cancelled automatically because of the expanded 
spectrum of the theory and so the model is no longer considered to be
unnatural.  In the next section, we  will briefly discuss the phenomenology 
of the Higgs bosons occurring in supersymmetric models
and emphasize the similarity of much of the phenomenology to that of the
Standard Model Higgs.\cite{hami,hk}

\section{Higgs Bosons in Supersymmetric Models}

Supersymmetry is a symmetry which relates particles of differing
spin, (in the above example,
fermions and scalars).
  The particles are combined into  ${\it superfields}$,
which contain fields differing by one-half unit of spin.\cite{wess}
  The
simplest example, the scalar superfield, contains a complex
scalar, $S$, and a two- component Majorana fermion, $\psi$.
The supersymmetry then  completely specifies the allowed interactions.

In a supersymmetric theory the scalars and fermions
in a superfield have the
same couplings to gauge bosons
 and so  the cancellation of quadratic divergences
occurs automatically as in the previous section.
This is one of the primary motivations for introducing
supersymmetric models. 
 The supersymmetric Lagrangian
contains scalar and  fermion pairs  ${\it of~equal ~mass}$ and  so the 
supersymmetry connects particles of different spin, but
with all other
characteristics the same. 
It is clear, then, that {\bf
supersymmetry must be a broken symmetry}.
There is no scalar particle, for example, with the mass and
quantum numbers of the electron.  In fact, there are no candidate
supersymmetric scalar partners for any of the fermions in the
experimentally observed 
spectrum.  We will take a non-zero
 mass splitting between the particles of
a superfield as a signal for supersymmetry breaking.

Supersymmetric theories are easily constructed according to the rules
of  supersymmetry.  
There are two types of superfields relevant for
our purposes:\footnote{
The superfields also contain ``auxiliary fields'', which
are fields with no kinetic energy terms in
the Lagrangian.\cite{wess}  These fields are not important
for our purposes.} 
\begin{enumerate}
\item  ${\it Chiral ~Superfields}$:  These consist of 
a complex scalar field, $S$, and a $2$-component Majorana fermion
field, ${\psi}$.
\item ${\it Massless Vector~Superfields}$:  These consist of a massless
gauge field with field strength $F_{\mu\nu}^A$
 and a $2$-component Majorana fermion field, $\lambda_A$, termed
a ${\it gaugino} $. 
The index $A$ is the gauge index.  
\end{enumerate}

The minimal supersymmetric model (MSSM) respects the same $SU(3)\times
SU(2)_L\times U(1)_Y$ gauge symmetries as does the
Standard Model.  
The particles necessary to construct the minimal  supersymmetric 
version of the Standard Model are shown in Tables 3 and 4
in terms of the superfields, (which are denoted by the
superscript ``hat'').  
Since there are no candidates for supersymmetric partners of
the observed particles, we must double the entire spectrum,
placing the observed particles in superfields with new
postulated superpartners.  
 There are, of course, quark and
lepton superfields for all $3$ generations and we have
listed  in Table 3
only the members of the first generation.  
The superfield ${\hat Q}$ thus consists of an $SU(2)_L$
doublet of quarks:
\beq
Q_L= 
\left( \begin{array}{c} u \\
d\end{array}\right)_L
\eeq
and their scalar partners which are also in
an $SU(2)_L$ doublet,
\beq
{\tilde Q}_L=
\left(\begin{array}{c}  {\tilde u}_L\\ 
{\tilde d}_L\end{array}\right)
\quad .  
\eeq   Similarly, the
superfield ${\hat U}^c$ (${\hat D}^c$)
 contains the right-handed
up  (down) anti-quark, ${\overline u}_R$ (${\overline d}_R$), 
 and its scalar partner, ${\tilde u}_R^*$ (${\tilde d}_R^*$).
The scalar partners of the quarks are fancifully called
squarks.  
We see that each quark has $2$ scalar partners, one corresponding
to each quark chirality.
The leptons are contained in the $SU(2)_L$ doublet superfield
${\hat L}$ which contains the left-handed fermions,
\beq
L_L=\left(\begin{array}{c} \nu \\
e\end{array}\right)_L
\eeq
and their scalar partners,
\beq
{\tilde L}_L=\left(\begin{array}{c}
{\tilde \nu}_L\\
{\tilde e}_L\end{array}\right)
\quad .
\eeq
Finally, the right-handed anti-electron, ${\overline e}_R$, is contained
in the superfield ${\hat E}^c$ and has a scalar partner
${\tilde e}_R^*$.  The scalar partners of the leptons
are termed sleptons.  

  The $SU(3)\times SU(2)_L\times U(1)$  gauge fields all obtain
Majorana fermion partners in a supersymmetric  model.
  The ${\hat G}^a$ superfield contains
the gluons, $g^a$, and their partners the gluinos, ${\tilde g}^a$;
${\hat W}_i$ contains the $SU(2)_L$ gauge bosons, $W_i$ and
their fermion partners, ${\tilde \omega}_i$ (winos);
and ${\hat B}$ contains the $U(1)$ gauge field, $B$,
and its fermion partner, ${\tilde b}$ (bino).  The usual
notation is to denote the supersymmetric partner of a fermion
or gauge field with the same letter, but with a tilde over it.   
\begin{table}[htb]
\begin{center}
{Table 3: Chiral Superfields of the MSSM}
\vskip6pt
\renewcommand\arraystretch{1.2}
\begin{tabular}{|lccrc|}
\hline
\multicolumn{1}{|c}{Superfield}& SU(3)& $SU(2)_L$& $U(1)_Y$
& Particle Content 
\\
\hline
${\hat Q}$   &    $3$          & $2$&  $~{1\over 3}$
& ($u_L,d_L$), (${\tilde u}_L,{\tilde d}_L$)\\
${\hat U}^c$ & ${\overline 3}$ & $1$& $-{4\over 3}$
&${\overline u}_R$, ${\tilde u}_R^*$\\
${\hat D}^c$ & ${\overline 3}$ & $1$&  $~{2\over 3}$
&${\overline d}_R$, ${\tilde d}_R^*$\\
${\hat L}$   & $1$             & $2$& $~-1$
& $(\nu_L,e_L)$, (${\tilde \nu}_L, {\tilde e}_L$)\\
${\hat E}^c$ & $1$             & $1$& $~2$ 
& ${\overline e}_R$, ${\tilde e}_R^*$\\
${\hat \Phi_1}$ & $1$             & $2$& $-1$ 
&($\Phi_1, {\tilde h}_1$)\\
${\hat \Phi_2}$ & $1$             & $2$& $~1$
& $(\Phi_2, {\tilde h}_2)$ \\ 
\hline
\end{tabular}
\end{center}
\end{table}
 
\begin{table}[htb]
\begin{center}
{Table 4: Vector Superfields of the MSSM}
\vskip6pt
\renewcommand\arraystretch{1.2}
\begin{tabular}{|lcccc|}
\hline
\multicolumn{1}{|c}{Superfield}&SU(3)&$SU(2)_L$&$U(1)_Y$
& Particle Content\\
\hline
${\hat G^a}$  &  $8$  &  $1$  &  $0$ 
&$g$, ${\tilde g}$ \\
${\hat W^i}$  &  $1$  &  $3$  &  $0$ 
& $W_i$, ${\tilde \omega}_i$  \\
${\hat B}$  &  $1$  &  $1$  &  $0$ 
& $B$, ${\tilde b}$  \\
\hline
\end{tabular}
\end{center}
\end{table}

In the standard (non-supersymmetric) model of electroweak 
interactions, the fermion masses are generated by 
the Yukawa terms in the Lagrangian
\beq
{\cal L}=-\lambda_d {\overline Q}_L \Phi d_R-
\lambda_u {\overline Q}_L \Phi^c u_R + h.c.
\eeq
In a supersymmetric theory,  however, a term proportional to 
$\Phi^c$ is not allowed
and so another scalar doublet must be 
added in order to give the $\tau_3=1$ components of the $SU(2)_L$
fermion doublets mass. 
The minimal supersymmetric model (MSSM) therefore
has two Higgs doublet,
$\Phi_1$ and $\Phi_2$.

The Higgs sector of the MSSM is very similar to that of
a general $2$ Higgs doublet model.\cite{hks}
The first 
contribution to the Higgs potential, $V_D$,
\beqn
V_D=& \sum_a{1\over 2} D^a D^a\nonumber \\
D^a \equiv & -g \Phi_i^*T^a_{ij}\Phi_j
\eeqn
is called the ``D''-term.  
The D-terms corresponding to the $U(1)_Y$ and $SU(2)_L$ gauge
groups are given by,
\beqn
U(1):\quad D^1=& - {g^\prime\over 2} \biggl(
\mid 
\Phi_2\mid^2-\mid \Phi_1\mid^2\biggr)
\nonumber \\
SU(2):\quad D^a=& - {g\over 2}\biggl
(\Phi_1^{i*} \tau^a_{ij}\Phi_1^j
    +\Phi_2^{i*} \tau^a_{ij}\Phi_2^j\biggr),
\eeqn
(where $T^a={\tau^a\over 2}$).  The D terms then contribute
to the scalar potential:
\beq
V_D={g^\prime \over 8}\biggl(\mid \Phi_2\mid^2 -\mid 
\Phi_1\mid^2\biggr)^2
    +{g^2\over 8}\biggl( \Phi_1^{i*}
\tau_{ij}^a \Phi_1^j+ \Phi_2^{i*}\tau^a_{ij}\Phi_2^j\biggr)^2
\quad .
\eeq   
Using the $SU(2)$ identity,
\beq
\tau_{ij}^a\tau_{kl}^a=2 \delta_{il}\delta_{jk}-\delta_{ij}\delta_{kl}
\eeq
we find,
\beqn
V_D=&{g^2\over 8}\biggl\{ 4\mid \Phi_1^*\cdot \Phi_2\mid^2
-2 (\Phi_1^*\cdot \Phi_2)(\Phi_2^*\cdot \Phi_2)+
\biggl(\mid \Phi_1\mid^2+\mid \Phi_2\mid^2\biggr)^2\biggr\}                                           
\nonumber \\
& +{g^{\prime 2}\over 8}\biggl( \mid \Phi_2\mid^2-\mid 
\Phi_1\mid^2\biggr)^2
.
\eeqn
The remainder of the scalar potential is given in terms of a
function, $W$ (the superpotential),
 which can be at most cubic in the scalar superfields.
The $SU(2)_L\times U(1)_Y$ gauge invariance allows only one interaction
involving only the Higgs scalar fields,
\beq
W=\mu {\hat \Phi_1}{\hat \Phi_2}
\quad .
\eeq
The supersymmetry algebra requires that $W$ give a contribution
to the scalar  potential,\cite{wess}
\beq
V_i=\sum_o\mid {\partial W\over \partial z_i}\mid ^2
\quad , 
\eeq 
where $z_i$ is a superfield.  To obtain the interactions, we take the
derivative of $W$ with respect to $z$ and then evaluate the
result in terms of the scalar component of $z$.
The supersymmetric scalar potential is then,  
\beq
V=\mid \mu\mid^2\biggl(\mid \Phi_1\mid^2+\mid \Phi_2\mid^2\biggr)
+{g^2+g^{\prime 2}\over 8}\biggl(\mid \Phi_2\mid^2
-\mid \Phi_1\mid^2\biggr)^2+{g^2\over 2}\mid \Phi_1^*\cdot
 \Phi_2\mid^2
 .
\eeq
This potential has its minimum at $\langle \Phi_1^0\rangle=
\langle \Phi_2^0\rangle=0$, giving $\langle V\rangle=0$ and
so represents a model with no electroweak symmetry breaking.

It is difficult to break supersymmetry (and we know that it
must be broken since there are  no scalars degenerate in mass
with the known fermions).  The simplest solution is simply to add
all possible soft supersymmetry breaking mass terms.  In the
Higgs sector, this amounts to adding masses for each doublet,
along with an arbitrary mixing term.\footnote{These soft supersymmetry
breaking terms do not generate quadratic divergences.}  
  The scalar potential
involving the Higgs bosons becomes,  
\beqn
V_H&=&
\biggl(\mid \mu\mid^2 +m_1^2\biggr)\mid \Phi_1\mid^2
+\biggl(\mid \mu \mid^2+m_2^2\biggr)\mid \Phi_2\mid^2
-\mu B \epsilon_{ij}\biggl(\Phi_1^i \Phi_2^j+{\rm h.c.} \biggr)
\nonumber \\
&&
+{g^2+g^{\prime 2}\over 8}\biggl(
\mid \Phi_1\mid^2 - \mid \Phi_2\mid^2\biggr)^2
+{1\over 2} g^2 \mid \Phi_1^*\Phi_2\mid^2
\quad .
\label{higgspot} 
\eeqn 
The Higgs potential of the SUSY model can be seen
to depend on $3$ independent combinations of parameters, 
\beqn 
&& \mid\mu\mid^2+m_1^2,
\nonumber \\
&& 
\mid \mu \mid^2+m_2^2,
\nonumber \\
&&~~~  \mu B~~, 
\eeqn   
where $B$ is a new mass parameter.  
This is in 
contrast to the general $2$ Higgs doublet model where 
there are $6$ arbitrary coupling constants (and a phase)
 in the potential.  
From Eq. \ref{higgspot}, it is clear that the quartic couplings
are fixed in terms of the gauge couplings and so they are
not free parameters.  
Note that $V_H$ automatically conserves CP since
any complex phase in $\mu B$ can be absorbed into the
definitions of the Higgs fields.    
  
Clearly, if $\mu B=0$ then all the terms in the potential
are positive and the minimum  of the potential
occurs with $V=0$ and $\langle \Phi_1^0\rangle=\langle \Phi_2^0
\rangle=0$.
  Hence all $3$ parameters
must be non-zero in order for the electroweak symmetry to be
broken.  
\footnote{
We assume that the parameters are arranged in such
a way that the scalar partners of the quarks and leptons
do not obtain vacuum expectation values.  Such vacuum
expectation values would spontaneously break the $SU(3)$ 
color  gauge symmetry or lepton number.
}
The symmetry is broken when the neutral components of the Higgs doublets
get vacuum expectation values,
\beqn
\langle \Phi_1^0\rangle & \equiv & v_1
\nonumber \\
\langle \Phi_2^0\rangle & \equiv & v_2
\quad . 
\eeqn  
By redefining the Higgs fields, we can always 
choose $v_1$ and $v_2$ positive.

\begin{figure}[t]
\vskip -.5in 
\centering   
\epsfxsize=4.in
\leavevmode
\epsffile{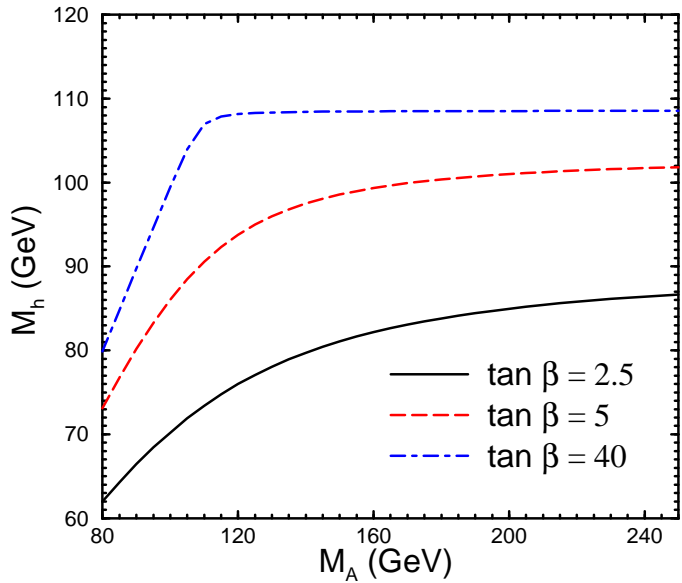}
\caption{Mass of the neutral Higgs bosons as
a function of the pseudoscalar mass, $M_A$,
and $\tan\beta$.
 This figure assumes a common scalar mass of
$1~TeV$, and neglects mixing effects, ($A_i=\mu=0$).}
\vspace*{.5in}
\label{mssmhm}  
\end{figure} 

In the MSSM, the Higgs mechanism works in the same manner
as in the Standard Model.  
When the electroweak symmetry is broken,
the $W$ gauge boson gets a mass which
is fixed by $v_1$ and $v_2$,
\beq
M_W^2={g^2\over 2}(v_1^2+v_2^2)\quad .\eeq
Before the symmetry was broken, the $2$ complex $SU(2)_L$ 
Higgs doublets had $8$ degrees of freedom.  Three of
these were  absorbed to give the $W$ and $Z$ gauge bosons
their masses, leaving $5$ physical degrees of freedom.
There is now a charged Higgs boson, $H^\pm$, a CP -odd neutral
Higgs boson, $A^0$, and $2$ CP-even neutral Higgs bosons, $h$ and $H$.
It is a general prediction of supersymmetric models that there will be
an expanded sector of physical Higgs bosons. 
After fixing $v_1^2+v_2^2$ such that the $W$  boson gets the correct
mass, the Higgs sector is  then described by $2$ additional
parameters.  The
usual choice is
\beq
\tan\beta\equiv {v_2\over v_1}\eeq
and $M_A$, the mass of the pseudoscalar Higgs boson.
Once these two parameters are given, then the  masses of
the remaining Higgs bosons can be calculated in terms
of $M_A$ and $\tan\beta$.
Note that we can chose $0 \le \beta\le {\pi\over 2}$ since we have
chosen $v_1, v_2 > 0$.

In the MSSM, the $\mu$ parameter is a source of concern. 
It cannot be 
set to zero because then
there would be no symmetry breaking.   The $Z$-boson
mass can be written in terms of the radiatively corrected neutral
Higgs boson masses and $\mu$:
\beq
M_Z^2=2\biggl[ {M_h^2-M_H^2\tan^2\beta\over 
\tan^2 \beta-1}\biggr] - 2 \mu^2
.
\eeq
This requires a delicate cancellation between the Higgs masses
and $\mu$.  This is unattractive, since much of the motivation for 
supersymmetric  
theories is the desire to avoid unnatural cancellations.
The $\mu$ parameter can, however, be generated naturally in theories with
additional Higgs singlets.

\begin{figure}[t]
\vskip -.5in 
\centering   
\epsfxsize=4.in
\leavevmode
\epsffile{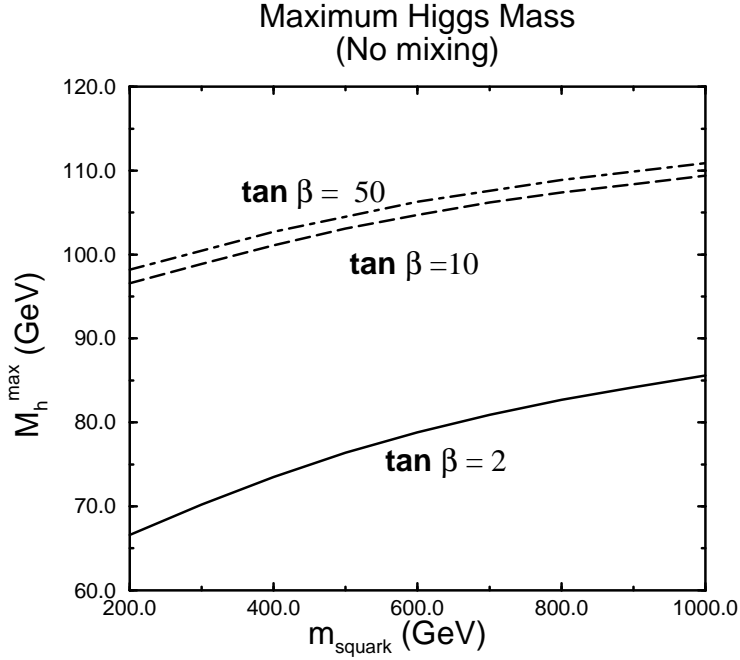}
\caption{Maximum value of the lightest Higgs boson  mass as a
function of the  squark mass.
This figure includes 2-loop radiative corrections and renormalization
group improvements.
(We have assumed degenerate squarks and set  the
mixing parameters $A_i=\mu=0$.)}  
\end{figure}

It is straightforward to find the physical Higgs bosons 
and their masses in terms of the parameters of 
Eq. \ref{higgspot}.  Details can be found in Ref. 1.  
The neutral Higgs masses are found by diagonalizing
the $2\times 2$ Higgs mass matrix and    
by convention, $h$ is taken to be the lighter of the neutral
Higgs.   At tree level, the masses of the neutral Higgs
bosons are given by,
\beq
M_{h,H}^2={1\over 2}\biggl\{
M_A^2 +M_Z^2 \mp \biggl( (M_A^2+M_Z^2)^2-4 M_Z^2 M_A^2\cos^2
2 \beta \biggr)^{1/2}\biggr\}
.
\eeq
The pseudoscalar mass is given by,
\beq 
M_A^2={2 \mid \mu B \mid \over \sin 2 \beta},
\eeq
and the charged scalar mass is,
\beq
M_{H^\pm}^2=M_W^2+M_A^2
\quad .  
\eeq
We see that   at tree level,
 Eq. \ref{higgspot}  gives important predictions about
the relative masses of  the Higgs bosons,
\beqn
M_{H^+} &>& M_W \nonumber \\
M_H &>& M_Z \nonumber \\
M_h &<& M_A\nonumber \\
M_h &<& M_Z\mid \cos 2 \beta\mid 
 \quad .  
\label{higgmass}
\eeqn
These relations yield  the attractive
prediction that the lightest neutral
Higgs boson is lighter than the $Z$  boson. 
  However,
 loop corrections to the
relations of Eq. \ref{higgmass} are large.  
In fact the corrections  to $M_h^2$ grow like $G_F M_t^4$
and receive contributions   from loops with both top
quarks 
and squarks.  In a model with unbroken supersymmetry,
these contributions would cancel. Since the supersymmetry
has been broken by splitting the masses of the
fermions and their scalar partners, 
 the neutral Higgs boson masses become
at one- loop,\cite{susyrad}
\beqn 
M_{h,H}^2&=& {1\over 2}\biggl\{ M_A^2+M_Z^2+{\epsilon_h\over \sin^2
\beta}\pm\biggl[
\biggl(M_A^2-M_Z^2)\cos 2 \beta +
{\epsilon_h\over \sin^2\beta}\biggr)^2
\nonumber  \\ &&~~
+\biggl(M_A^2+M_Z^2\biggr)^2\sin^2 2 \beta\biggr]^{1/2}\biggr\}
\eeqn 
where  $\epsilon_h$ is the contribution
of the one-loop  corrections,
\beq
\epsilon_h\equiv {3 G_F\over \sqrt{2}\pi^2}M_t^4
\log \biggl(1+ {{\tilde m}^2\over M_t^2}\biggr)
\quad .
\eeq 
We have assumed that all of  the squarks
have equal masses,  ${\tilde m}$, and have 
 neglected the smaller effects from the mixing parameters,
$A_i$ and $\mu$.  In Fig. \ref{mssmhm}, we show the 
masses of the neutral Higgs bosons
as a function of the pseudoscalar mass
and for three values of $\tan\beta$.  
For $\tan\beta > 1$, the mass eigenvalues increase monotonically
with increasing $M_A$ and give an upper bound to the
mass of the lightest Higgs boson,
\beq
M_h^2 < M_Z^2 \cos^2 2 \beta +\epsilon_h
\quad .
\eeq  
The corrections from $\epsilon_h$ are always positive and
increase the mass of the lightest neutral Higgs boson with
increasing top quark mass.  
From Fig. \ref{mssmhm}, we see that $M_h$ obtains its maximal value for
rather modest values of the pseudoscalar mass, $M_A > 300~GeV$.  
The radiative corrections to the charged Higgs mass-squared
are proportional to $M_t^2$ and so are much smaller than
the corrections to the neutral masses.

\begin{figure}[t]
\vskip -.5in 
\centering   
\epsfxsize=4.in
\leavevmode
\epsffile{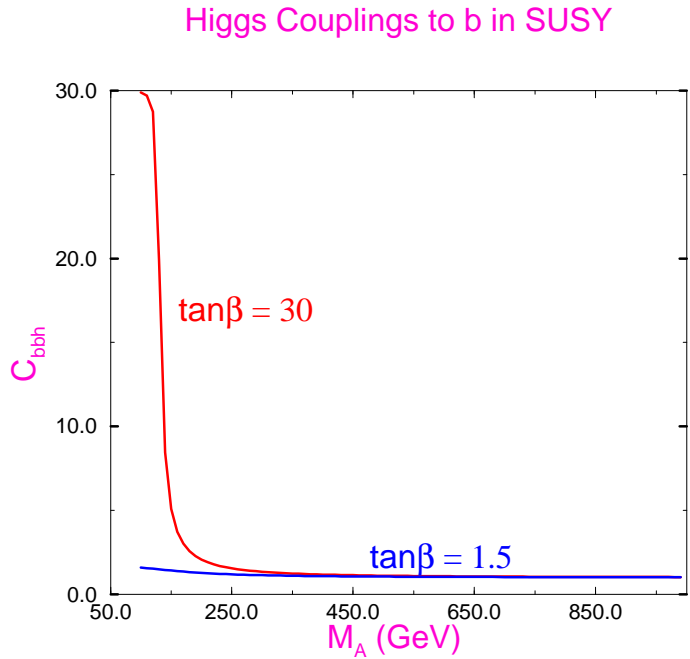}
\caption{Coupling of the lightest Higgs boson
to  charge $-1/3$  quarks.
The value  
$C_{bbh}=1$ corresponds to the Standard Model coupling
of the Higgs boson to charge $-1/3$ quarks.}
\vspace*{.5in}  
\end{figure}  

There are many analyses\cite{susyrad}
which include a variety of two-loop effects, renormalization
group effects, etc., but the important point is that for
 given values  of $\tan\beta$ and the squark masses,
 there is an upper bound
on the lightest neutral Higgs boson mass.  
  For large values of $\tan\beta$
the limit is relatively insensitive to the value of
$\tan\beta$ and with  a squark mass less than about $1~TeV$,
the upper limit on the Higgs mass is about $110~GeV$ if mixing in
the top squark sector is negligible ($A_T\sim 0$).
For large mixing, this limit is raised to
around $130~GeV$.
\begin{itemize}
\item  
The minimal supersymmetric   model predicts a neutral Higgs
boson with a mass less than around $130~GeV$.
\end{itemize}
Such a mass scale may be accessible at LEP2, an
upgraded Tevatron or the
LHC and provides a definitive test of the MSSM.     

In a more complicated supersymmetric  model with a richer Higgs
structure, the upper bound on the lightest Higgs boson
mass will be  changed.  However,
the requirement that the Higgs self coupling remain
perturbative up to the Planck scale gives an upper
bound on the lightest supersymmetric  Higgs boson of around  
$150~GeV$ in all models with only singlets and doublets
of Higgs bosons.\cite{quiros}
This is a very strong statement.  It implies that either
there is a relatively light Higgs boson
or else there is some new physics between the weak scale
and the Planck scale which causes the Higgs
couplings to become non-perturbative.

Another feature of  the MSSM is that the fermion- Higgs couplings 
are no longer strictly proportional to mass. 
  From
the superpotential can be found both the scalar potential and
the Yukawa interactions of the fermions with the scalars:
\beq 
{\cal L}_{W}=-\sum_i \mid {\partial W\over
\partial z_i}\mid ^2 -{1\over 2}\sum_{ij}
\biggl[ {\overline \psi}_{iL} {\partial^2 W
\over \partial z_i \partial z_j}\psi_j+{\rm
h.c.}\biggr]
, 
\label{lagw} 
\eeq
where $z_i$ is a chiral superfield.
This form of the Lagrangian is dictated by the supersymmetry
and by the requirement that it be renormalizable.
  To obtain the interactions, we take the derivatives of $W$ with
respect to the superfields, $z_i$, and then evaluate the result in 
terms of
the scalar component of $z_i$.   
  
The usual approach is to write the most general
$SU(3)\times SU(2)_L\times U(1)_Y$ invariant 
superpotential with arbitrary coefficients for the interactions,  
\begin{eqnarray}
W &=& \epsilon_{ij} \mu {\hat H}_1 ^i {\hat H}_2^j 
+\epsilon_{ij}
\biggl[ \lambda_L  {\hat H}_{1 }^i {\hat L}^{cj}{\hat E}^c +
\lambda_D  {\hat H}_1^i {\hat Q}^j {\hat D}^c
+\lambda_U  {\hat H}_2^j {\hat Q}^i {\hat U}^c\biggr]
\nonumber \\
&& + \epsilon_{ij}\biggl[ {\lambda_1} {\hat L}^i 
{\hat L}^j {\hat E}^c +
\lambda_2 {\hat L}^i {\hat Q}^j {\hat D}^c
\biggr] 
+\lambda_3 {\hat U}^c {\hat D}^c {\hat D}^c
, 
\label{superpot} 
\eeqn  
(where  $i,j$ are $SU(2)$ indices).
We have written the superpotential in terms of the fields of the first
generation.  In principle, the $\lambda_i$
 could all be  matrices which
mix the interactions of the $3$ generations.

The
terms in the square brackets proportional
to $\lambda_L$, $\lambda_D$, and $\lambda_U$  give the usual
Yukawa interactions of the fermions with the
Higgs bosons from the term
\beq
{\overline \psi}_i \biggl({\partial^2 W\over \partial z_i 
\partial z_j}\biggr) \psi_j.
\eeq
 Hence   these coefficients are
determined  in terms of the fermion masses and
the 
vacuum expectation values of the neutral
members of the scalar components of the Higgs doublets and are not
free parameters at all.

It is convenient to write  
  the couplings
for the  neutral Higgs  bosons to the fermions
in terms of the Standard  Model Higgs couplings,
\beq
{\cal L}=-{g m_i\over 2 M_W} \biggl[C_{ffh}{\overline f}_i f_i h
+C_{ffH} {\overline f}_i f_i H
+C_{ffA}{\overline f}_i \gamma_5 f_i A\biggr],  
\eeq
where $C_{ffh}$ is $1$ for a Standard Model Higgs
boson.
The $C_{ffi}$ are given in Table 5 and plotted in Figs. 33 and 34
as a function of $M_A$. 
We see that for small $M_A$ and large $\tan\beta$, the couplings of
the neutral Higgs boson to fermions can be significantly different
from the Standard Model couplings; the $b$-quark coupling becomes
enhanced, while the $t$-quark coupling to 
the lightest Higgs boson  is suppressed.  
 When $M_A$ becomes large
the Higgs-fermion couplings approach their standard model
values, $C_{ffh}\rightarrow 1$. In fact even for $M_A\sim 300~GeV$,
the Higgs-fermion couplings are very close to their 
Standard Model values.

\begin{table}[t]
\begin{center}
{Table 5: Higgs Boson  Couplings to fermions}\vskip6pt
\renewcommand\arraystretch{1.2}
\begin{tabular}{|lccc|}
\hline
\multicolumn{1}{|c}{$f$}& $C_{ffh}$& $C_{ffH}$
 & $C_{ffA}  $
\\
\hline
$u$   &    ${\cos\alpha\over \sin\beta}$ &
    ${\sin\alpha\over\sin\beta}$
& $\cot\beta$ \\ 
$d$   &    $-{\sin\alpha\over\cos\beta}$ & 
     ${\cos\alpha\over\cos\beta}$ 
& $\tan\beta$  \\
\hline
\end{tabular}
\end{center}
\end{table}

\begin{figure}[t]
\vskip .5in 
\centering   
\epsfxsize=3.5in
\leavevmode
\epsffile{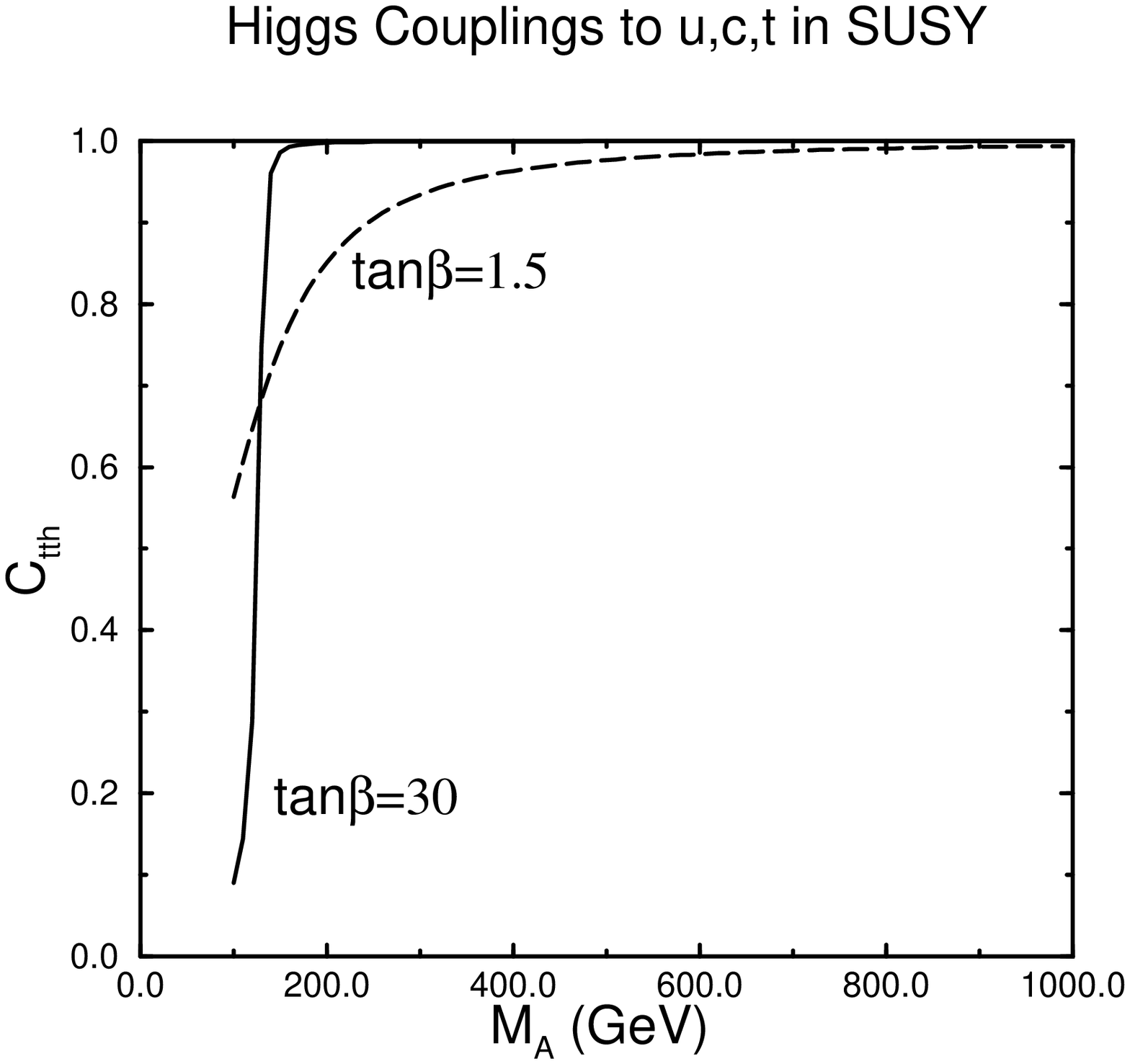}
\caption{Coupling of the lightest Higgs boson
to charge $2/3$ quarks.
The value  
$C_{tth}=1$ yields the Standard Model coupling of the Higgs
boson  
to charge $2/3$ quarks.}
\vspace*{.5in}
\label{ctth}
\end{figure}

The Higgs boson couplings to gauge bosons are fixed by the
$SU(2)_L\times U(1)$ gauge invariance.  
Some of the  phenomenologically important 
couplings  
 are:
\beqn
Z^\mu Z^\nu h:&&{igM_Z\over\cos\theta_W} \sin(\beta-\alpha)g^{\mu\nu}
\nonumber \\   
Z^\mu Z^\nu H:&&{igM_Z\over \cos\theta_W}\cos(\beta-\alpha) g^{\mu\nu}
\nonumber \\
W^\mu W^\nu h:&& igM_W \sin(\beta-\alpha)g^{\mu\nu}
\nonumber \\   
W^\mu W^\nu H:&& igM_W \cos(\beta-\alpha)g^{\mu\nu}
\nonumber \\
Z^\mu h(p)A(p^\prime):&&{g\cos(\beta-\alpha)\over 2 \cos\theta_W}
(p+p^\prime)^\mu\nonumber \\
Z^\mu H(p)A(p^\prime):&&-{g\sin(\beta-\alpha)\over 2 \cos\theta_W}
(p+p^\prime)^\mu
\quad . 
\label{vvhcoup}  
\eeqn  
We see that the couplings of the Higgs bosons to the gauge bosons
all depend on the same angular factor, $\beta-\alpha$. 
  The pseudoscalar, $A^0$, has no tree level coupling to pairs of
gauge bosons.   The angle $\beta$ is a
free parameter while the neutral Higgs mixing angle,
$\alpha$, which enters
into many of the couplings, can be found  at leading logarithm 
in terms of 
$M_A$ and $\beta$: 
\beq
\tan 2 \alpha={(M_A^2+M_Z^2)\sin 2\beta
\over (M_A^2-M_Z^2)\cos 2 \beta+\epsilon_h/\sin^2\beta}
\quad . 
\eeq
With our conventions, $-{\pi\over 2}\le \alpha\le 0$. 
It is clear that the couplings of the neutral scalars to vector
bosons ($V=W^\pm,Z$) are suppressed relative to those of the standard 
model
\beq
g^2_{H_1VV}+g^2_{H_2VV}=g^2_{H VV}(SM), 
\eeq
where $g_{HVV}$ is the coupling of the Higgs boson to vector
bosons.
Because of this sum rule, the $WW$ scattering production mechanism
tends not to be as important in supersymmetric
models as in the Standard
Model.

  A complete set of couplings
for the Higgs bosons (including the charged and pseudoscalar Higgs)
 at tree level 
can be found in Ref. 1. 
These couplings completely determine the decay modes
of the supersymmetric  Higgs bosons and their  experimental signatures.
 The important point is that 
(at lowest order)  all of the couplings are completely determined
in terms of $M_A$ and $\tan\beta$.  When radiative corrections are
included there is a   dependence on the squark masses and 
the mixing parameters in the mass matrices. This dependence is explored
in detail in Ref. 34.

It is an important feature  of the MSSM  that for large $M_A$,
the Higgs sector looks exactly  like that of the Standard Model.  As
$M_A\rightarrow \infty$, the masses of the charged Higgs
bosons, $H^\pm$,
and the heavier neutral Higgs, $H$, become large leaving
only the lighter Higgs boson, $h$, in the spectrum.  In this limit,
the couplings of the lighter Higgs
boson, $h$, to fermions and gauge bosons take
on their Standard Model values. We have, 
\beqn
\sin(\beta-\alpha)&& 
\rightarrow 1~ {\hbox {for}}~M_A\rightarrow \infty
\nonumber \\ 
\cos(\beta-\alpha)&&\rightarrow  0 
\quad . 
\eeqn
From Eq. \ref{vvhcoup}, we see that the heavier Higgs
boson, $H$, decouples from the gauge bosons in the heavy $M_A$ limit,
while
the lighter Higgs boson, $h$, has Standard Model couplings.
The Standard Model limit is also
 rapidly approached in the fermion-Higgs couplings for $M_A > 300~GeV$.
  In the
limit of large $M_A$, it will thus 
 be exceedingly difficult to differentiate a supersymmetric 
Higgs sector from the  Standard Model Higgs boson.
\begin{itemize}
\item
The supersymmetric Higgs sector with large $M_A$ looks like the
Standard Model Higgs sector.
\end{itemize}                                 
In this case, it will be difficult to discover supersymmetry  through the
Higgs sector.  Instead, it will be necessary to find some
of the other supersymmetric  partners of the observed particles.

At a hadron collider, the neutral  Higgs bosons of a supersymmetric
model  can 
be searched for using  many of the same techniques as in the Standard 
Model.\cite{fptasi,mren}  For most choices of the parameter space, gluon fusion
is the dominant production mechanism.  In the Standard
Model, it was only the top quark contribution to gluon
fusion which was important. In a supersymmetric  model, however,
the coupling to the $b$ quark can be important for small values
of $\cos\beta$, as can be seen
from Table 5.

Supersymmetric  models have a rich particle spectrum in
which to search for evidence of the Higgs mechanism. 
The various decays such as $h,H\rightarrow \gamma\gamma$, $H^+\rightarrow
l^+\nu$, $A^0\rightarrow \tau^+\tau^-$, etc, are sensitive to
different regions in the $M_A-\tan\beta$ parameter space.
It takes the combination of many decay channels
in order to be able to cover the  parameter space completely with out
any holes.  Discussions of the capabilities of the LHC detectors
to experimentally observe evidence for the Higgs bosons of supersymmetric
models can be found in the ATLAS\cite{atlas} and CMS\cite{cms}
studies.
An upgraded Tevatron will also have the capability to obtain meaningful
limits on the symmetry breaking sector of a supersymmetric model.\cite{tevlims} 

Both the Tevatron and the LEP and LEP2 colliders have searched for 
the Higgs bosons and other new particles occuring in a SUSY model
and have ruled out large portions of the $tan\beta$- $M_A$ parameter
space.\cite{dj}  
\section{Conclusions}
\begin{table}[t]
\begin{center}
{Table 6: Higgs Mass Reach of Future Accelerators}
\vskip6pt
\renewcommand\arraystretch{1.2}
\begin{tabular}{|lcl|}
\hline
\multicolumn{1}{|c}{Accelerator}& Luminosity& Higgs Mass Reach
\\
\hline
LEP2 ($192~GeV$) &  $150~ pb^{-1}$  &   $95~GeV$\\
Tevatron         &   $5-10~fb^{-1}$ & $80-100~GeV$\\
TEV-33           &  $25-30~fb^{-1}$ &   $120~GeV$ \\
LHC              &  $100~fb^{-1}$   &  $800~GeV$  \\          
NLC ($500~GeV$)  &  $50~fb^{-1}$   & $350~GeV$
\\
\hline
\end{tabular}
\end{center}
\end{table}

Our current experimental
 knowledge  of the Standard Model
 Higgs boson gives only the limits $M_h> 90~GeV$
and $M_h< 280~GeV$ from direct search experiments and
 precision measurements
at the LEP  and LEP2 experiments.
From here, we must wait until the advent of 
an upgraded Tevatron and the
LHC for further limits.  Through the decay $h\rightarrow \gamma \gamma$
and the production process $pp\rightarrow Z l^+l^-$, the LHC will 
probe the mass region between $100< M_h< 180~GeV$,
while the Tevatron is sensitive to $M_h<130~GeV$ with $30~fb^{-1}$
through the $Wh$ production process.
For the higher mass region, $180 < M_h < 800~GeV$, the LHC
will be able
to see the Higgs boson through the gold plated decay mode, $h\rightarrow
ZZ\rightarrow l^+l^-l^+l^-$.
The expected sensitivity of future colliders is summarized in Table 
6.\cite{snowhiggs}

One of the important yardsticks for all current and future accelerators is their
ability to discover (or to definitively exclude) the Higgs boson
of the Standard Model.
We  hope that at the time of the LHC, we will be able to
probe all mass scales up to $M_h\sim 800~GeV$.   Having
found the Higgs boson, the next goal will be to determine if
it is a Standard Model Higgs boson, or a Higgs boson of 
some more complicated theory such as the MSSM.

If the Higgs boson is not
  found below this mass scale then we are in the regime where
perturbative unitarity has broken down
and we are led to the
exciting conclusion that there must be new physics beyond the Standard Model 
waiting to be discovered.
\section*{Acknowledgements}
I am grateful to G. Senjanovic, A. Masiero, and D. Smirnov for organizing
such a successful and enjoyable summer school.  
 

\begin{thebibliography}{999}

\bibitem{hhg}{J.~Gunion, H.~Haber, G.~Kane, and S.~Dawson,
{\it The Higgs Hunters Guide}, (Addison-Wesley, Menlo Park, 1990).}

\bibitem{bag}{Similar material can be found in J.~Bagger, 
{\it Physics Beyond the Standard Model},
lectures given at the 1991 TASI summer school, Boulder, CO,
2-28 June, 1991, 
(World Scientific, Singapore, 1992);
S.~Dawson, {\it Introduction to the Physics of Higgs Bosons},
lectures given at the 1994 TASI summer school, Boulder, CO,
29 May -24 June, 1994, 
(World Scientific, Singapore, 1994), hep-ph/9411325;
 A. Djouadi, {\it Int. J. Mod. Phys.} {\bf A10}
(1995) 1;
 R.~Chivukula, {\it Models of Electroweak Symmetry Breaking: Course},
 lectures given at  the Les Houches Center of 
Physics, Les Houches, France, 16-26 June, 1998, hep-ph/9803219;
M.~Spira and P.~Zerwas, {\it Electroweak Symmetry Breaking and
Higgs Physics}, lectures given at the International University School
of Nuclear and Particle Physics, Schladming, Austria, 1-8 March 1997,
hep-ph/9803257.}


\bibitem{quigg}{An introduction to the Standard Model can be
found in C. ~Quigg, {\it Gauge Theories of the Strong, Weak,
and Electromagnetic Interactions}, (Benjamin-Cummings, Reading, MA.,
1983).}

\bibitem{ablee}{E.~Abers and B.~Lee, {\it Phys. Rep.} {\bf 9} (1975) 1.}


\bibitem{gold}{P.~W.~ Higgs, {\it Phys. Rev. Lett.} {\bf 13} (1964)508;
 {\it Phys. Rev.} {\bf 145} (1966) 1156;
F. ~Englert and R.~Brout, {\it Phys. Rev. Lett.} {\bf 13}
(1964) 321; G.~S.~Guralnik, C.~R. Hagen, and T.~Kibble,
{\rm Phys. Rev. Lett.} {\bf 13} (1965) 585;
T.~Kibble, {\it Phys. Rev.} {\bf 155} (1967) 1554.}


\bibitem{ws}{S.~Glashow, {\it Nucl. Phys.} {\bf 22} (1961) 579;
S.~Weinberg, {\it Phys. Rev. Lett.} {\bf 19} (1976) 1264;
A.~Salam, in {\it Elementary Particle Theory}, ed. N.~Svartholm
(Almqvist and Wiksells, Stockholm, 1969), p. 367.}


\bibitem{triv}{K.~Wilson, {\it Phys. Rev.} {\bf B4} (1971) 3184;
K.~Wilson and J.~Kogut, {\it Phys. Rep.} {\bf 12} (1974) 75;
R.~Dashen and H.~Neuberger, {\it Phys. Rev.
Lett.} {\bf 50} (1983) 1897;  P.~Hasenfratz and J.~Nager,
{\it Z.~Phys.} {\bf C37} (1988);
J.~Kuti, L.~Lin, and Y.~Shen, {\it Phys. Rev. Lett.}
{\bf 61} (1988) 678;
M.~Luscher and P.~Weisz, {\it Phys. Lett.} {\bf B212} (1988)
(472).}

\bibitem{chivtriv}{R. Chivukula and E.~Simmons, {\it
Phys. Lett.} {\bf B388} (1996) 788.}

\bibitem{quirrev}{M.~Quiros, {\it Perspectives in Higgs Physics},
Ed. G.~Kane, (World Scientific, Singapore, 1997).}


\bibitem{hasen}{A.~Hasenfratz, {\it Quantum Fields on the Computer},
Ed. M.~Creutz, (World Scientific, Singapore, 1992),p. 125.}



\bibitem{rge} {T.~Cheng, E.~Eichten and L.~Li, {\it Phys. Rev.}
{\bf D9} (1974) 2259;
B.~Pendleton and G.~Ross, {\it Phys. Lett.} {\bf B98} (1981) 291;
C.~Hill, {\it Phys. Rev.} {\bf D24} (1981) 691;
J.~Bagger, S.~Dimopoulos and E.~ Masso,
{\it Nucl. Phys.} {\bf B253} (1985) 397;
M.~Beg, C.~Panagiotakopoulos, and A.~Sirlin, {\it Phys. Rev.
Lett.} {\bf 52} (1984) 883;
M.~Duncan, R.~Philippe, and M.~Sher, {\it Phys. Lett.}
{\bf B153} (1985) 165;
K.~Babu and E.~Ma, {\it Phys. Rev. Lett.} {\bf 55} (1985) 3005.}

\bibitem{cab}{
N.~Cabibbo {\it et.al.}, {\it Nucl. Phys.} {\bf B158} (1979) 295.}

\bibitem{linde}{A.~Linde, {\it JETP Lett} {\bf 23} (1976) 64;
{\it Phys. Lett.} {\bf B62} (1976) 435;
S.~Weinberg, {\it Phys. Rev. Lett.} {\bf 36} (1976) 294;
S.~Coleman and E.~Weinberg, {\it Phys. Rev.} {\bf D7} (1973) 188.}


\bibitem{vacbounds}{M.~Lindner, M.~Sher, and H. Zaglauer,
{\it Phys. Lett.} {\bf B228} (1989) 139;
C.~Ford {\it et. al.}, {\it Nucl. Phys.} {\bf B395} (1993) 62;
M. Sher, {\it Phys. Rep.} {\bf 179} (1989)274;
 F.~del Aguila, M~Martinez,
M.~Quiros, {\it Nucl. Phys.} {\bf B381} (1992) 451;
J. ~Casas, J.~Espinosa, and M.~Quiros, {\it Phys. Lett.} {\bf B324}
(1995) 171; {\it Phys. Lett.} {\bf B382} (1996) 374;
J.~Espinosa and M.~Quiros, {\it Phys. Lett.} {\bf B353} (1995) 257.}

\bibitem{sher}{M. Sher, {\it Phys. Lett.} {\bf B317} (1993)159;
addendum, {\bf B331} (1994) 448.}


\bibitem{rhoh}{W.~Marciano and A.~Sirlin, {\it Phys. Rev. Lett.}
{\bf 46} (1981) 163;
W. Marciano, S.~Sarantakos, and A.~Sirlin, {\it Nucl. Phys. }
 {\bf B217} (1988) 84.}

\bibitem{abl} {T.~Appelquist and C.~Bernard, {\it Phys. Rev.}
{\bf D22} (1980) 200; A.~Longhitano, {\it Nucl. Phys.} {\bf B188}
(1981) 118; T.~Appelquist and G.~Wu, {\it Phys. Rev.}
{\bf D51} (1995) 240.}


\bibitem{veltman} {M.~Veltman, {\it Acta. Phys. Pol.} {\bf B8} (1977) 475.}

\bibitem{einhorn}{M.~Einhorn and J.~Wudka, {\it Phys. Rev. }{\bf D39}
(1989) 2758.}

\bibitem{kar}{D.~Karlen, results presented at {\it XXIX International
Conference on High Energy Physics}, Vancouver, Canada, July 23-29, 1998.}

\bibitem{spirrev}{M.~Spira, {\it Fortsch. Phys.} {\bf 46} (1998) 203.} 

\bibitem{hbqcd}{ E.~Braaten and J.~Leveille, {\it Phys. Rev.} {\bf D22}
(1980) 715; and M.~Drees and K.~Hikasa, {\it Phys. Lett.} {\bf B240}
(1990) 455.}

\bibitem{h2l}{K.~Melnikov, {\it Phys. Rev.} {\bf D53} (1996) 5020.}


\bibitem{ewh}{A.~Djouadi and P.~Gambino, {\it Phys. Rev. Lett.}
{\bf 73} (1994) 2528.}  


\bibitem{hdecay}{A.~Djouadi, J.~Kalinowski, and M.~Spira,
{\it Comput. Phys. Comm. } {\bf 108} (1998) 56.}

\bibitem{wkwm}{W.-Y.~Keung and W.~Marciano, {\it Phys. Rev.} {\bf D30}
(1984) 248.}


\bibitem{hggg}{T.~Inami, T.~Kubota, and Y.~Okada, {\it Z.~Phys.}
{\bf C18} (1983) 69; M.~Spira, A.~Djouadi, D.~Graudenz, and P.~Zerwas,
{\it Nucl. Phys.} {\bf B453} (1995) 17.} 

\bibitem{hzg}{R. ~Cahn, M.~Chanowitz, and N.~Fleishon, {\it Phys.
Lett.} {\bf B82} (1979)113; G.~Gamberini, G.~Giudice and G.~Ridolfi,
{\it Nucl. Phys. } {\bf B292} (1987) 237.} 

\bibitem{hgg}{A.~Vainshtein, M.~Voloshin, V.~Sakharov, and M.~Shifman,
{\it Sov. J. Nucl. Phys.} {\bf 30} (1979) 711.} 


\bibitem{spgg} {S.~Dawson and R.~Kauffman, {\it Phys. Rev.}
{\bf D49} (1993) 2298; A.~Djouadi, M.~Spira, and P.~Zerwas, {\it
Phys. Lett.} {\bf B311} (1993) 255.}
 


\bibitem{sigee}{B.~Ioffe and V. ~Khoze, {\it Sov. J. Part. Nucl.
Phys.} {\bf 9} (1978) 50.}


\bibitem{kniehl}{B.~Kniehl, {\it Phys. Rep.} {\bf 240C} (1994) 211.}


\bibitem{berends}{F.~Berends, W.~van Neerven, and G.~Burgers,
{\it Nucl. Phys.} {\bf B297} (1988) 429; erratum
{\bf B304} (1988) 921.}

\bibitem{lepstud}{M.~Carena and P.~Zerwas, {\it Higgs Physics},
CERN Yellow Report, CERN-96-01, hep-ph/9602250.}
                  
\bibitem{bbgh}{V.~Barger, M.~Berger, J.~Gunion, and T.~Han,
{\it Phys. Rev. Lett.} {\bf 78} (1997) 3991.}


\bibitem{spin}{V.~Barger, K.~Cheung, A.~Djouadi, B.~Kniehl, and
P.~Zerwas, {\it Phys. Rev.} {\bf D49} (1994) 49.} 

\bibitem{zhnsm}{W.~Kilian, M.~Kramer, and P.~Zerwas, {\it Phys.
Lett.} {\bf B381} (1996) 243.}

\bibitem{glue}{F.~Wilczek, {\it Phys. Rev. Lett.} {\bf 39}
(1977) 1304; J.~Ellis {\it et.al.}, {\it Phys. Lett.}
{\bf 83B} (1979) 339;
H.~Georgi {\it et. al.}, {\it Phys. Rev. Lett.}
{\bf 40} (1978) 692; T.~Rizzo, {\it Phys. Rev.} {\bf D22} (1980) 178.}


\bibitem{qcd1}{M.~Spira, A.~Djouadi, D.~Graudenz, and
P.~Zerwas, {\it Nucl. Phys.} {\bf B453} (1995) 17.}

\bibitem{qcd2}{S.~Dawson, {\it Nucl. Phys.} {\bf B359} (1991) 283.}

\bibitem{russ} {A.~Vainshtein {\it et.al.}, {\it Sov. J. Nucl. Phys.};
 M.~Voloshin, {\it Sov. J.~Nucl. Phys. }{\bf B44} (1986) 478.}


\bibitem{qcdsum}{R.~Kauffman, {\it Phys. Rev.} {\bf D45} (1992) 1512.}

\bibitem{atlas}{LOI of the ATLAS Collaboration, CERN/LHCC/92-4,
Oct., 1992.}

\bibitem{cms} {LOI of the CMS Collaboration, CERN/LHCC/92-3,
Oct., 1992.} 

\bibitem{fptasi}{F.~Paige,
{\it Supersymmetry Signatures at the Cern LHC},
lectures given at the 1997 TAS7 summer school, Boulder, CO,
2-28 June, 1997, 
(World Scientific, Singapore, 1998).}

\bibitem{agel}{P.~Agrawal and S.~Ellis, {\it Phys. Lett.} {\bf B229}
(1989) 145.}

\bibitem{gg}{J.~Gunion, G.~Kane, and J.~Wudka, {\it Nucl. Phys.}
{\bf B299} (1988) 231.}


\bibitem{keith}{K.~Ellis ${\it et.al.}$,
 {\it Nucl. Phys.} {\bf B297}(1988) 221.}

\bibitem{rain}{D.~Rainwater, D.~Zeppenfeld, and K.~Hagiwara,
hep-ph/9808468.}


\bibitem{stange}{W.~Marciano,
 A.~Stange, and S.~Willenbrock, {\it Phys.
Rev.} {\bf D49} (1994) 1354; {\it Phys. Rev.} {\bf D50} (1994) 4491.}

\bibitem{wwrc}{T.~Han and S.~Willenbrock, {\it Phys. Lett.} {\bf B273}
(1991) 167.}

\bibitem{tev2000} { {\it Future Electroweak Physics at the Fermilab Tevatron;
Report of the TeV 2000 Study Group}, Ed. D.~Amidei and R.~Brock, 
Fermilab-PUB-96-082, April, 1996.}

\bibitem{mren}{S.~Mrenna and C.~P.~Yuan, {\it Phys.
Lett.} {\bf B416} (1998) 200.}

\bibitem{tevwhstud}{S.~Kim, S.~Kuhlmann, and W.~-M.~Yao, in
{\it Proceedings of the 1996 DPF/DPB Summer Study on New Directions
for High Energy Physics}, 1996; W.~-~M.~Yao, in
{\it Proceedings of the 1996 DPF/DPB Summer Study on
New Directions for High Energy Physics}, (Snowmass, Colorado, July, 1996.}


\bibitem{effw}{S.~Dawson, {\it Nucl. Phys.} {\bf B249} (1985) 42;
G.~Kane, W.~Repko, and W.~Rolnick, {\it Phys. Lett} {\bf B148} (1984) 367;
M~Chanowitz and M.~Gaillard, {\it Phys. Lett.} {\bf B142}
(1984) 85.}

\bibitem{effgam}{S.~Brodsky, T.~Kinoshita, and H.~Terazawa,
{\it Phys. Rev.} {\bf D4} (1971) 1532.}

\bibitem{zzgold}{U.~Baur and E.~Glover, {\it Nucl. Phys.} {\bf B347}
(1990) 12; {\it Phys. Rev.} {\bf D44} (1991) 99.}


\bibitem{barger}{R.~Cahn {\it et. al.} {\it Phys. Rev.} {\bf D35}
(1987) 1626;
V.~Barger, T.~Han, and R.~Phillips,
{\it Phys. Rev.} {\bf D37} (1988) 2005;
J.~Gunion and M.~Soldate, {\it Phys. Rev.} {\bf D34} (1986) 826.}

\bibitem{dr}{A~Djouadi, D.~Haidt, B.~Kniehl, B.~Mele, and P.~Zerwas,
{\it Proceedings of $e^+e^-$ Collisions at $500~GeV$:
The Physics Potential}, (Munich-Annecy-Hamburg), ed. P.Zerwas,
DESY 92-123A.}

\bibitem{dawros}{S.~Dawson and J.~Rosner, {\it Phys. Lett.},
{\bf B148} (1984) 497.} 


\bibitem{tth}{A.~Djouadi, J.~Kalinowski, and P.~Zerwas,
{\it Zeit. fur Phy.} {\bf C54} (1992) 255; K. Gaemers and 
G.~Gounaris, {\it Phys. Lett.} {\bf B77} (1978) 379.}

\bibitem{ttqcd}{S.~Dawson and L.~Reina, hep-ph/9808443;
	S.~Dittmaier {\it et.al.}, hep-ph/9808433.}
\bibitem{marwil}
{W.~Marciano and S.~Willenbrock, {\it Phys. Rev.} {\bf D37}
(1988) 2509.}


\bibitem{lqt} {B.~Lee, C.~Quigg, and H.~Thacker, {\it Phys. Rev.}
{\bf D16} (1977) 1519; D.~Dicus and V.~Mathur, {\it Phys. Rev.}
{\bf D7} (1973) 3111.}

\bibitem{duncan}{M.~Duncan, G.~Kane, and W.~Repko, {\it Nucl. Phys.}
{\bf B272} (1986) 517.}

\bibitem{et}{J.~Cornwall, D.~Levin, and G.~Tiktopoulos,
{\it Phys. Rev. }{\bf D10} (1974) 1145; {\bf D11} (1975) 972E;
B.~Lee, C.~Quigg, and H.~Thacker, {\it Phys. Rev.} {\bf D16} (1977) 1519;
M.~Chanowitz and M.~Gaillard, {\it Nucl. Phys.} {\bf B261} (1985) 379;
Y.-P.~Yao and C.~Yuan, {\it Phys. Rev.} {\bf D38} (1988) 2237;
J.~Bagger and C. Schmidt, {\it Phys. Rev.} {\bf D41} (1990) 264;
H.~Veltman, {\it Phys. Rev.} {\bf D41} (1990) 2294.}

\bibitem{sdsw}{S.~Dawson and S.~Willenbrock, {\it Phys.
Rev.} {\bf D40} (1989); M.~Veltman and F.~Yndu
rain,
{\it Nucl. Phys.} {\bf B163} (1979) 402.}



\bibitem{chan}{M.~Chanowitz and M.~Gaillard, {\it Nucl.
Phys.} {\bf B261} (1985) 379.}


\bibitem{cgg}{M.~Chanowitz, H.~Georgi, and M.~Golden,
{\it Phys. Rev. Lett.} {\bf 57} (1986) 2344;
{\it Phys. Rev.}{\bf D36} (1987) 1490.}

\bibitem{leut} {J.~Gasser and H.~Leutwyler, {\it Ann. Phys.}
{\bf 158} (1984) 142; {\it Nucl. Phys.} {\bf B250} (1985) 465.}


\bibitem{wb}{S.~Weinberg, {\it Phys. Rev. Lett.} {\bf 17} (1966) 616.}
\bibitem{dvlep}{S. Dawson and G.~Valencia, {\it Nucl. Phys. } {\bf B439}
(1995) 3.}

\bibitem{dawsz}{S.~Alam, S.~Dawson and R.~Szalapski, {\it Phys. Rev.}
{\bf D57} (1998) 1577.}
      
\bibitem{sz}{R.~Szalapski, {\it Phys. Rev.} {\bf D57}
(1998) 5519; A. De Rujula, M.~Gavela, P.~Hernandez, and E.~Masso,
{\it Nucl. Phys.} {\bf B384} (1992)3.}

\bibitem{pt}{M.~Peskin and T.~Takeuchi, {\it Phys. Rev. Lett.}
{\bf 65} (1990) 964; D.~Kennedy and B.~Lynn, {\it Nucl.
Phys.} {\bf B322} (1989) 1.}


\bibitem{hpzh87}{K.~Hagiwara, S. Ishihara, R.~Szalapski,
	and D.~Zeppenfeld, {\it Phys. Rev. } {\bf D48} (1993) 2182.}

\bibitem{fls}{A.~Falk, M.~Luke, and E.~Simmons, {\it Nucl. Phys.}
{\bf B365} (1991) 523.}


\bibitem{bdv}{J.~Bagger, S.~Dawson, and G.~Valencia,
{\it Nucl. Phys.} {\bf B399} (1993) 364.}

\bibitem{baghan} {J.~Bagger {\it et.al.}, {\it Phys. Rev.}
{\bf D49} (1994) 1246; R.~Chivukula, M.~Dugan, and M.~Golden,
{\it Ann. Rev. Nucl. Part. Sci.} {\bf 45} (1995) 255.}

\bibitem{drv}{J.~Donoghue, C.~Ramirez, and G.~Valencia, {\it Phys.
Rev. }{\bf D38} (1988) 2195; {\it Phys. Rev.} {\bf D39} (1989) 1947;
 M.~Herrero and E.~Morales,
{\it Nucl. Phys.} {\bf B418} (1993) 364.}

\bibitem{apc} {T.~Appelquist and J.~Carrazone, {\it Phys.
Rev.} {\bf D11} (1975) 2856.}

\bibitem{hami} {See lectures in this school by N. Arkani-Hamed.}

\bibitem{hk}{For a review of SUSY phenomenology, see G.~Kane and H.~Haber,
{\it Phys. Rep. }{\bf 117C} (1985) 75;
H.~Haber, TASI 1992 (World Scientific, Singapore, 1992).}

\bibitem{wess}{J.~Bagger and J.~Wess, {\it Supersymmetry and 
Supergravity} (Princeton University Press, 1983).} 

\bibitem{hks}{G.~Kane, H.~Haber, and T.~Stirling, {\it Nucl. Phys.}
{\bf B161} (1979) 493.}


\bibitem{susyrad}{H.~Haber and R.~Hempfling, {\it Phys. Rev. Lett.}
{\bf 66} (1991) 1815;  J.~Ellis, G.~Ridolfi and F.~Zwirner,
{\it Phys. Lett.} {\bf B257} (1991) 83;
M.~Berger, {\it Phys. Rev.} {\bf D41} (1990) 225;
Y.~Okada, M.~Yamaguchi, and T. ~Yanagida, {\it Prog. Theor.
Phys. Lett.} {\bf 85} (1991)1; M. Carena, M. Quiros, and C. Wagner,
{\it Nucl. Phys.} {\bf B461} (1996) 407.}

\bibitem{quiros}{G.~Kane, C.~Kolda,M.~Quiros, and J.~ Wells,
{\it Phys. Rev. Lett.} {\bf 70} (1993) 2686.} 

\bibitem{tevlims}{M.~Carena, S.~Mrenna, and C.~Wagner, ANL-HEP-PR-98-54, 
hep-ph/9808312.} 

\bibitem{dj}{A. Djouadi, report of the MSSM working group for the
Workshop on GDR-Supersymmetry, hep-ph/9901246; Daniel Treille,
results  presented
at {\it XXIX International Conference on High Energy Physics},
Vancouver, Canada, July 23-29, 1998.} 

\bibitem{snowhiggs}{H.~Haber, {\it et. al.}, in {\it
Proceedings of the 1996 DPF/DPB Summer Study on New Directions for
High Energy Physics}, 1996.}   
\end{thebibliography}
\end{document}